\documentclass[pdflatex,sn-mathphys-num]{sn-jnl}

\usepackage{graphicx}%
\usepackage{graphics}%
\usepackage{multirow}%
\usepackage{amsmath,amssymb,amsfonts}%
\usepackage{mathrsfs}%
\usepackage{xcolor}%
\usepackage{listings}%
\usepackage{subfig}%
\usepackage{natbib}%
\usepackage{epstopdf}%
\usepackage{anyfontsize}%
\usepackage{enumitem}%
\usepackage{ulem}%

\begin{document}

\title[Article Title]{Fractal Scaling of Moffatt Vortices in Triangular Cavity Flow}

\author[1]{\fnm{Rathindra Nath} \sur{Basak}}\email{rnbasak10@gmail.com}
\equalcont{These authors contributed equally to this work.}

\author*[1]{\fnm{Sougata} \sur{Biswas}}\email{sougata.biswas@visva-bharati.ac.in}
\equalcont{These authors contributed equally to this work.}

\author[2]{\fnm{Jiten C.} \sur{Kalita}}\email{jiten@iitg.ac.in}
\equalcont{These authors contributed equally to this work.}

\affil[1]{\orgdiv{Department of Mathematics}, \orgname{Visva-Bharati (A Central University)}, \orgaddress{\street{Siksha Bhavana}, \city{Santiniketan}, \postcode{731235}, \state{West Bengal}, \country{India}}}

\affil[2]{\orgdiv{Department of Mathematics}, \orgname{Indian Institute of Technology Guwahati}, \orgaddress{\city{Guwahati}, \postcode{781039}, \state{Assam}, \country{India}}}

\abstract{This study examines the formation, quantification, and fractal characterization of corner vortices in slow viscous incompressible flow within a triangular cavity. The governing Navier--Stokes equations are solved numerically using a pressure-based coupled solver, and the resulting vortex cascade is analyzed through the size and intensity ratios of successive eddies in the spirit of Moffatt’s theory of corner vortices. The fractal properties of the vortex sequence are then investigated using the area--perimeter method. An empirical relation is proposed to estimate the fractal dimension of any successive vortex in the cascade for arbitrary grid resolution. The results demonstrate that the corner vortices possess non-integer fractal dimensions between 1 and 2, and that this dimension is systematically linked to vortex size and intensity. The influence of Reynolds number on the fractal scaling is also examined. Finally, a comparative analysis of self-similarity in triangular and square cavities confirms that the observed corner-vortex cascade exhibits robust fractal behavior across geometries and flow regimes.}

\keywords{Triangular cavity flow; Corner vortices; Area-perimeter method; Fractal dimension; Self-similarity; Vorticity distribution}



\maketitle

\section{Introduction}
Over the last few decades, the formation, evolution, characterization, identification, and dynamics of vortices in viscous incompressible internal fluid flows \cite{bruneau20062d,burggraf1966analytical,ghia1982high,jeong1995identification} have received considerable interest amongst the scientific community. Lid-driven square or rectangular cavities (both two- and three-dimensional) represent some of the simplest geometries for studying vortex formation and their characteristics in incompressible viscous flows. Owing to the lid motion and flow disturbances, many vortical structures such as primary and secondary vortices, corner vortices, Taylor–Görtler–like (TGL) vortices, mushroom-shaped vortices, U-shaped vortices, and ring-like vortices \cite{koseff1984lid,freitas1985numerical,chiang1996finite,romano2020lagrangian,babor2023lagrangian,basak2025formation,basak2026formation} are observed in these cavities. Irregular shaped cavities, such as triangular cavities, also serve as primary configurations for studying flow characteristics and vortex dynamics, and they find numerous engineering applications \cite{japar2020effect,javed2015mhd,cherif2022hydrothermal}. The flow in a triangular cavity exhibits distinct features that Moffatt \cite{moffatt1964viscous} first studied analytically in the Stokes regime. In particular, creeping flows (slow flows) in the neighborhood of the corner of the solid boundary are of fundamental interest. Such flows are referred to as \lq\lq Stokes flow\rq\rq\hspace{0.025cm} in the existing literature \cite{papanastasiou2021viscous,white2006viscous}. The governing equations for Stokes flow are the linear approximation of the non-linear N-S equations, wherein the non-linear inertia term is negligible compared to the linear viscous term. In other words, Stokes flow can be thought of as a fluid flow occurring at very low Reynolds numbers ($Re \to 0$) \cite{deville2022introduction}. Such flows are crucial for analyzing the behavior of highly viscous fluids.
\par
The existence of a so-called infinite sequence of vortices formed in Stokes flow between two rigid boundaries meeting at a corner dates back to the pioneering work of Dean and Montagnon \cite{dean1949steady} and was later theoretically established by Moffatt \cite{moffatt1964viscous,moffatt1964viscous2}. A flow visualization experiment by
Taneda \cite{taneda1979visualization} in a V-notch confirmed the existence of these vortices, which are now commonly referred to as \lq\lq Moffatt vortices\rq\rq\hspace{0.025cm} by the fluid dynamicists. The strength and size of each successive vortex diminish rapidly as
one moves towards the corner. Typically, the primary vortex is generated by a moving boundary (such as a rotating cylinder or a sliding lid) close to the rigid walls; this primary vortex, in turn, induces the next smaller vortex, and the process continues iteratively. The direction of rotation of each induced vortex is opposite to that of its predecessor, and the dividing streamlines between them are concave outwards.
\par
Of late, there has been considerable research on the existence of Moffatt vortices in slow, viscous, incompressible flows within various geometries \cite{kirkinis2014moffatt,malhotra2005nested,malyuga2005viscous,shankar2005moffatt,shtern2014moffatt}, established primarily through theoretical studies using eigenvalue analysis. These theoretical investigations are based on the solution of the biharmonic form of the N-S equations for Stokes flow \cite{papanastasiou2021viscous},
     \begin{equation}
         \Delta^2 \psi = 0
         \label{EA}
     \end{equation}
in plane polar co-ordinates $(r, \theta)$. Here, $\psi$ is the stream function and $\Delta^2$ is the biharmonic operator. The solution is assumed to take the form $\psi = r^{\nu} f_{\nu}(\theta)$, where $\nu$ is any real or complex number, referred to as the ``exponent" of the corresponding solution. This assumption transforms the governing equation (\ref{EA}) into an eigenvalue problem \cite{shtern2014moffatt,papanastasiou2021viscous}:
     \begin{equation}
         f^{(iv)} + \{(\nu + 1)^2 + (\nu - 1)^2\} f^{''} + (\nu^2 - 1)^2 f = 0.
         \label{EB}
     \end{equation}
This linear, homogeneous, fourth-order ordinary differential equation yields a solution of the form \cite{moffatt1964viscous,moffatt1964viscous2,shtern2014moffatt,papanastasiou2021viscous}:
     \begin{equation}
         f(\theta) = A \hspace{0.1cm } \sin(\nu - 1) \theta + B \hspace{0.1cm } \cos(\nu - 1) \theta + C \hspace{0.1cm } \sin(\nu + 1) \theta + D \hspace{0.1cm } \cos(\nu + 1) \theta.
         \label{EC}
     \end{equation}
It was found that $\nu$ becomes a complex number when the angle between the two planes is sufficiently acute, implying infinite oscillations, i.e., an infinite sequence of counter-rotating vortices as the corner is approached \cite{moffatt1964viscous,moffatt1964viscous2}. Nonetheless, very few numerical studies on this topic are available in the literature \cite{biswas2004backward,heaton2008appearance,kirkinis2022odd, biswas2016moffatt,biswas2018moffatt}. While recent studies have also explored Moffatt vortices in screw extrusion and electrohydrodynamic flows \cite{polychronopoulos2018computer,naumov2020velocity,he2022moffatt,taylor2022viscoplastic}, Kalita et al. \cite{kalita2018finiteness} utilized topological theorems to provide a rigorous proof establishing that the sequence of these Moffatt vortices cannot be infinite; rather, they are at most finite at the corners. Motivated by these competing theoretical perspectives, we attempt to numerically quantify these vortices in a triangular cavity by computing their sizes and intensities.
\par
Building on this quantification, we further investigate the fractal nature of Moffatt vortices within the triangular cavity. Traditionally, Euclidean geometry characterizes regular objects such as lines, planes, and spheres using integer topological dimensions. However, determining the dimension of highly complex or irregular structures is inherently challenging, as they defy these standard integer classifications. This limitation motivated the development of fractal geometry as a framework for describing and analyzing complex, irregular structures. A ``fractal" is a complex geometric structure that exhibits self-similarity across different scales. Classic examples of fractal include the Sierpi\'{n}ski triangle, Koch curves, and natural coastlines. Since the complexity of these objects cannot be captured by standard integer values, they are characterized by a non-integer or fractional metric known as the fractal dimension. This dimension provides a precise mathematical description of an object's irregularity and complexity. For planar curves and boundaries, this fractal dimension typically lies between $1$ and $2$; a higher value indicates a greater degree of complexity, representing a state where the curve tends to fill the two-dimensional plane \cite{mandelbrot1983fractal}.
\par
In the context of natural phenomena, the fractal dimension serves as a robust tool for characterizing complex sizes and geometries \cite{mandelbrot1983fractal}. Several classical methods exist for computing this dimension, primarily the self-similarity and box-counting techniques. Additionally, empirical approaches such as the structural function, the root-mean-square and the area-perimeter methods have proven highly effective. Notably, Lovejoy \cite{lovejoy1982area} pioneered the use of the area-perimeter relation to demonstrate the fractal nature of clouds in the Earth’s atmosphere. Subsequent studies have successfully applied this method to various atmospheric structures; for instance, Rys and Waldvogel \cite{rys1986fractal} determined similar dimensions for hail clouds, Henderson-Sellers \cite{henderson1986martian} analyzed cloud formations in the Mars atmosphere, and Bazell and Desert \cite{bazell1988fractal} investigated the fractal structure of infrared cirrus clouds.
\par
Much like the atmospheric clouds, vortical structures in nature and engineering, ranging from tornadoes and whirlpools to the turbulent wakes behind aircraft wings exhibit highly complex and intricate boundaries. While the role of fractals in various macroscopic fluid flows has been extensively studied over the last few decades \cite{turcotte1988fractals,constantin1985determining,sreenivasan1991fractals,ueki1999fractal,mazzi2004fractal,balankin2012map,lanotte2016vortex,el2022fractal}, quantifying the fractal aspects of specific vortical structures using the area-perimeter method remains a relatively unexplored area within the fluid dynamics community. 
Although the area-perimeter method has been employed in recent years to investigate the fractal properties of helical vortex ropes in hydraulic turbines by Li et al. \cite{li2022temporal} and K\'{a}rm\'{a}n vortices behind a NACA0009 hydrofoil by Zhang et al. \cite{zhang2023investigation}, its application has remained strictly confined to these specific configurations. To the best of our knowledge, the fractal characteristics of Moffatt-type corner vortices have not yet been investigated using this approach to uncover the physical insights associated with their fractal dimensions.

\par
The primary objective of the current study is to quantify Moffatt-type corner vortices in a triangular cavity by simulating low Reynolds number flows. While these structures have been studied analytically, rigorous numerical characterization utilizing a pressure-based coupled solver remains largely unexplored for this specific triangular geometry. We validate our approach by comparing our computational results both qualitatively and quantitatively with established experimental \cite{taneda1979visualization} and theoretical findings \cite{moffatt1964viscous,moffatt1964viscous2}. Building upon this, we explore the fractal characteristics of the corner vortices utilizing the area-perimeter method \cite{lovejoy1982area,zhang2023investigation}. Our analysis reveals that these vortices are inherently self-similar. By determining the approximate fractal dimension of each corner vortex, our computations demonstrate that this dimension correlates with changes in vortex shape and intensity, ultimately aiding in the identification of the exact number of vortices present at low Reynolds numbers. We further analyze the effect of increasing Reynolds numbers ($Re$) on the fractal nature of these structures. Finally, to demonstrate the robustness of our findings, we present a comparative study of self-similarity in the vortex sequences between triangular and square cavities, corroborating the conclusion that corner vortices possess a distinct fractal nature across different geometries.

\section{Problem description and numerical methods}
\begin{figure}[!ht]
    \centering
    \includegraphics[width=5cm]{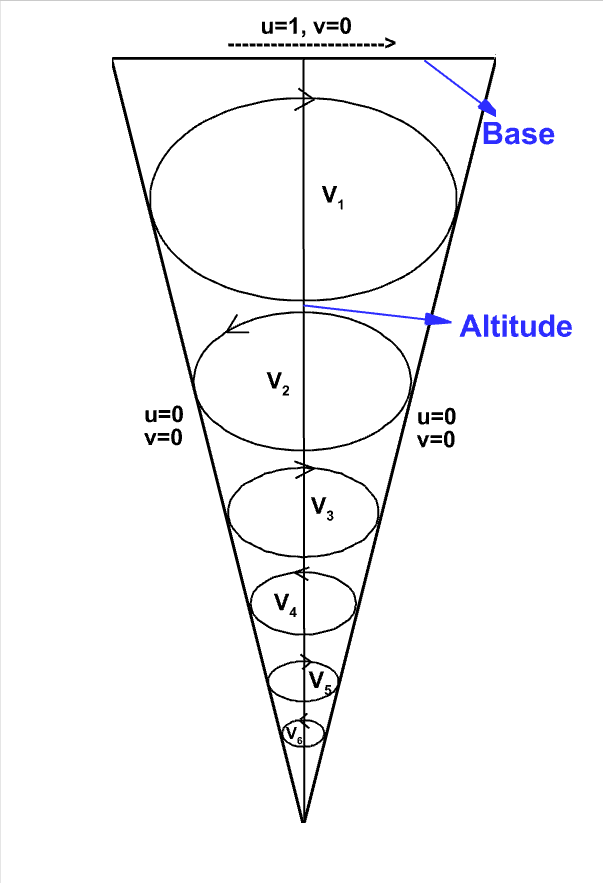}
    \caption{Schematic of flow configuration in the driven triangular cavity.}
    \label{Fig0}
\end{figure}

The present study considers the steady, two-dimensional, incompressible, and viscous flow inside an isosceles triangular cavity with an altitude twice its base, as shown in Fig. \ref{Fig0}.
\par
Let $L$ denote the length of the base of the cavity and $U$ be the uniform velocity of the moving top lid. The flow inside the triangular cavity is governed by the non-dimensional, primitive variable form of the Navier-Stokes (N-S) equations (neglecting gravitational effects):
     \begin{equation}
         \frac{\partial u}{\partial x} + \frac{\partial v}{\partial y} = 0,
         \label{Equa}
     \end{equation}
     \begin{equation}
        u \frac{\partial u}{\partial x} + v \frac{\partial u}{\partial y} = - \frac{\partial p}{\partial x} + \frac{1}{Re} \nabla^2 u,
        \label{Equb}
     \end{equation}
     \begin{equation}
         u \frac{\partial v}{\partial x} + v \frac{\partial v}{\partial y} = - \frac{\partial p}{\partial y} + \frac{1}{Re} \nabla^2 v,
         \label{Equc}
     \end{equation}
where $u$, $v$ are the velocity components along the $x$, $y$-directions respectively, $p$ is the pressure, $\displaystyle Re = \frac{UL}{\nu}$ is the Reynolds number with $\nu$ being the kinematic viscosity of the fluid.

\par
In the triangular cavity under consideration (refer to Fig. \ref{Fig0}), a non-dimensional horizontal velocity of $u =U= 1$ is prescribed at the moving top lid, while the vertical velocity is zero ($v = 0$). For the stationary inclined walls, the standard no-slip conditions ($u = v = 0$) are imposed. The shear stress generated by the uniform motion of the top lid drives the adjacent fluid, transferring momentum deep into the interior of the cavity. As depicted in Fig. \ref{Fig0}, this mechanism generates a distinct flow structure characterized by a dominant primary circulation cell near the top lid. This primary vortex kinematically drives a cascade of progressively smaller and weaker secondary and tertiary eddies that propagate downward, with the sequence continuing until the viscous effects completely dissipate the motion near the bottom corner.
\begin{figure}[!ht]
    \centering
    \includegraphics[width=5.3cm]{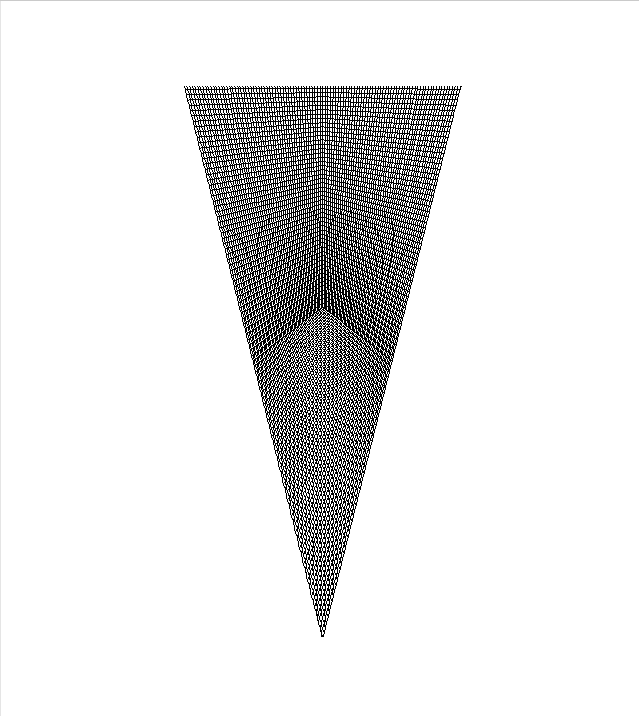}
    \caption{Layout of the uniform quadrilateral grid employed for the spatial discretization of the triangular cavity.}
    \label{FigD}
\end{figure}

The numerical simulations were performed using the finite-volume-based commercial solver ANSYS Fluent (Release 21.1). Though Fluent inherently solves the governing equations in dimensional form, a non-dimensional framework is adopted through the appropriate selection of geometric and fluid properties. The characteristic length scale of the computational domain and the fluid density ($\rho$) were set to unity. Consequently, to satisfy the non-dimensional framework, the dynamic viscosity was defined as $\mu = \frac{1}{Re}$. The convective terms in equations (\ref{Equa})-(\ref{Equc}) were discretized using a second-order upwind scheme, while the diffusive terms were discretized using a central difference scheme. The pressure-velocity coupling was resolved using the coupled algorithm. All computations in the present study were executed on a workstation equipped with an Intel Core i7-3770 processor and 16 GB of RAM.

\section{Domain discretization and grid independence study}
\subsection{Domain discretization}
The governing equations (Eqs. \ref{Equa}-\ref{Equc}) were discretized using a uniform grid system as depicted in Fig. \ref{FigD}. 
The computational domain was divided into smaller quadrilateral elements, with systematic refinement applied to evaluate spatial convergence. Four distinct grid resolution levels were generated for the computational analysis. The details of these grid systems are summarized in Table \ref{Table6}. As the mesh is refined from the coarser to the finer grid, the element sizes decrease geometrically, resulting in a proportional increase in the number of divisions along the upper moving lid, as well as in the total number of nodes and elements.
\par

\begin{table}[!ht]
    \centering
    \begin{tabular}{ccccc} 
    \hline
    Grid & Element Size & No of Divisions in Upper Lid & Total Nodes & Total Elements \\
    \hline
    Coarser & 0.04 & 52 & 2107 & 2028 \\
    Medium1 & 0.02 & 104 & 8269 & 8112 \\
    Medium2 & 0.01 & 206 & 32137 & 31827 \\
    Finer & 0.005 & 412 & 127927 & 127308 \\
    \hline
    \end{tabular}
    \vspace{0.25cm}
    \caption{Total number of nodes and elements for the four grid resolutions evaluated.}
    \label{Table6}
\end{table}
\par
\subsection{Grid independence study}
\begin{figure}[!ht]
    \centering
    \includegraphics[width=5.5cm]{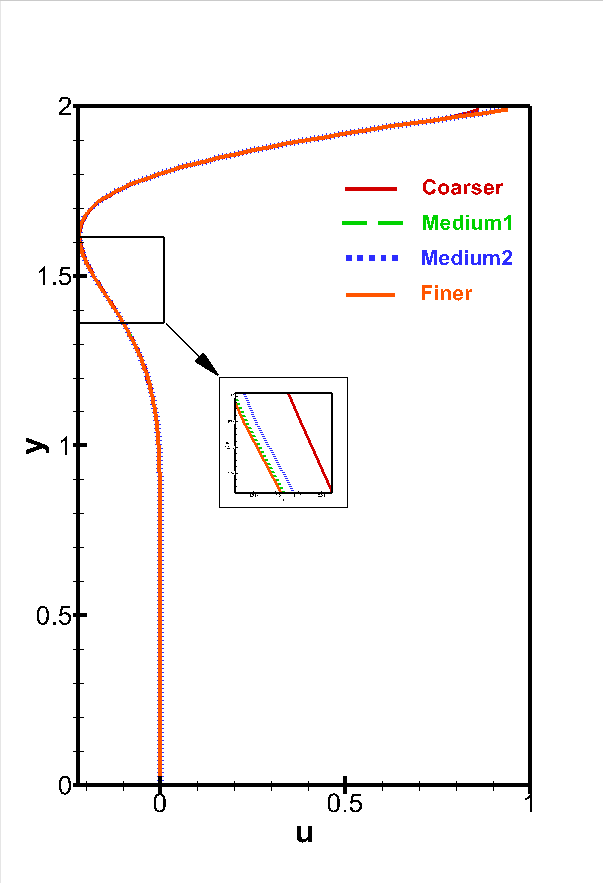}
    \caption{Comparison of $u$-velocity profiles along the vertical centerline ($x = 0.5$) for four different grid systems at $Re = 1$.}
    \label{Fig5}
\end{figure}

To establish grid independence, the horizontal velocity components ($u$) along the vertical centerline of the cavity ($x = 0.5$) were plotted for all four grids, as depicted in Fig. \ref{Fig5}. The resulting velocity profiles demonstrate excellent agreement across the different mesh resolutions. Furthermore, a quantitative comparison of the spatial coordinates of the vortex centers for each vortex is provided in Table \ref{Table5}, further confirming the spatial convergence of the numerical solution. Consequently, to ensure maximum physical accuracy, only the results obtained from the finer grid are utilized for the subsequent flow quantification and vortex dynamics analysis. However, to rigorously evaluate the grid-scaling effects on the fractal characteristics of these corner vortices, data from all four spatial grid systems are retained and analyzed in the later sections of this study.

\begin{table}[!ht]
    \centering
    \begin{tabular}{ccccc} 
    \hline
    Vortex Center & Coarser & Medium1 & Medium2 & Finer \\
    \hline
    $V_1$ & (0.5000, 1.8006) & (0.5000, 1.8019) & (0.5000, 1.8016) & (0.5000, 1.8017) \\
    $V_2$ & (0.5000, 0.9039) & (0.5000, 0.9052) & (0.5000, 0.9053) & (0.5000, 0.9052) \\
    $V_3$ & (0.5000, 0.4359) & (0.5000, 0.4496) & (0.5000, 0.4487) & (0.5000, 0.4502) \\
    $V_4$ & (0.5000, 0.1935) & (0.5000, 0.2218) & (0.5000, 0.2136) & (0.5000, 0.2236) \\
    $V_5$ & (0.5000, 0.0736) & (0.5000, 0.1031) & (0.5000, 0.0916) & (0.5000, 0.1099) \\
    $V_6$ & - & - & (0.5000, 0.0404) & (0.5000, 0.0521) \\
    $V_7$ & - & - & - & (0.5000, 0.0316) \\
    \hline
    \end{tabular}
    \vspace{0.25cm}
    \caption{Comparison of vortex center coordinates across the four spatial grid systems at $Re = 1$.}
    \label{Table5}
\end{table}

\newpage
\section{Results and Discussions}
The following sections discuss the quantification of corner vortices as Moffatt eddies and their fractal characteristics in a triangular cavity flow.

\subsection{Quantification of corner vortices}
A thorough investigation of the literature reveals that the studies of Moffatt eddies are mainly limited to the Stokes flow regime. Keeping this in mind, we choose $Re$ to be of value $1$ in our numerical computation.

\begin{figure}[!ht]
    \centering
    \includegraphics[width=7.4cm]{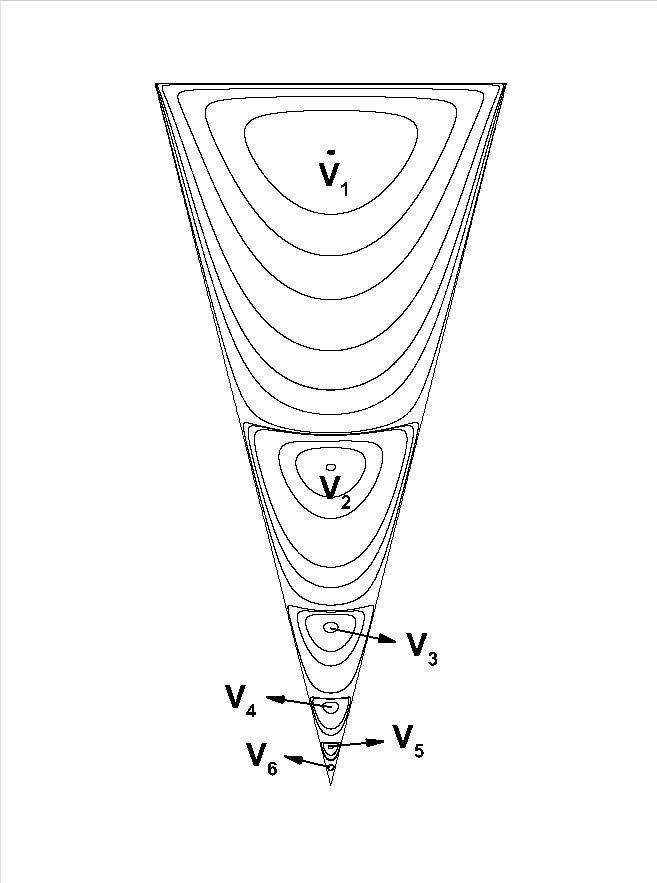}
    \caption{Streamline patterns of the corner vortices in the triangular cavity at $Re = 1$.}
    \label{FigA}
\end{figure}

In Fig. \ref{FigA}, we depict corner vortices in the triangular cavity by plotting the streamlines. These eddies appear in a sequential stack, cascading in a self-similar fashion as they approach the corner tip. To characterize the sequence approaching the bottom stationary corner, we adopt the nomenclature $V_1, V_2, V_3, \cdots V_n$ for vortices in succession. 

\begin{figure}[!ht]
	\centering
	\begin{tabular}{cc}
		{\includegraphics[width=4.5cm]{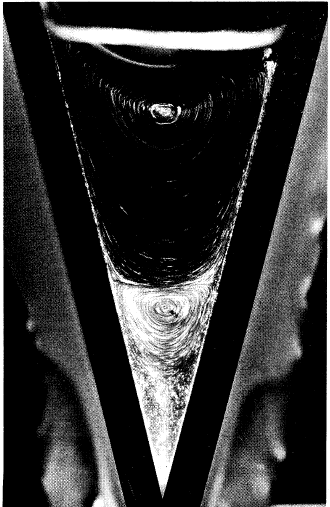}}
		&
		\hspace{.4cm}
		{\includegraphics[width=5.7cm]{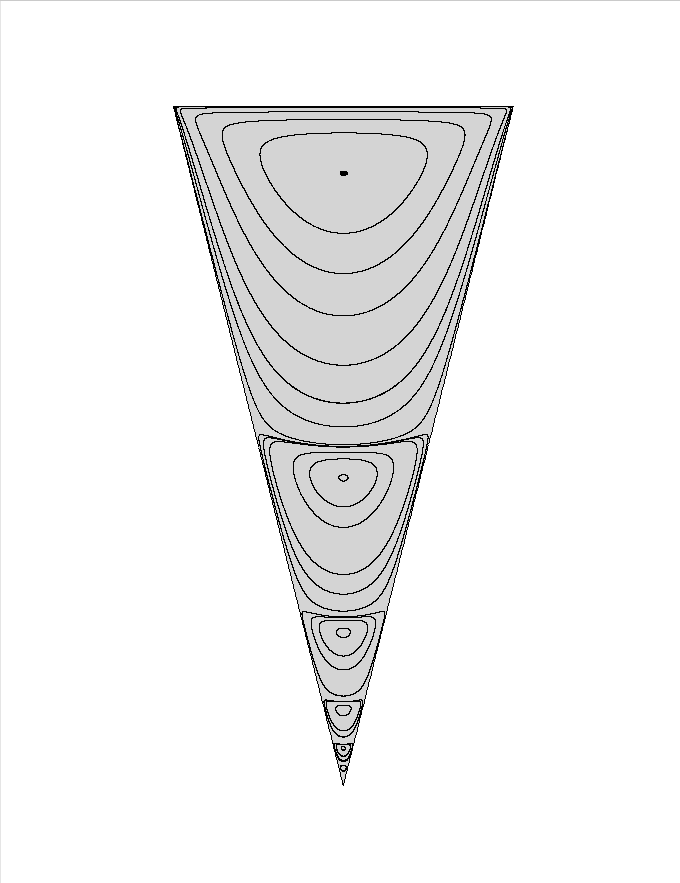}}
		\vspace{.4cm}
		\\ \hspace{1cm} (a) & (b) \
	\end{tabular}
\caption{ Comparison of the corner vortex structures: (a) experimental flow visualization by Taneda \cite{taneda1979visualization}, and (b) streamline contours from the present numerical simulation at $Re = 1$.}
\label{FigB}
\end{figure}

Through our computation, we are able to trace seven members of the sequence of corner eddies. In figure \ref{FigB}, we  compare our results with the only experimental result available in the literature, namely, by Taneda \cite{taneda1979visualization}. As can be seen from the figure, they are excellent match.

\par
Table \ref{Table1} presents the center coordinates, intensity, and characteristic size for each member of the corner vortex sequence. Eddy intensity is quantified by the stream function ($\psi$) value at its core, while its size is characterized by the vertical distance ($r$) from the vortex center to the cavity tip. Both the size and intensity of these eddies decrease rapidly as they approach the corner, following a distinct geometric progression. Based on the present computations, the asymptotic common ratios for size and intensity are determined to be:
     \begin{align}
          \frac{r_{n+1}}{r_n} \approx 0.49 \hspace{0.15cm} \text{and} \hspace{0.15cm}  \frac{\psi_{n+1}}{\psi_n}  \approx 0.0012,
     \end{align}
where, $n = 1, 2, \cdots, 5$ (for details refer to Table \ref{Table2}). These computed ratios are in excellent agreement with the theoretical predictions of Moffatt \cite{moffatt1964viscous,moffatt1964viscous2}. Furthermore, despite the fine grid resolution, the seventh vortex ($V_7$) is only marginally captured and remains incompletely developed within the computational domain. The extremely small magnitude of the stream function for this final eddy indicates that further grid refinement would be required to fully resolve its structure.

\begin{table}[!ht]
\centering
\begin{tabular}{cccc} 
 \hline
 \textbf{Vortex} & \textbf{Intensity ($\psi$)} & \textbf{Center Location} & \textbf{Size ($r$)} \\ 
 \hline
  $V_1$ & $-8.478 \times 10^{-2}$ & (0.5000, 1.8017) & 1.8017 \\
  $V_2$ & $1.116 \times 10^{-4}$ & (0.5000, 0.9052) & 0.9052 \\
  $V_3$ & $-1.362 \times 10^{-7}$ & (0.5000, 0.4502) & 0.4502 \\
  $V_4$ & $1.669 \times 10^{-10}$ & (0.5000, 0.2236) & 0.2236 \\
  $V_5$ & $-2.07 \times 10^{-13}$ & (0.5000, 0.1099) & 0.1099 \\ 
  $V_6$ & $2.675 \times 10^{-16}$ & (0.5000, 0.0521) & 0.0521 \\ 
  $V_7$ & - & (0.5000, 0.0316) & 0.0316 \\
 \hline
\end{tabular}
\caption{Characteristics of the successive corner vortices in the triangular cavity at $Re = 1$.}
\label{Table1}
\end{table}

\begin{table}[!ht]
\centering
\begin{tabular}{ccc} 
 \hline
 \textbf{Eddy/Vortex Ratio} & \textbf{Intensity} & \textbf{Size} \\ 
 \hline
  $V_2:V_1$ & 0.001316 & 0.5024 \\
  $V_3:V_2$ & 0.001221 & 0.4973 \\
  $V_4:V_3$ & 0.001224 & 0.4967 \\
  $V_5:V_4$ & 0.001240 & 0.4915 \\
  $V_6:V_5$ & 0.001292 & 0.4734 \\
 \hline
\end{tabular}
\caption{Asymptotic geometric and intensity ratios of successive corner vortices within the triangular cavity at $Re = 1$.}
\label{Table2}
\end{table}

\par
\begin{figure}[!ht]
    \centering
    \includegraphics[width=5.8cm]{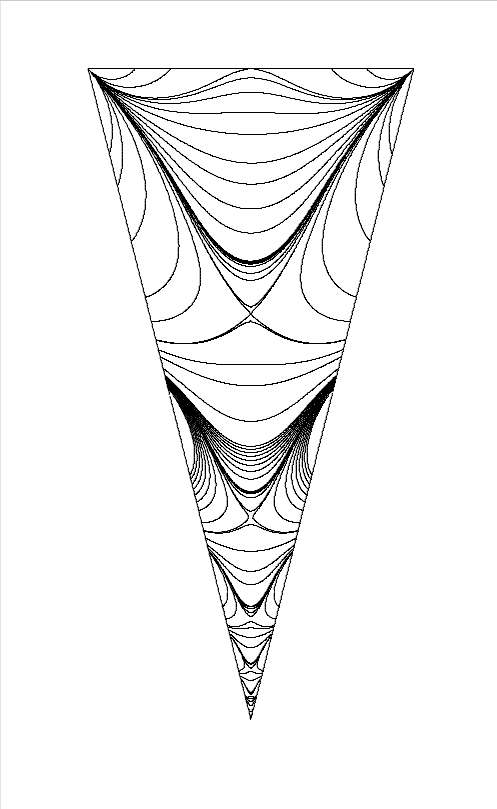}
    \caption{Vorticity distribution of corner vortices within a triangular cavity at $Re=1$.}
    \label{FigC}
\end{figure}

Figure \ref{FigC} displays the constant vorticity contours for $Re = 1$. As depicted, the vorticity field within the cavity exhibits clear symmetry about the vertical centerline. Notably, the characteristics of both the eddy sequence and the resulting vorticity distribution remain entirely consistent with the analytical predictions of Moffatt \cite{moffatt1964viscous,moffatt1964viscous2}.

\subsection{Fractal nature of corner vortices}
In incompressible viscous flows, vortical structures play a pivotal role in shaping the structural dynamics and macroscopic behavior of the fluid. In corner flows, such as the triangular cavity driven by an upper lid, direct numerical simulation (DNS) reveals a nested sequence of Moffatt vortices. To study the fractal nature of this vortex sequence, we employ the area-perimeter method to determine its fractal dimension. While the outermost stagnation streamline of an individual vortex may appear geometrically simple, closely resembling a semi-ellipse, the fractal characteristics do not arise from a convoluted, continuous boundary. Rather, they emerge from the discrete, self-similar scaling of the nested eddies as they approach the corner. Therefore, we use the semi-elliptical representation as a geometric approximation to estimate the perimeter and area of successive vortex envelopes, allowing us to quantify the multiscale properties of the entire vortex cascade.

The fractal perimeter dimension, $D$, of any vortex, which is derived from the power-law relationship between its outer edge perimeter, $P$, and its area, $A$, given by the following equation \cite{zhang2023investigation}:
\begin{equation}
    P = k A^{\frac{D}{2}},
\end{equation}
where $k$ is a scaling constant. Assuming $k$ is negligible for pure scaling purposes,
the fractal perimeter dimension $D$ for each vortex can be expressed as \cite{zhang2023investigation}:
\begin{equation}
    D \approx 2 \frac{\log P}{\log A}.
    \label{Equ1}
\end{equation}
The fractal dimension serves as a robust metric to describe the correlation between large-scale and small-scale self-similar vortices. According to previous statistical observations, larger macroscopic objects tend to exhibit a smaller fractal dimension than their smaller counterparts. In this study, we aim to characterize the relationship between vortical structures of varying scales and their inherent fractal nature within the triangular lid-driven cavity. Furthermore, we seek to statistically quantify the total number of vortices captured across different grid resolutions using this area-perimeter approach, subsequently validating these estimates against direct flow visualization.
\par
\begin{figure}[!ht]
	\centering
	\begin{tabular}{cc}
		{\includegraphics[width=6.2cm]{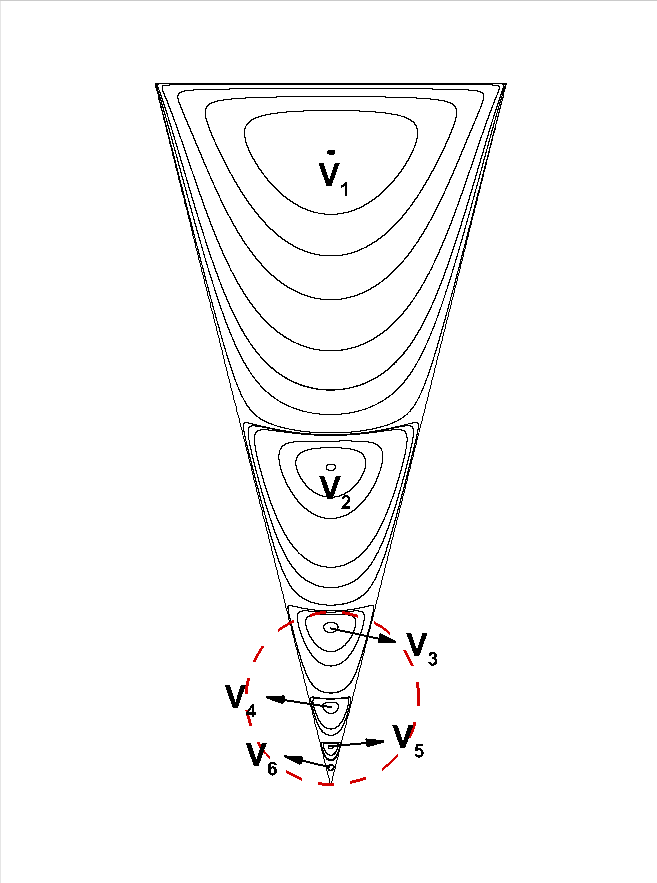}}
		&
		\hspace{.4cm}
		{\includegraphics[width=6.2cm]{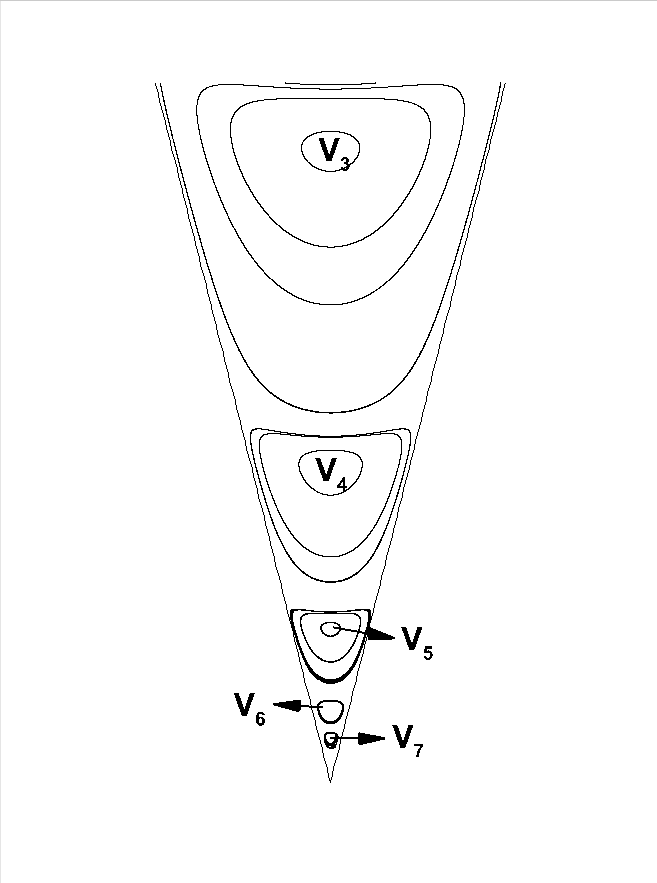}}
		\vspace{.4cm}
		\\ \hspace{1cm} (a) & (b) \
	\end{tabular}
\caption{Resolution of seven corner vortices in a triangular cavity using a fine grid mesh at $Re=1$.}
\label{Fig1}
\end{figure}

In the finer grid resolution with an element size of 0.005, a detailed examination of the cavity’s corner region reveals that the outermost streamline of each vortex is nearly semi-elliptical. A total of seven successive corner vortices are resolved extending toward the lower apex of the cavity, as depicted in Fig. \ref{Fig1}. These structures exhibit clear geometric self-similarity across varying scales of magnification. Consequently, to apply the aforementioned area-perimeter method for determining the fractal perimeter dimension, the specific perimeter and area of each individual vortex in this sequence must be quantified. Assuming that the outermost streamline of a vortex can be geometrically approximated as a semi-ellipse with a semi-major axis $a$ and a semi-minor axis $b$, its enclosed area is calculated as
\begin{equation}
    A = \displaystyle\frac{\pi}{2} ab.
    \label{Equ3}
\end{equation}

\par
Since the exact perimeter of a semi-ellipse cannot be expressed in closed form using elementary functions, we utilize Ramanujan's approximation \cite{ramanujan1914modular,moscato2024new} to estimate the arc length. By adding the length of the straight boundary, the highly accurate approximation for the total perimeter of the semi-ellipse is given by:
\begin{equation}
    P \approx \frac{\pi}{2} \Big[3(a+b) - \sqrt{(3a+b)(a+3b)}\Big] + 2b.
    \label{Equ4}
\end{equation}
\par
Our simulation results demonstrate that the vortex size decreases progressively by a fixed geometric ratio towards the lower corner. Because the simulated triangular cavity has an altitude twice its base, the semi-major and semi-minor axes of the vortices naturally follow the relation
\begin{equation}
    a = 2b.
    \label{Equ2}
\end{equation}  
Using this geometric constraint, we calculate the perimeter ($P$) and area ($A$) for each vortex in the sequence; these computed values are summarized in Table \ref{Table3}. 

\begin{figure}[!ht]
    \centering
    \includegraphics[width=8cm]{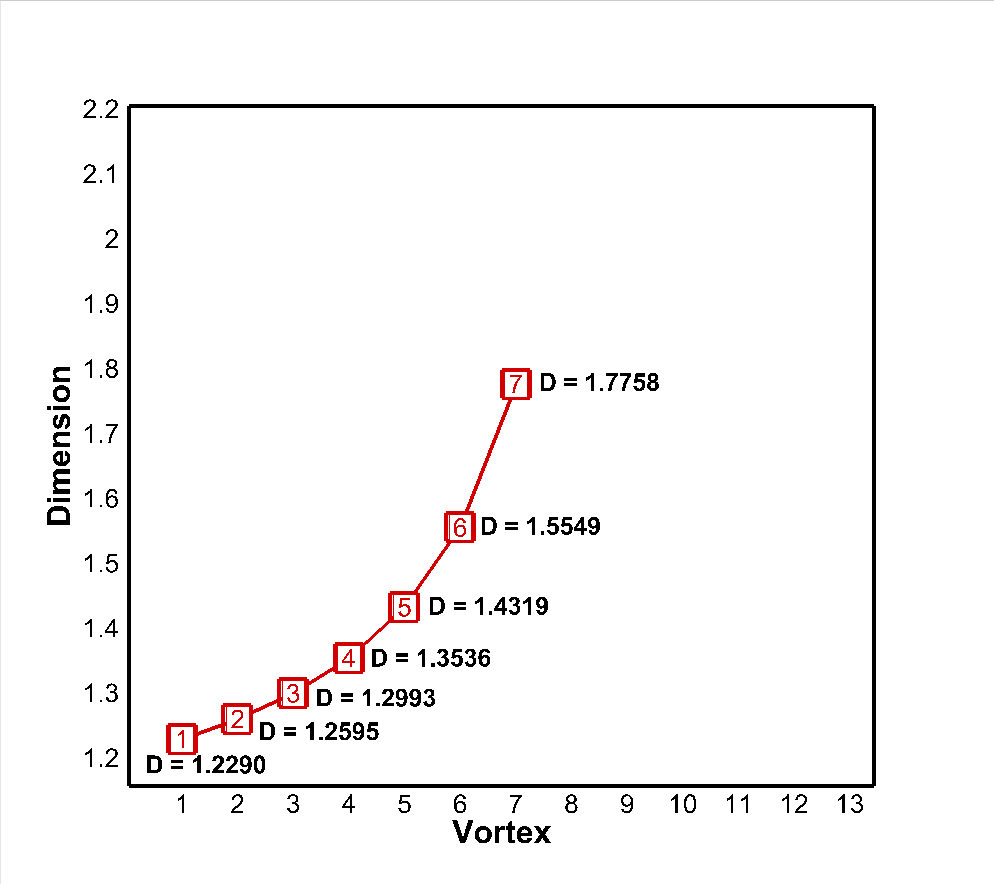}
    \caption{Fractal dimension of the sequence of corner vortices in a triangular cavity at $Re = 1$.}
    \label{Fig2}
\end{figure}

It should be noted that for these calculations, the local grid spacing ($1/412$) is normalized to unity. Consequently, the derived values for $a$, $b$, $P$, and $A$ are expressed in discrete grid units rather than physical dimensions. Because the upper lid is discretized into $412$ elements in this finest grid resolution, the semi-minor axis of the primary corner vortex ($V_1$) is initialized as $b = 206$. The spatial dimensions of all subsequent vortices diminish geometrically as they approach the corner. Finally, the scaling dimensions of the complete corner vortex cascade within our computed grid system are represented graphically in Fig. \ref{Fig2}.

\begin{table}[!ht]
\centering
\begin{tabular}{cccc} 
 \hline
 \textbf{Vortex} & \textbf{Perimeter} $(P)$ & \textbf{Area} $(A)$ & \textbf{Fractal Dimension} \\ 
 \hline
  $V_1$ & 1409.9074 & 133316.6258 & 1.2290 \\
  $V_2$ & 704.9536 & 33329.1564 & 1.2595 \\
  $V_3$ & 352.4768 & 8332.2891 & 1.2993 \\
  $V_4$ & 176.2384 & 2083.0723 & 1.3536 \\
  $V_5$ & 88.1192 & 520.7681 & 1.4319 \\
  $V_6$ & 44.0596 & 130.1920 & 1.5549 \\ 
  $V_7$ & 22.0298 & 32.5480 & 1.7758 \\
 \hline
\end{tabular}
\caption{Perimeter, area, and fractal dimension of corner vortices in a triangular cavity at $Re = 1$.}
\label{Table3}
\end{table}

\par

As shown in Table \ref{Table3}, the computed fractal perimeter dimension for each vortex lies between $1$ and $2$, confirming the fractal characteristics of the cascade generated by the lid movement. Notably, an inverse relationship exists between vortex size and its fractal dimension: the largest primary vortex exhibits the smallest dimension, whereas the smallest resolved vortex approaches a dimension of nearly $2$. This trend is driven by the severe spatial confinement as the cavity narrows toward the lower apex. As the available cross-sectional area shrinks, the sequential vortices are compressed into an increasingly restrictive geometry, limiting the sequence to a finite number of resolvable macroscopic vortices.

Furthermore, the perimeter and area of successive vortices maintain fixed reduction ratios of approximately $0.50$ and $0.25$, respectively, as they progress from the largest to the smallest structures. This scaling is strictly consistent with the vortex size ratios presented in Table \ref{Table2} and inherently follows from our semi-elliptical geometric approximation. By employing these scaling relations, we derive a generalized analytical expression for the fractal perimeter dimension of the $n$-th vortex in the sequence. The perimeter and area of the $n$-th vortex can be expressed in terms of the primary vortex properties as $P_n \approx P_1(0.5)^{n-1}$ and $A_n \approx A_1(0.25)^{n-1}$, respectively. Substituting these into equation (\ref{Equ1}), the fractal perimeter dimension for any successive vortex in the cascade can be dynamically estimated as:
\begin{equation}
    D(n) \approx 2 \bigg[\frac{\log P_1 - (n-1) \log2}{\log A_1 - 2(n-1) \log2}\bigg],
    \label{Equ5}
\end{equation}
where $n = 1, \dots, 7$ represents the sequential vortex index ($n = 1$ corresponding to the largest vortex, $V_1$, and $n = 7$ to the smallest resolved vortex, $V_7$), with $P_1$ and $A_1$ denoting the perimeter and area of the primary vortex, respectively. This proposed relation enables to find the fractal dimension of any successive vortex solely from the geometric scaling of the primary vortex $V_1$.

\par
To extend the proposed scaling law of area and perimeter beyond the mesh elements under consideration, we generalize this relationship across varying grid resolutions. Suppose the upper lid is discretized into $m$ elements rather than $412$. By introducing a grid-spacing normalization factor, $\Delta = 412/m$, it can be shown that the fractal perimeter dimension continues to strictly obey equation (\ref{Equ5}). Under this generalized discretization, the perimeter ($P_1$) and area ($A_1$) of the primary vortex $V_1$ scale according to the following exact relations:$$P_1 = m\Delta \left[ \frac{\pi}{4}(9 - \sqrt{35}) + 1 \right], \quad A_1 = \frac{\pi m^2 \Delta^2}{4},$$where these values represent the perimeter and area of the initial vortex ($V_1$) in the cascade, adjusted for the arbitrary grid resolution $m$.
\begin{figure}[!ht]
    \centering
    \includegraphics[width=7.5cm]{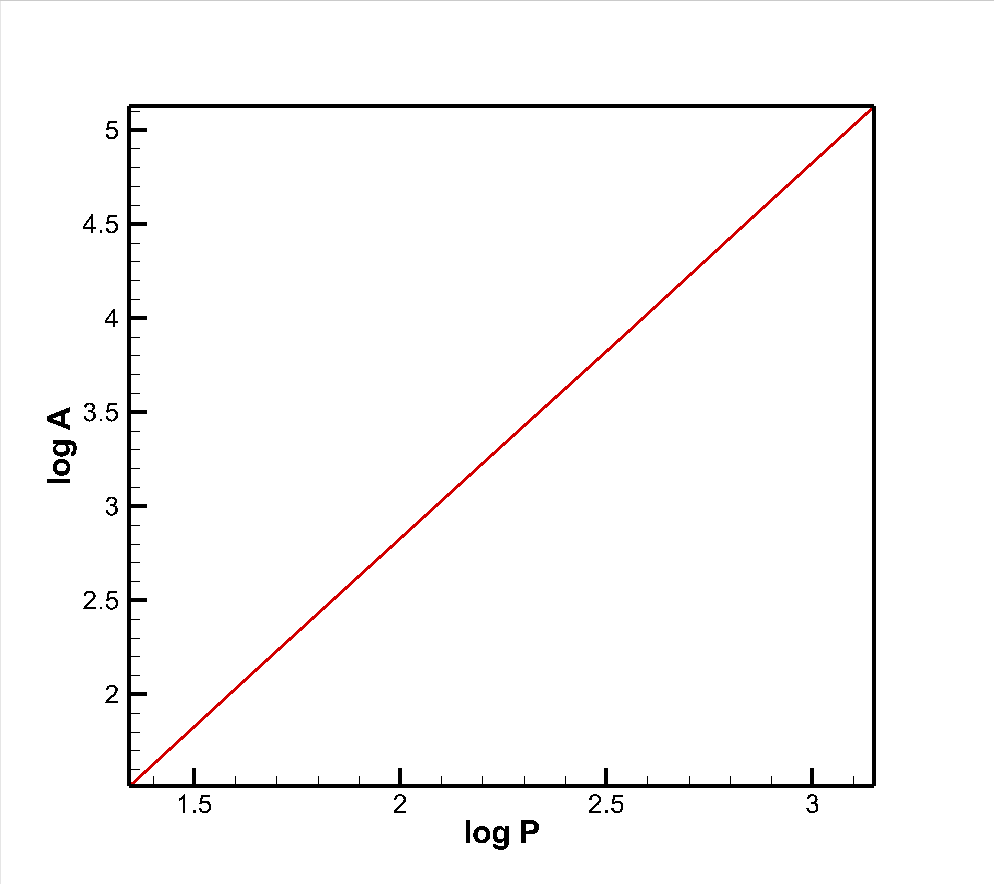}
    \caption{$\log-\log$ plot of perimeter (P) versus area (A) for the sequence of corner vortices in the triangular cavity at $Re=1$.}
    \label{Fig6}
\end{figure}

\par
Furthermore, by applying the area-perimeter method across the entire sequence of vortices, we establish a definitive linear relationship between $\log P$ and $\log A$. As illustrated in Fig. \ref{Fig6}, the extracted data points collapse tightly onto a linear trend line with a correlation coefficient of approximately 0.99. The slope of this line, which corresponds to $D/2$, yields a non-integer fractal dimension that deviates from the standard Euclidean topological dimension of $1$ for a simple closed curve. This non-integer dimension highlights the topological complexity of the flow near the corner, where the nested cascade maintains consistent geometric scaling despite a rapid spatial decay in size. Crucially, this geometric self-similarity indicates that the fundamental fluid dynamic mechanisms, specifically momentum diffusion and viscous dissipation remain qualitatively identical across scales, even as the intensity of the successive vortices diminishes by several orders of magnitude.
\par

Notably, the constant ratios of size ($r_{n+1}/r_{n}=\xi$) and intensity ($\psi_{n+1}/\psi_{n}=\eta$) presented in Table \ref{Table2} satisfy the self-similar relation $\psi(r_{n+1})=\eta\psi(r_{n+1}/{\xi})$, confirming the structural consistency of the flow field across the entire vortex cascade. Therefore, the characterization of these nested corner vortices as a fractal structure is rigorously justified by their inherent discrete scale-invariance.

In summary, the computed fractal dimension and inherent discrete scale-invariance for the vortex sequence $V_1, V_2 \cdots V_n$ confirm the stability of the Moffatt scaling regime within the triangular cavity. Ultimately, these findings effectively bridge the gap between classical fluid topology and fractal geometry.

\subsection{Influence of Reynolds number}
It is of interest to examine whether the self-similar and fractal characteristics of the flow persist beyond the Stokes flow regime. In this subsection, we consider two moderate Reynolds numbers, $Re = 100$ and $500$,  to investigate the extent to which the flow structures retain their geometric form and symmetry within triangular cavity configurations. Since invariance under dilation is a defining feature of fractal geometry, establishing self-similarity in this regime is essential. Such validation provides the basis for assessing whether the vortical structures continue to exhibit fractal behavior, which is a necessary condition for the applicability of the area–perimeter method.

\begin{figure*}[!ht]
\centering
    \subfloat[\centering{\label{Fig1a}}]{{\includegraphics[width=4cm]{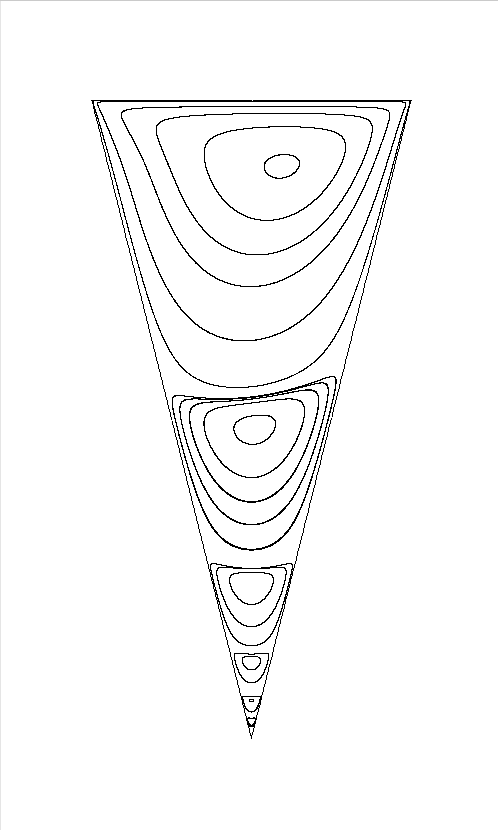} }}
    \qquad
    \subfloat[\centering{\label{Fig1b}}]{{\includegraphics[width=4.05cm]{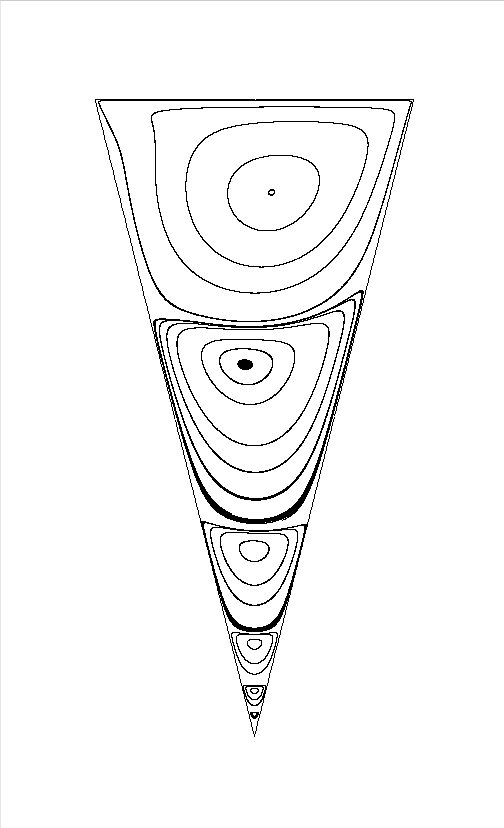} }}
\caption{Streamlines illustrating corner vortices in a triangular cavity at varying Reynolds numbers: (a) $Re = 100$ and (b) $Re = 500$.}
\label{FigE}
\end{figure*}

Figures \ref{FigA} and \ref{FigE} illustrate the evolution of streamlines within the triangular cavity for $Re = 1, 100,$ and $500$. In all cases, a sequence of vortices is observed, with their size progressively decreasing from the upper region of the cavity toward the bottom apex. At $Re=1$, the corner vortices exhibit a highly consistent and well-defined shape. As the Reynolds number increases to $Re=100$, the center of the primary vortex ($V_1$) shifts toward the right-hand altitude of the cavity. Despite this displacement, the outer boundaries of both $V_1$ and the secondary vortex $V_2$ retain a nearly semi-elliptical form, thereby supporting the applicability of the area–perimeter method for estimating their fractal dimensions.

At $Re=500$, the primary vortex contracts, whereas the secondary vortex expands and shifts upward toward the top wall. Notably, the higher-order vortices ($V_3, V_4$, etc.) preserve their characteristic shapes within the range of Reynolds numbers considered, indicating a persistent geometric structure across the regime.

\begin{figure*}[!ht]
\centering
    \subfloat[\centering{\label{figG1}}]{{\includegraphics[width=5.2cm]{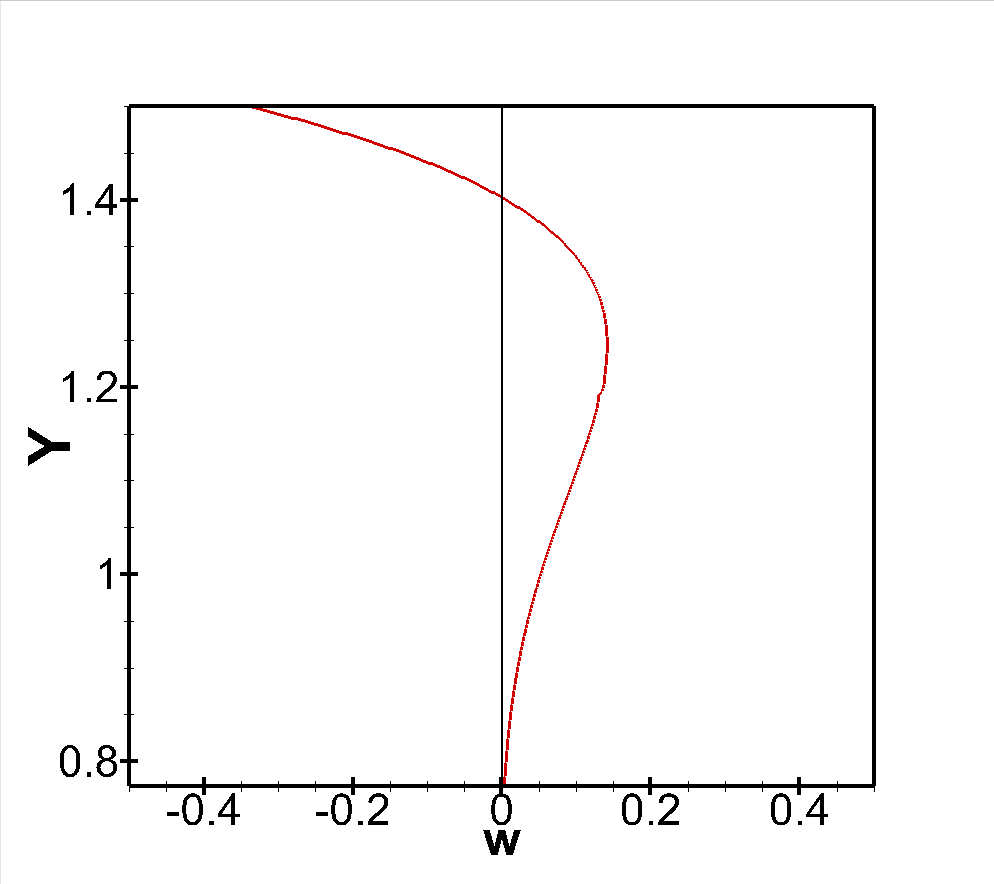} }}
    \qquad
    \subfloat[\centering{\label{figG2}}]{{\includegraphics[width=5.2cm]{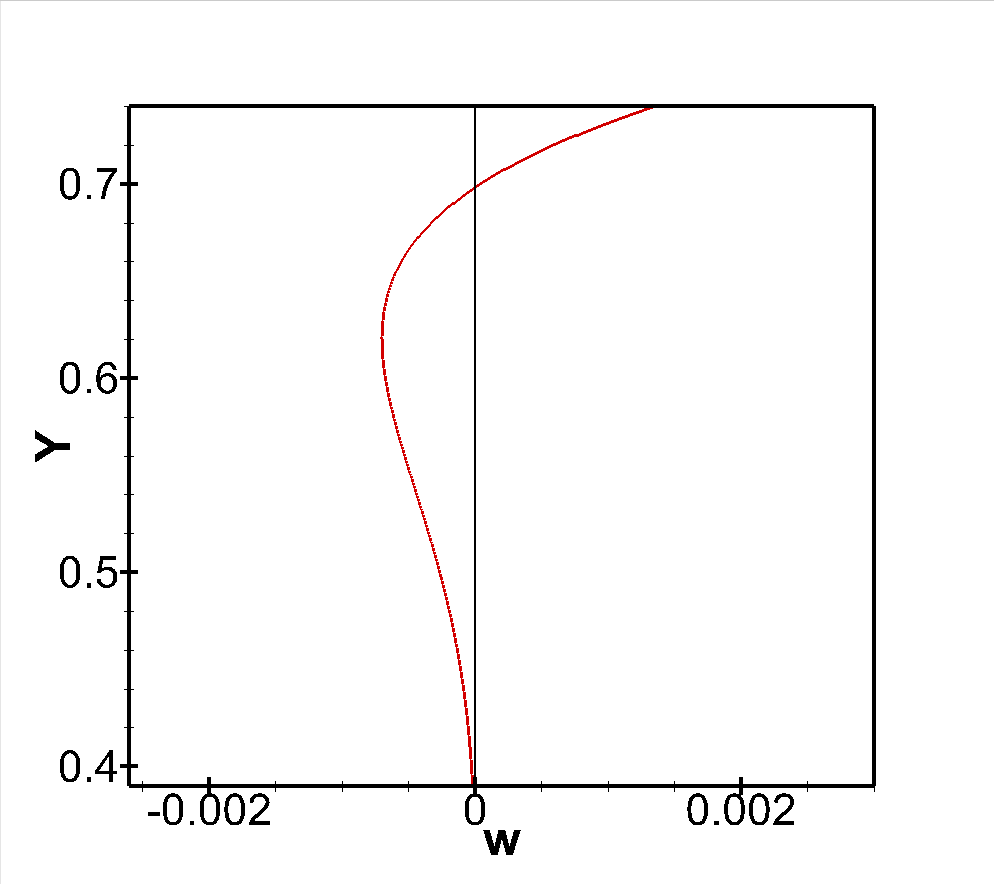} }}
    \qquad
    \subfloat[\centering{\label{figG3}}]{{\includegraphics[width=5.2cm]{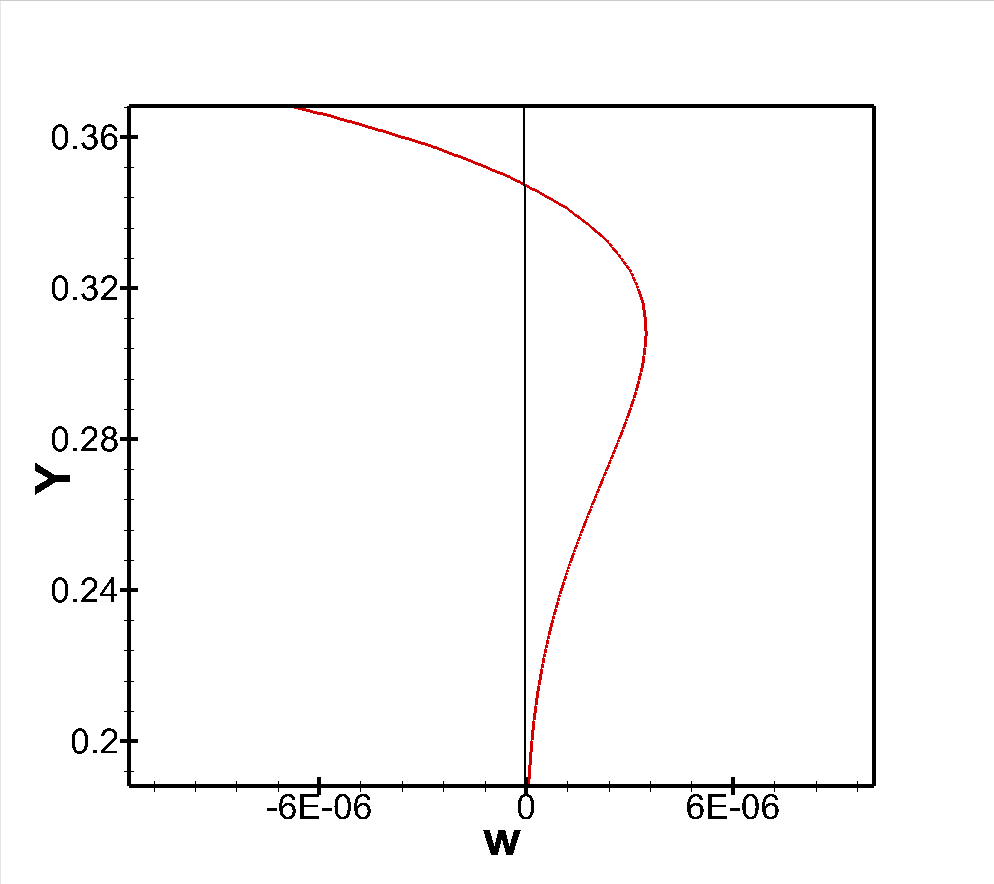} }}
    \qquad
    \subfloat[\centering{\label{figG4}}]{{\includegraphics[width=5.2cm]{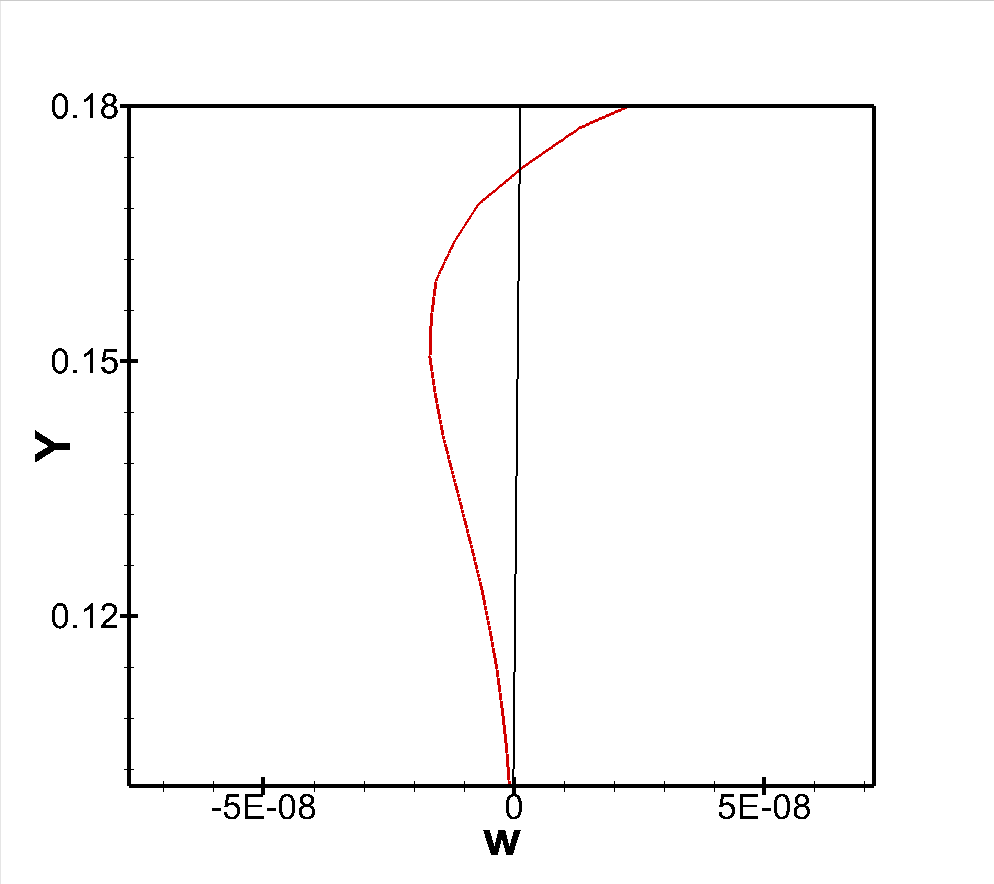} }}
    \qquad
    \subfloat[\centering{\label{figG5}}]{{\includegraphics[width=5.2cm]{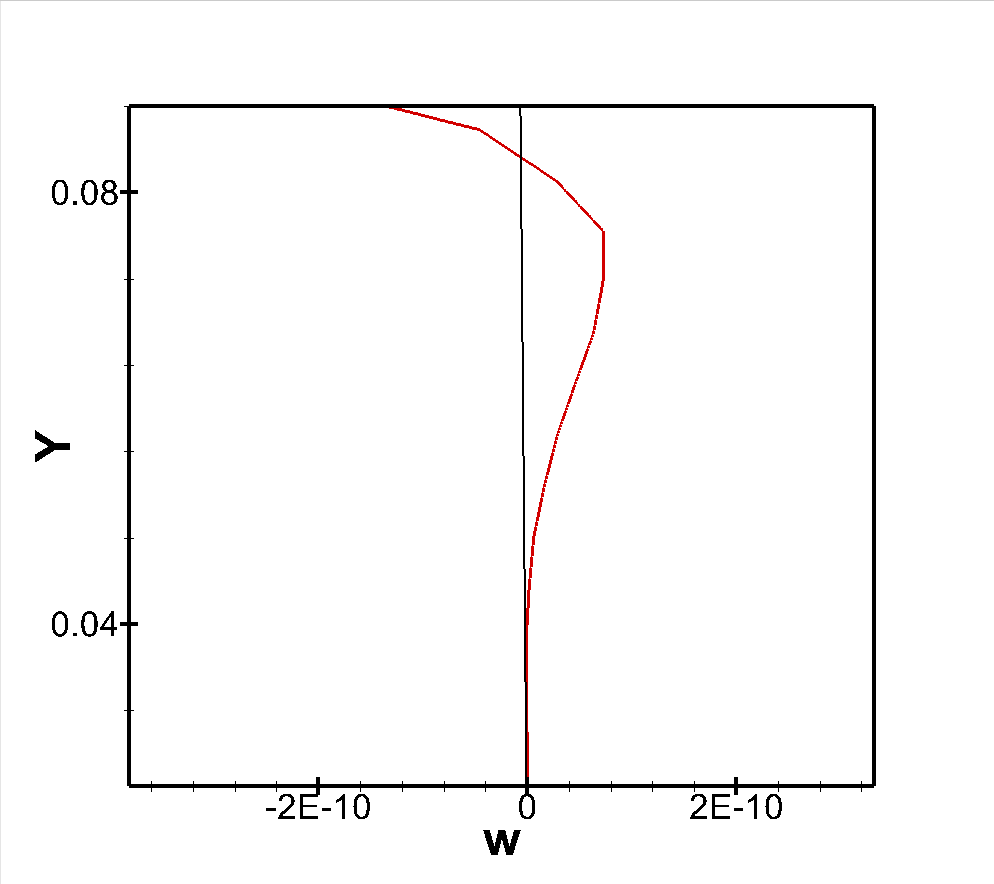} }}
    \qquad
    \subfloat[\centering{\label{figG6}}]{{\includegraphics[width=5.2cm]{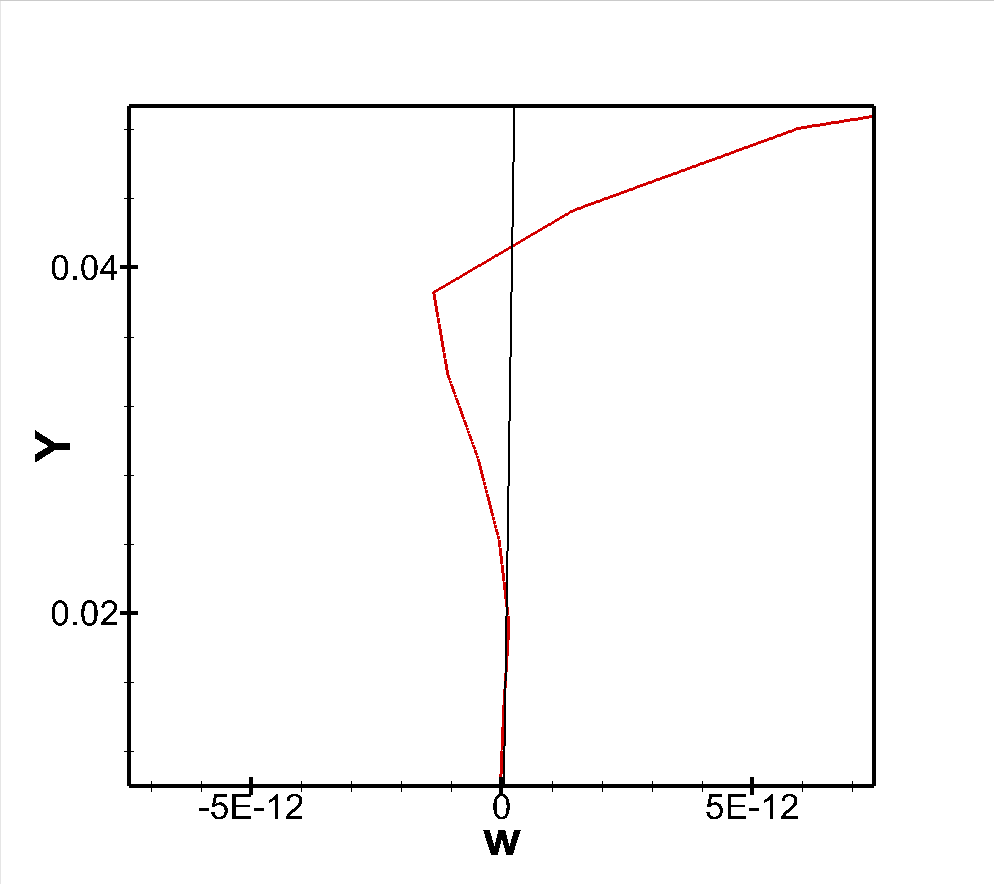} }}
\caption{Vorticity distribution in the triangular cavity for Stokes flow ($Re = 1$), demonstrating the self-similarity of the corner vortices.}
\label{figG}
\end{figure*}

\begin{figure*}[!ht]
\centering
    \subfloat[\centering{\label{figGa1}}]{{\includegraphics[width=5.2cm]{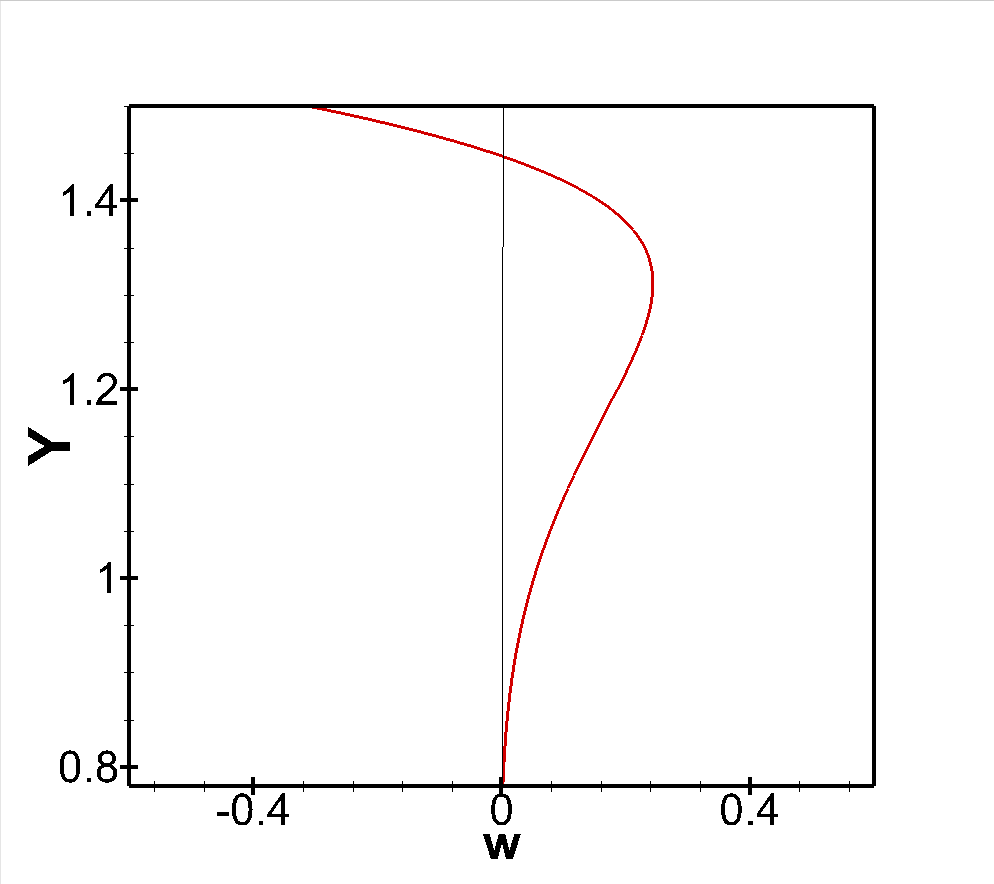} }}
    \qquad
    \subfloat[\centering{\label{figGa2}}]{{\includegraphics[width=5.2cm]{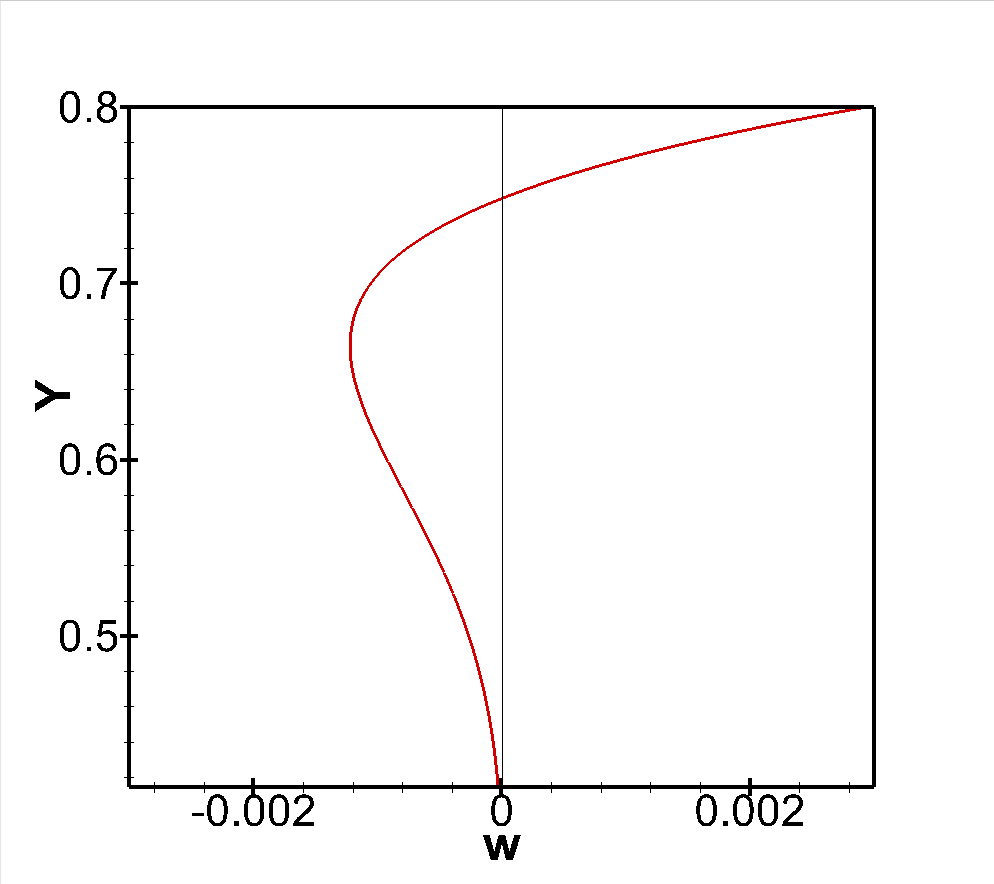} }}
    \qquad
    \subfloat[\centering{\label{figGa3}}]{{\includegraphics[width=5.2cm]{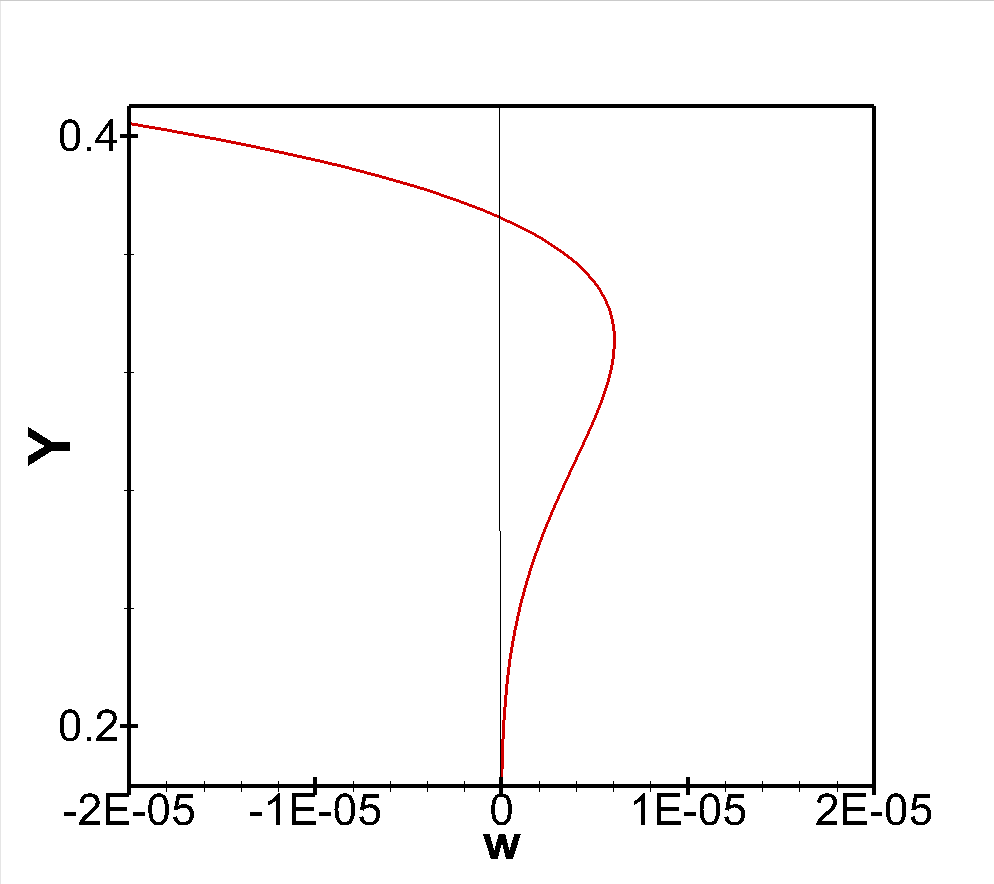} }}
    \qquad
    \subfloat[\centering{\label{figGa4}}]{{\includegraphics[width=5.2cm]{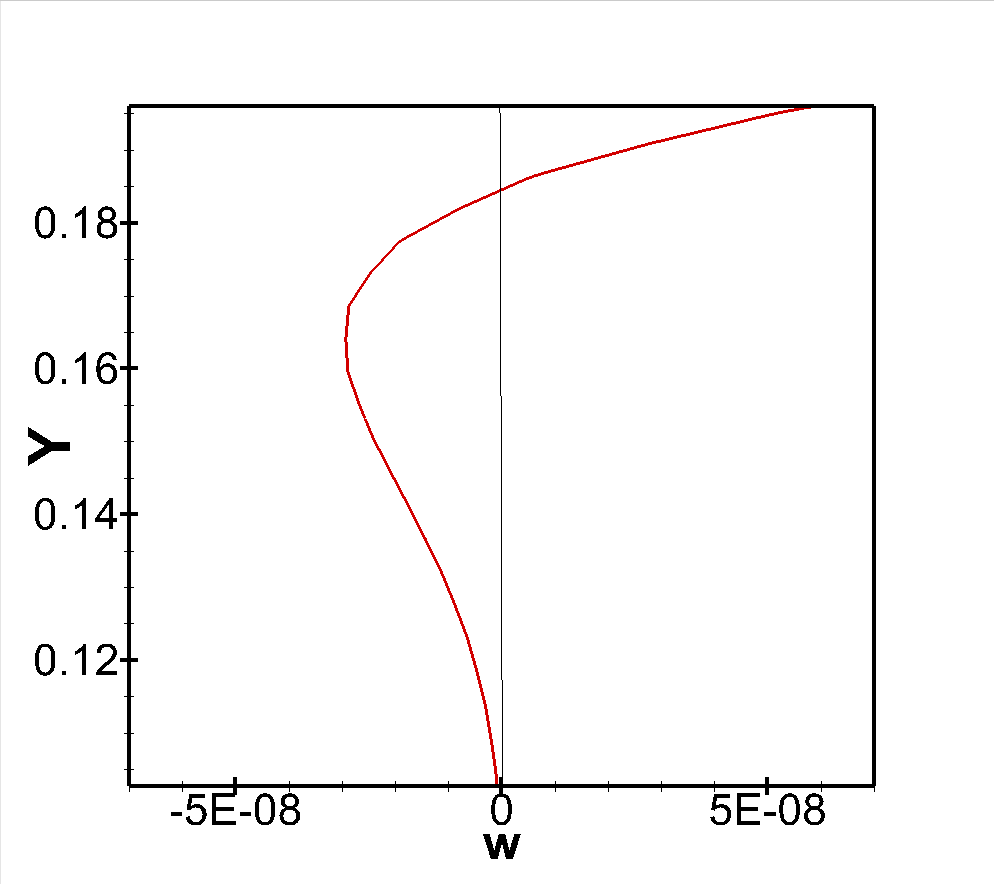} }}
    \qquad
    \subfloat[\centering{\label{figGa5}}]{{\includegraphics[width=5.2cm]{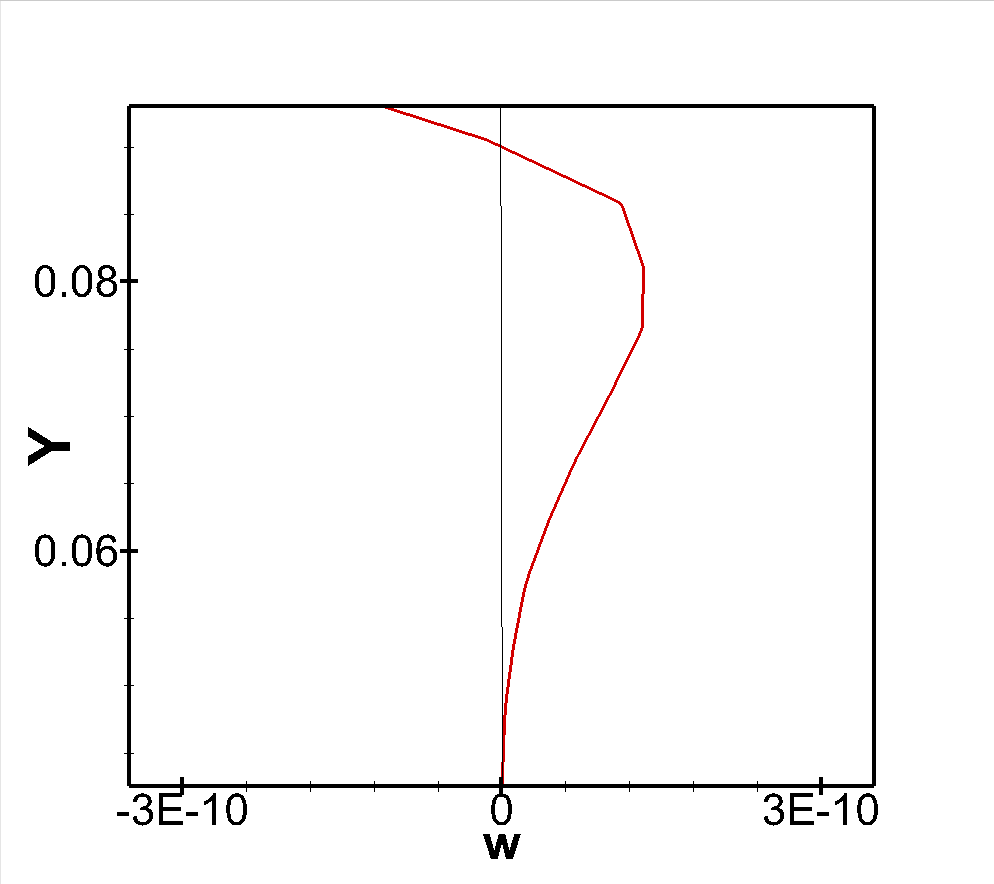} }}
    \qquad
    \subfloat[\centering{\label{figGa6}}]{{\includegraphics[width=5.2cm]{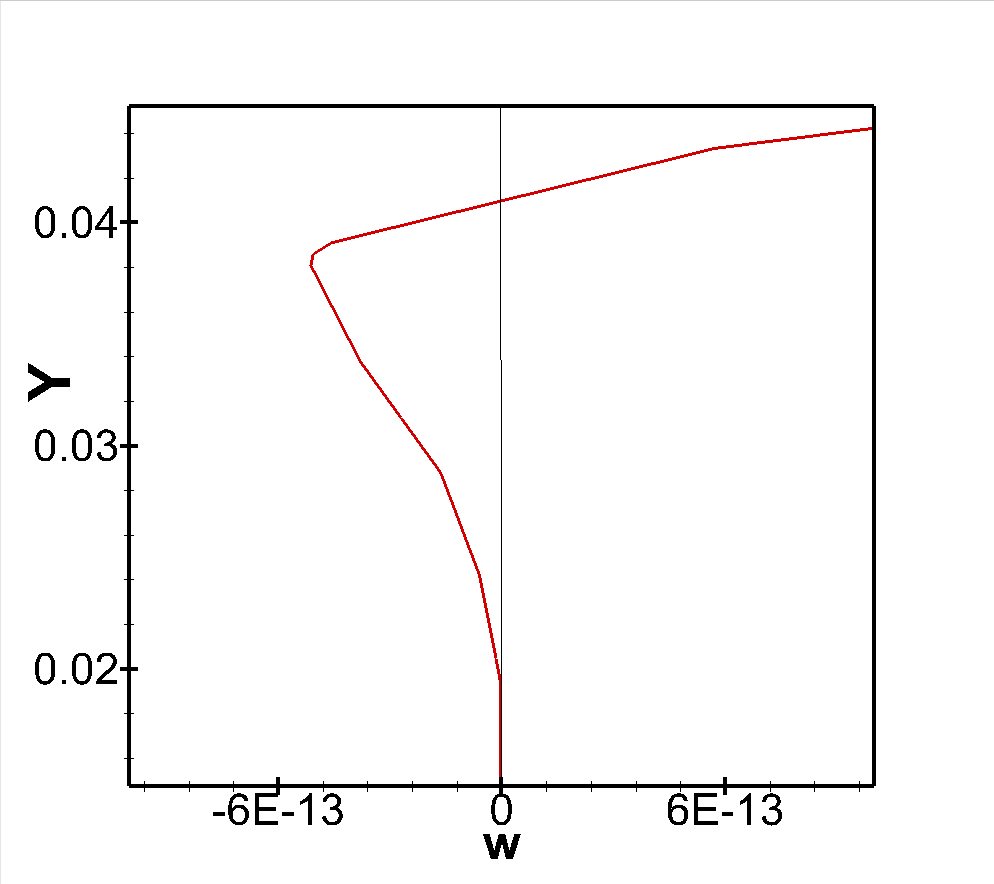} }}
\caption{Vorticity distribution in the triangular cavity at $Re = 100$, demonstrating the self-similarity of the corner vortices.}
\label{figGa}
\end{figure*}

\begin{figure*}[!ht]
\centering
    \subfloat[\centering{\label{figGb1}}]{{\includegraphics[width=5.2cm]{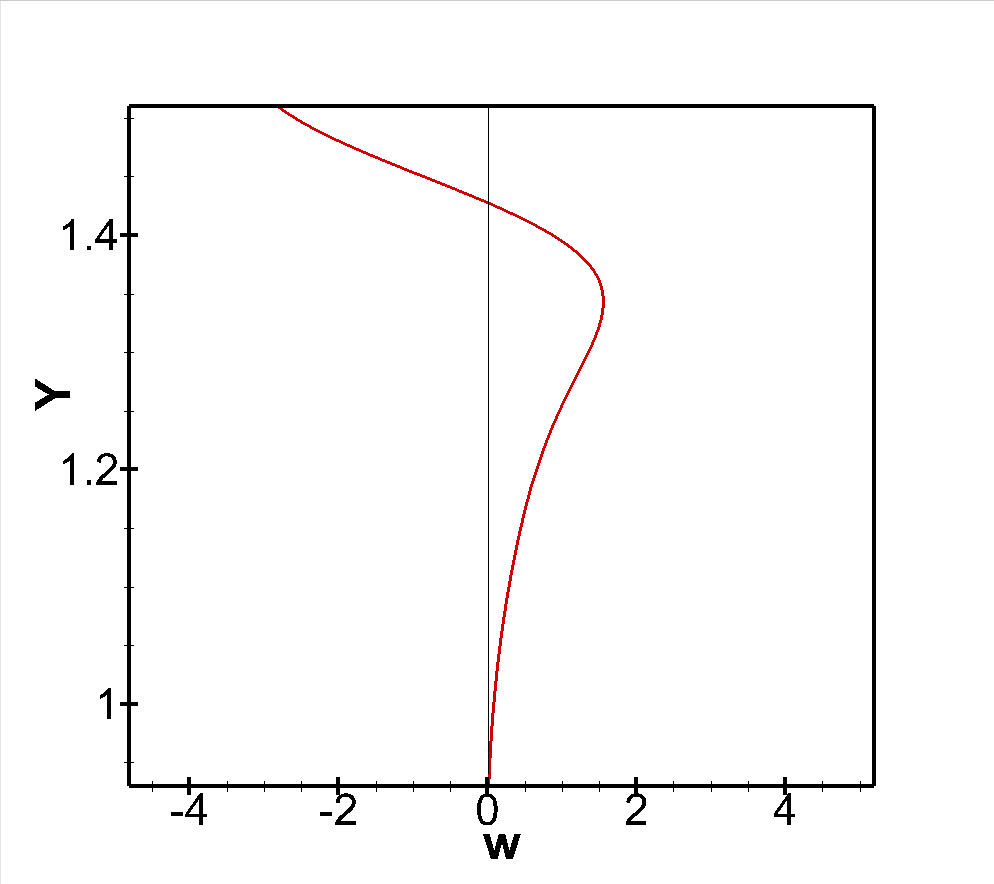} }}
    \qquad
    \subfloat[\centering{\label{figGb2}}]{{\includegraphics[width=5.2cm]{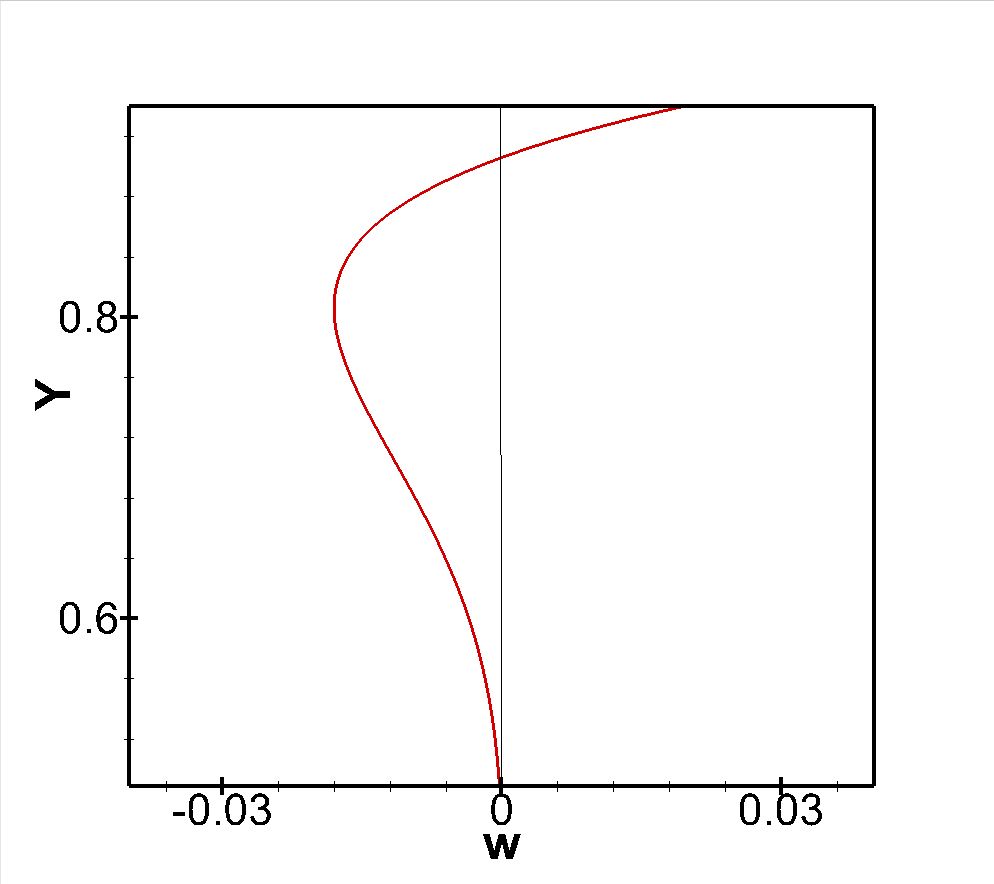} }}
    \qquad
    \subfloat[\centering{\label{figGb3}}]{{\includegraphics[width=5.2cm]{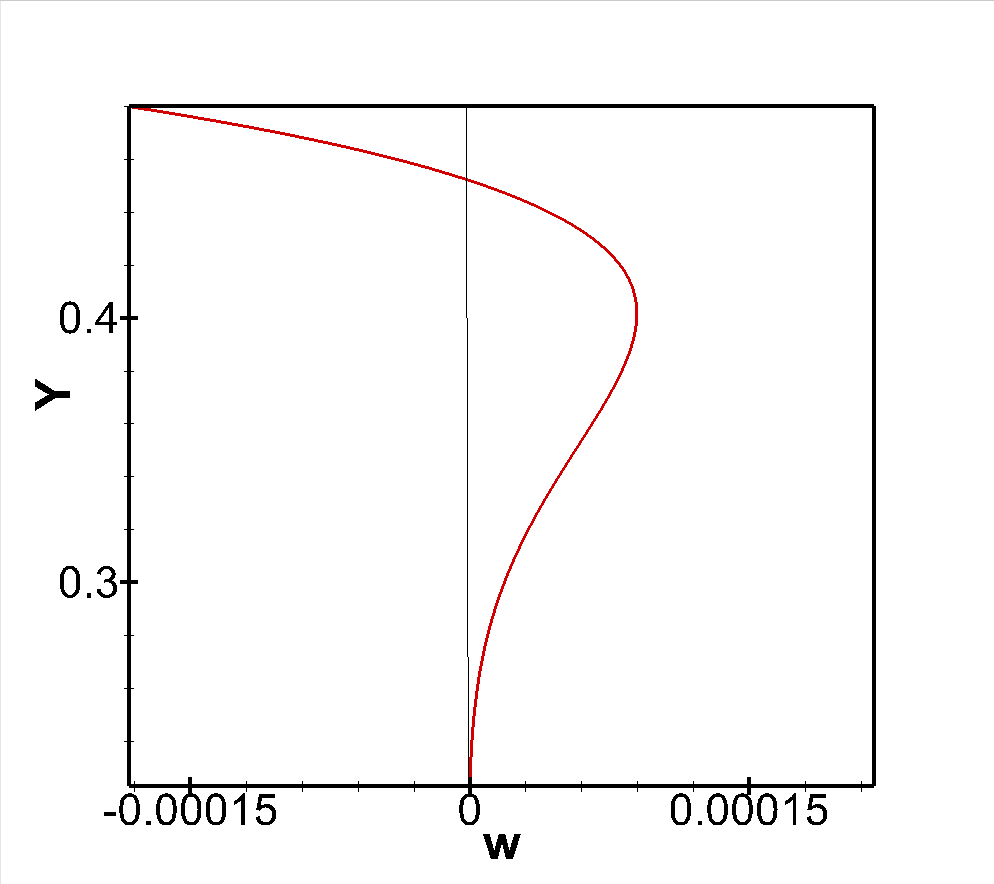} }}
    \qquad
    \subfloat[\centering{\label{figGb4}}]{{\includegraphics[width=5.2cm]{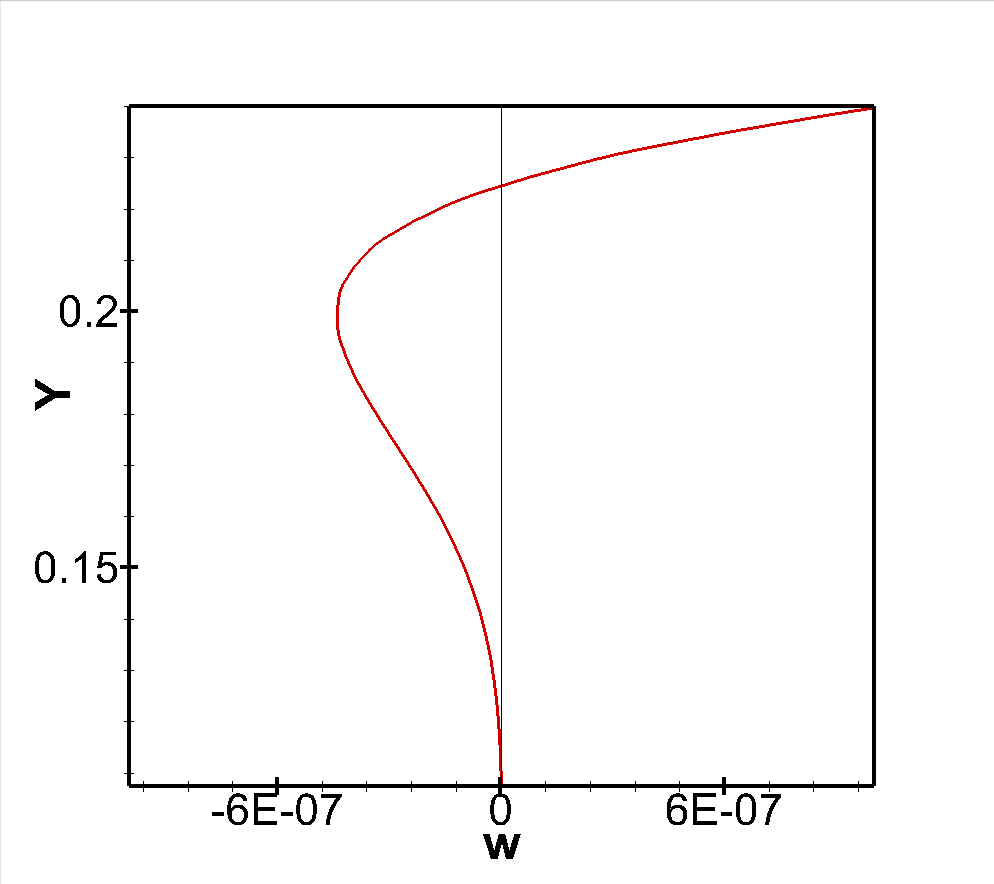} }}
    \qquad
    \subfloat[\centering{\label{figGb5}}]{{\includegraphics[width=5.2cm]{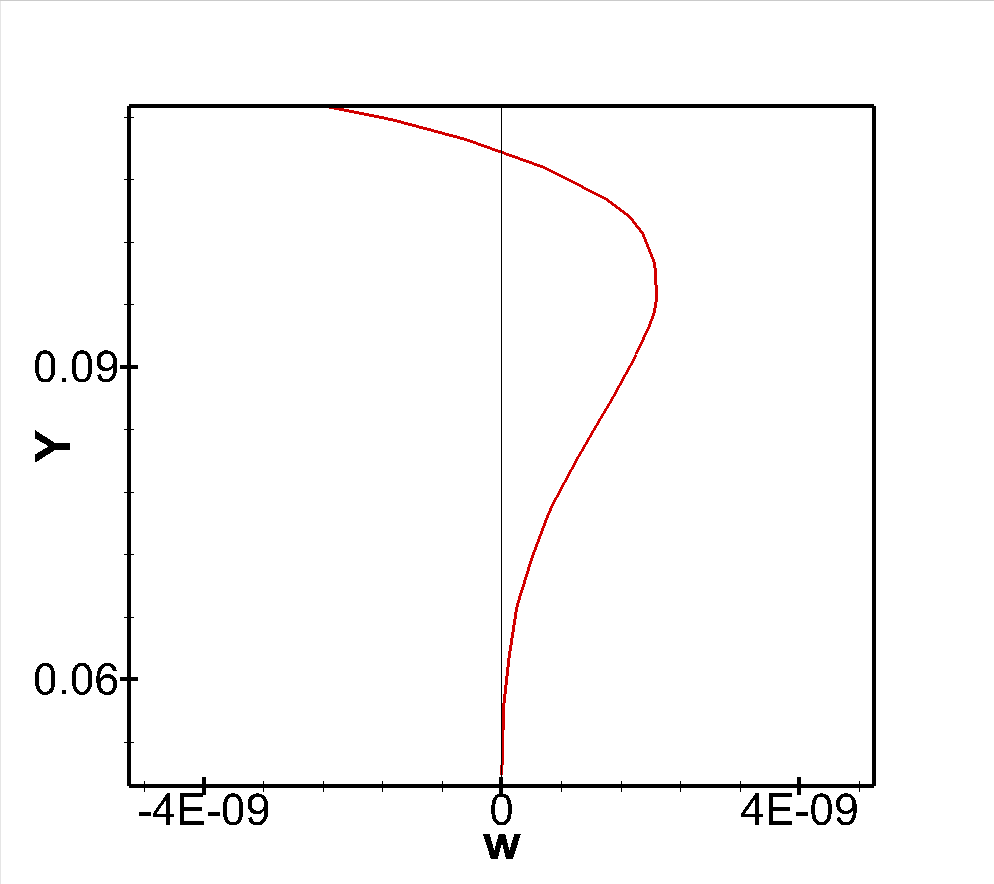} }}
    \qquad
    \subfloat[\centering{\label{figGb6}}]{{\includegraphics[width=5.2cm]{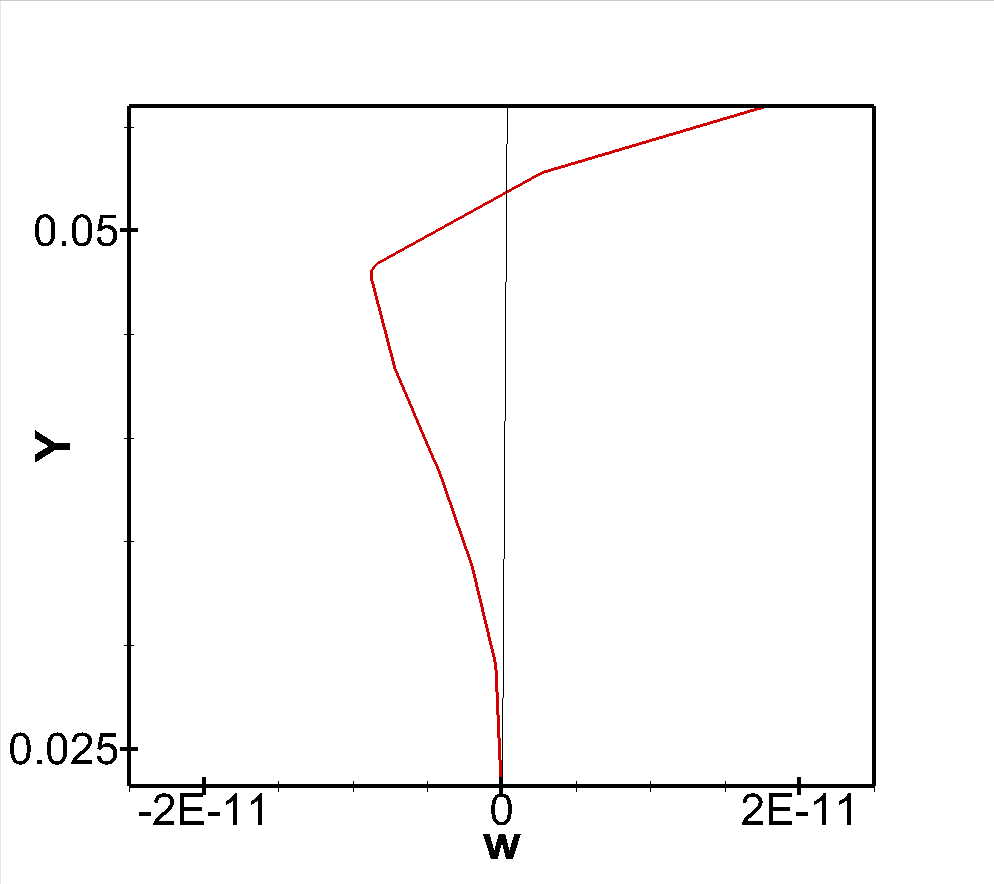} }}
\caption{Vorticity distribution in the triangular cavity at $Re = 500$, demonstrating the self-similarity of the corner vortices.}
\label{figGb}
\end{figure*}

Figures \ref{figG}-\ref{figGb} depict the self-similarity of these corner vortices for $Re = 1, 100,$ and $500$, respectively, by plotting the vorticity distribution along the cavity altitude where simultaneous vortices coexist. Because a change in the sign of vorticity denotes flow separation, the vorticity strictly alternates signs between any two adjacent vortices. These alternating positive and negative values signify the successive chain of co-rotating and counter-rotating vortices within the cavity.

The vorticity distribution curves exhibit an identical structural pattern across the region, differing only in orientation, which confirms the geometric self-similarity of these corner vortices. This recurrent pattern is evident in the curves corresponding to the sequence of the first six successive vortices ($V_1$ to $V_6$) shown in Figs. \ref{figG}-\ref{figGb}. Specifically, Figs. \ref{figG1},\ref{figGa1}, and \ref{figGb1} highlight the positive vorticity curve at the separation boundary between $V_1$ and $V_2$, while the subsequent subfigures present analogous curves for the interior higher-order vortices.
\par
Finally, Table \ref{Table7} lists the center coordinates of the first four vortices, quantitatively illustrating the displacement of the primary vortex center toward the right wall as $Re$ increases. Crucially, while the core of the primary vortex shifts with increasing inertia, its outer boundary profile remains structurally similar to its low-$Re$ counterparts, preserving the underlying fractal framework of the system.
\begin{table}[!ht]
    \centering
    \begin{tabular}{ccccc}
    \hline
    $Re$ & $V_1$ & $V_2$ & $V_3$ & $V_4$ \\
    \hline
    5 & (0.5059, 1.8016) & (0.5003, 0.9054) & (0.5000, 0.4503) & (0.5000, 0.2235) \\
    10 & (0.5119, 1.8015) & (0.5006, 0.9059) & (0.5000, 0.4506) & (0.5000, 0.2236) \\
    15 & (0.5178, 1.8015) & (0.5009, 0.9069) & (0.5000, 0.4510) & (0.5000, 0.2239) \\
    20 & (0.5237, 1.8013) & (0.5012, 0.9082) & (0.5000, 0.4517) & (0.5000, 0.2241) \\
    30 & (0.5352, 1.8010) & (0.5020, 0.9118) & (0.5000, 0.4536) & (0.5000, 0.2250) \\
    50 & (0.5569, 1.7998) & (0.5036, 0.9231) & (0.5000, 0.4590) & (0.5000, 0.2279) \\
    100 & (0.5963, 1.7923) & (0.5103, 0.9712) & (0.5000, 0.4825) & (0.5000, 0.2397) \\
    200 & (0.6069, 1.7597) & (0.5204, 1.0892) & (0.5000, 0.5393) & (0.5000, 0.2680) \\
    \hline
    \end{tabular}
    \vspace{0.25cm}
    \caption{Comparison of vortex center locations within the triangular cavity across varying Reynolds numbers.}
    \label{Table7}
\end{table}


\subsection{Comparative analysis of self-similarity in triangular and square cavities}

\begin{figure*}[!ht]
\centering
    \subfloat[\centering{\label{figI1}}]{{\includegraphics[width=6.2cm]{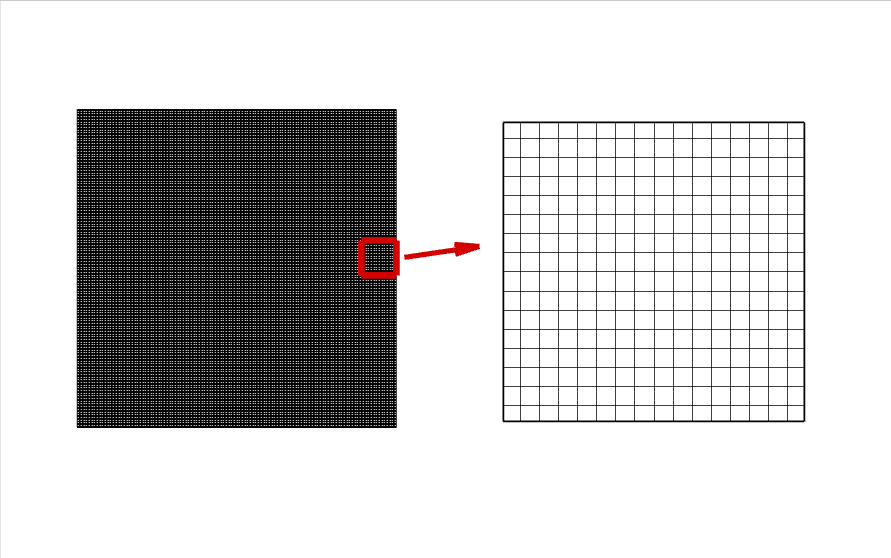} }}
    \qquad
    \subfloat[\centering{\label{figI2}}]{{\includegraphics[width=5.3cm]{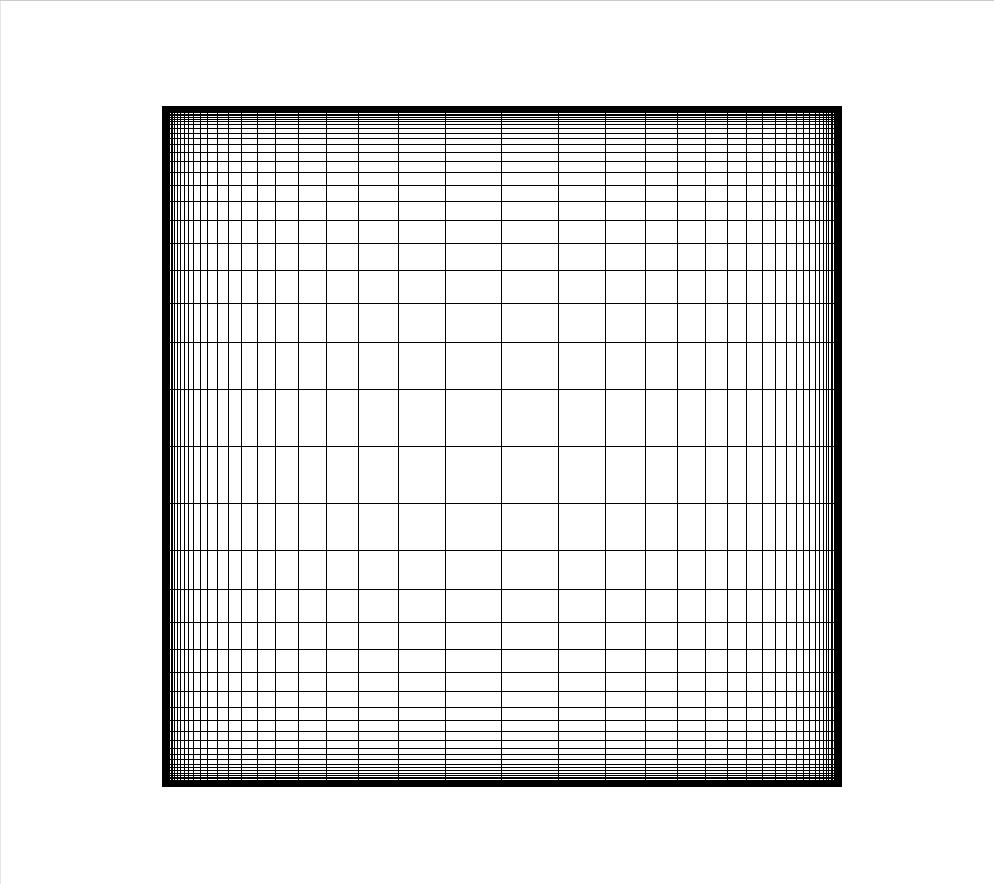} }}
\caption{Mesh configurations for the square cavity utilizing a $251 \times 251$ grid: (a) uniform and (b) clustered meshing strategies.}
\label{figI}
\end{figure*}

\begin{figure}[!ht]
    \centering
    \includegraphics[width=12.5cm]{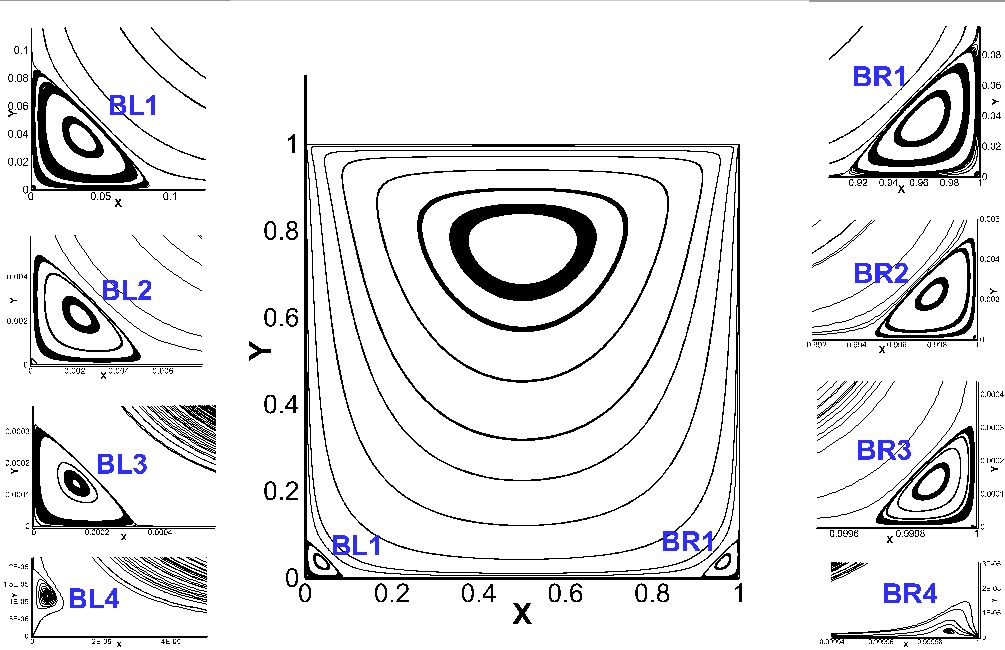}
    \caption{Sequence of the left (BL1-BL4) and right (BR1-BR4) corner vortices within the square cavity for Stokes flow ($Re = 1$).}
    \label{FigJ}
\end{figure}

To verify the universality of these results across different geometries, we compare our observations of the triangular cavity with those of a square cavity. Because the left and right corner vortices of the square cavity form self-similar sequences, we can determine specific fractal characteristics of these structures. We solved the square cavity problem for slow, incompressible viscous flow ($Re = 1$) in ANSYS Fluent, employing two grid resolutions of size $251 \times 251$: one uniform and one clustered at the corners (Fig. \ref{figI}). With the uniform mesh, we observed traces of only one left (BL1) and one right (BR1) corner vortex. However, the clustered mesh provided sufficient element density at the corners to reveal three left (BL1, BL2, BL3) and three right (BR1, BR2, BR3) successive corner vortices, along with slight traces of a fourth pair (BL4 and BR4). Figure \ref{FigJ} presents the traces of all corner vortices observed at the respective corners of the square cavity for Stokes flow. From the figure, one can easily observe the symmetry of the respective vortices at both corners about the vertical centerline.
\par 
Using the area-perimeter method, we calculated the fractal dimensions of BL1 and BR1 (Table \ref{Table8}), as the outer boundaries of these vortices within the square cavity closely resemble a right-angled isosceles triangle. For this calculation, the distance between two neighboring grid points ($1/250$) is considered to be unity. However, for the subsequent corner vortices, the grid elements are too small relative to the vortex size to determine their dimensions accurately. This highlights a major limitation of the area-perimeter method, which requires a sufficient number of grid points to capture the fractal dimension. Nevertheless, to the best of our knowledge, this is the first time the fractal dimensions of BL1 and BR1 have been reported in the literature.

\begin{table}[!ht]
\centering
\begin{tabular}{cccc} 
 \hline
 \textbf{Vortex} & \textbf{Perimeter} $(P)$ & \textbf{Area} $(A)$ & \textbf{Fractal Dimension} \\ 
 \hline
  BL1 & 78.5269 & 264.5 & 1.56456 \\
  BR1 & 88.7695 & 338.0 & 1.54079 \\
 \hline
\end{tabular}
\caption{Perimeter, area, and fractal dimensions of the BL1 and BR1 structures in the square cavity.}
\label{Table8}
\end{table}

\begin{figure}[!ht]
    \centering
    \includegraphics[width=11cm]{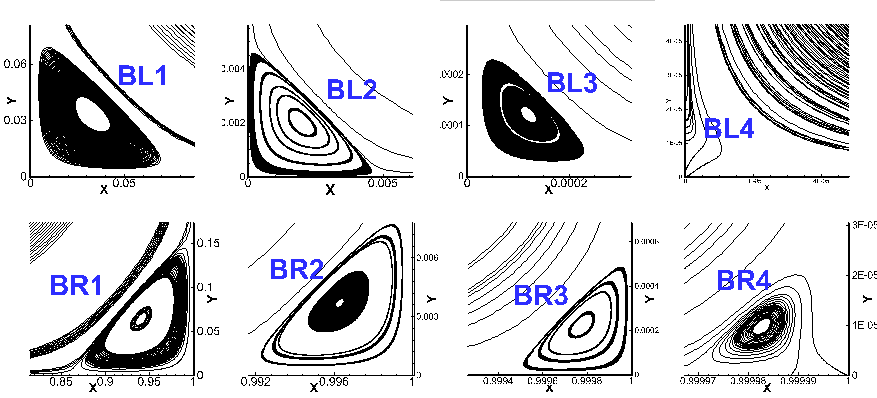}
    \caption{Sequence of the left (BL1-BL4) and right (BR1-BR4) corner vortices within the square cavity at $Re = 100$.}
    \label{FigH1}
\end{figure}

\begin{figure}[!ht]
    \centering
    \includegraphics[width=11cm]{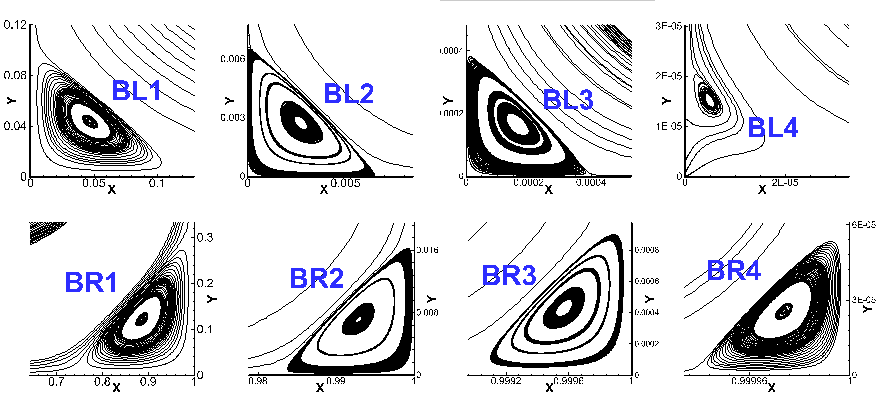}
    \caption{Sequence of the left (BL1-BL4) and right (BR1-BR4) corner vortices within the square cavity at $Re = 400$.}
    \label{FigH2}
\end{figure}

\begin{figure}[!ht]
    \centering
    \includegraphics[width=11cm]{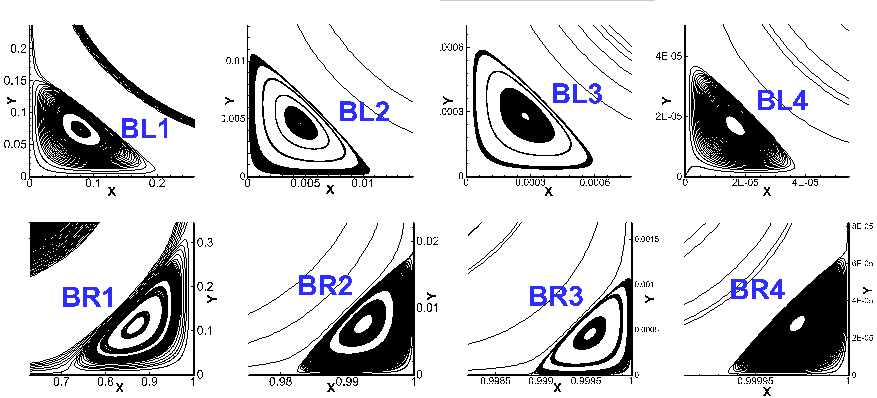}
    \caption{Sequence of the left (BL1-BL4) and right (BR1-BR4) corner vortices within the square cavity at $Re = 1000$.}
    \label{FigH3}
\end{figure}

\begin{figure}[!ht]
    \centering
    \includegraphics[width=11cm]{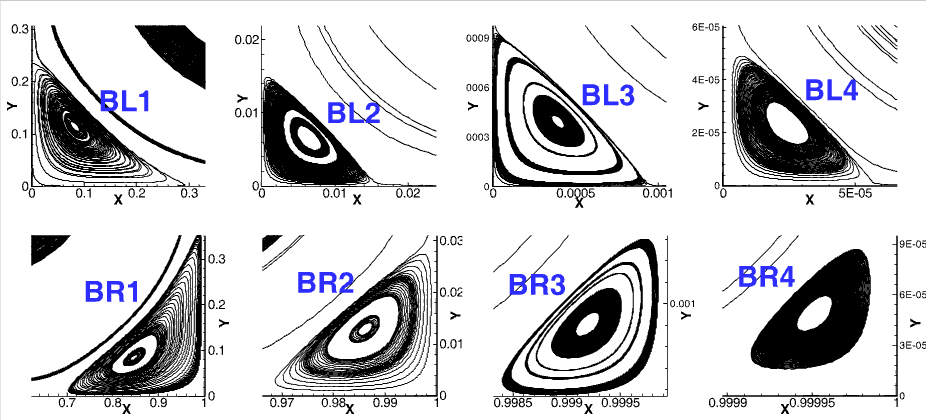}
    \caption{Sequence of the left (BL1-BL4) and right (BR1-BR4) corner vortices within the square cavity at $Re = 3200$.}
    \label{FigH4}
\end{figure}

\begin{figure*}[!ht]
\centering
    \subfloat[\centering{\label{figH1}}]{{\includegraphics[width=5cm]{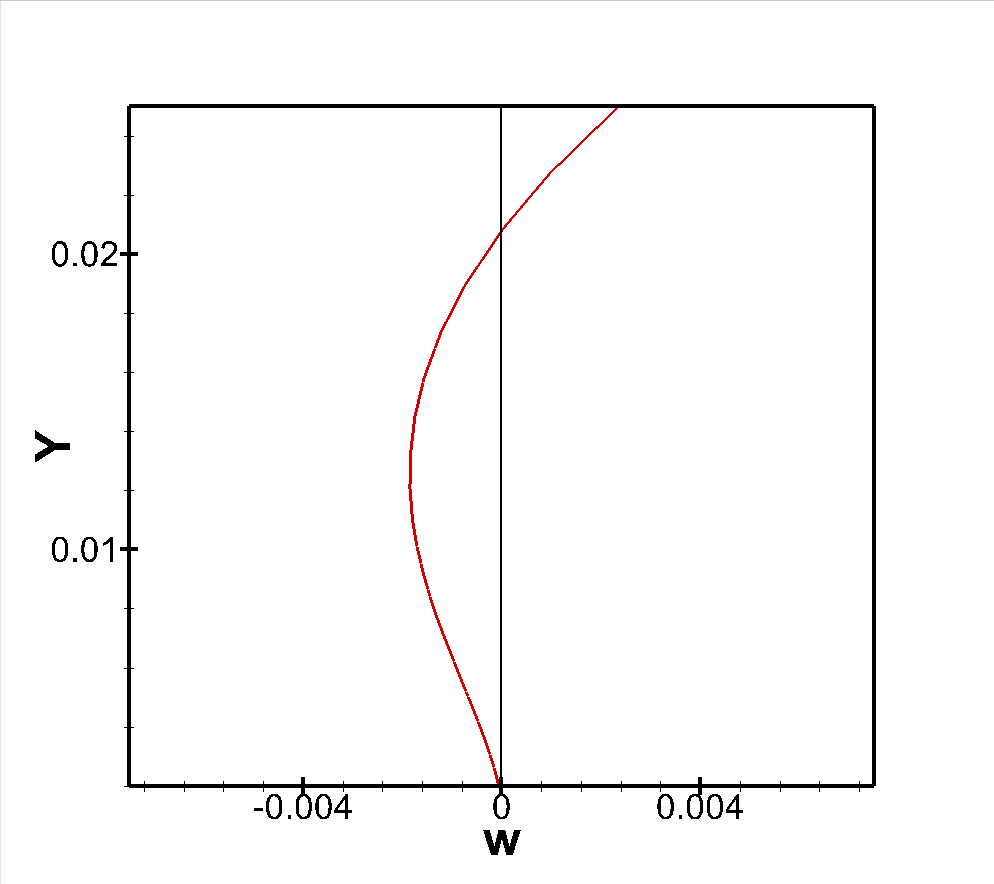} }}
    \qquad
    \subfloat[\centering{\label{figH2}}]{{\includegraphics[width=5cm]{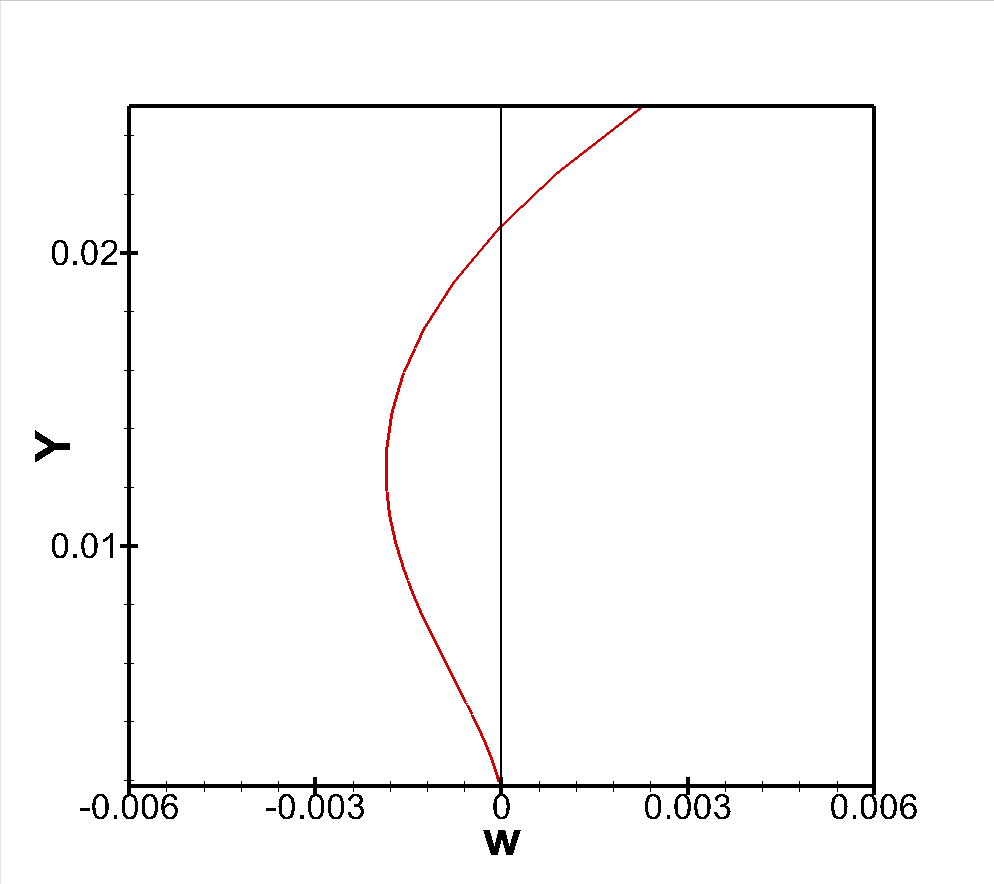} }}
    \qquad
    \subfloat[\centering{\label{figH3}}]{{\includegraphics[width=5cm]{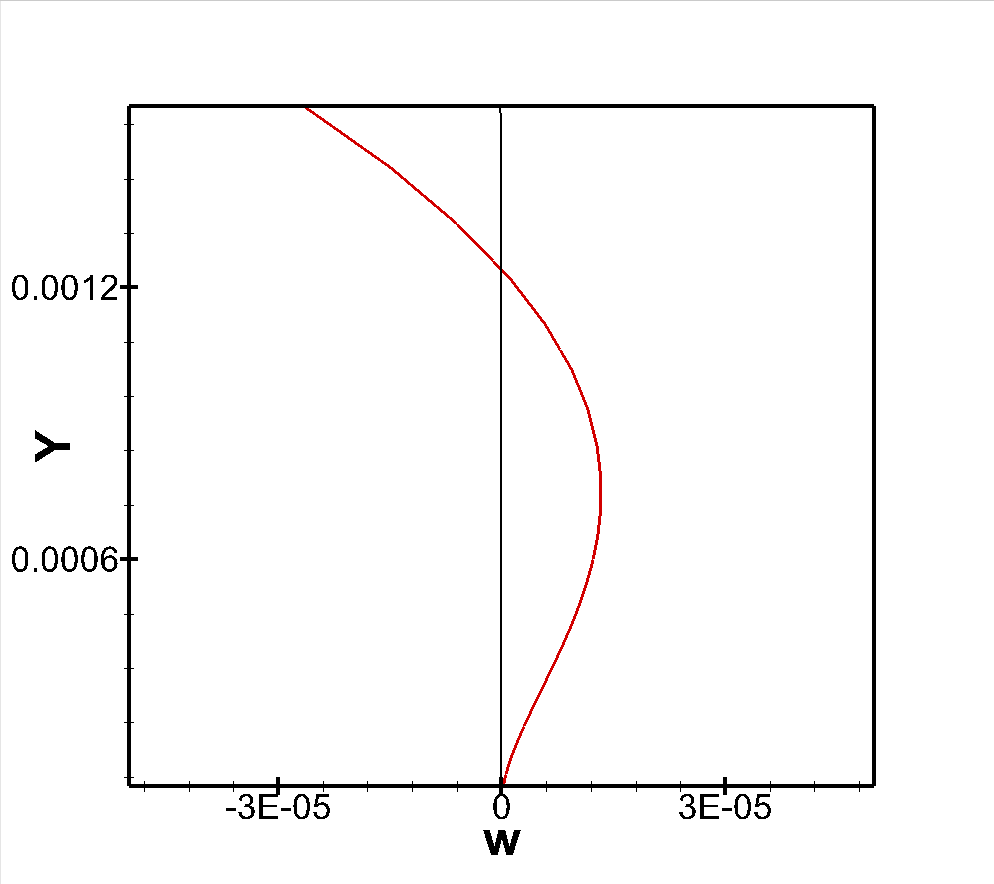} }}
    \qquad
    \subfloat[\centering{\label{figH4}}]{{\includegraphics[width=5cm]{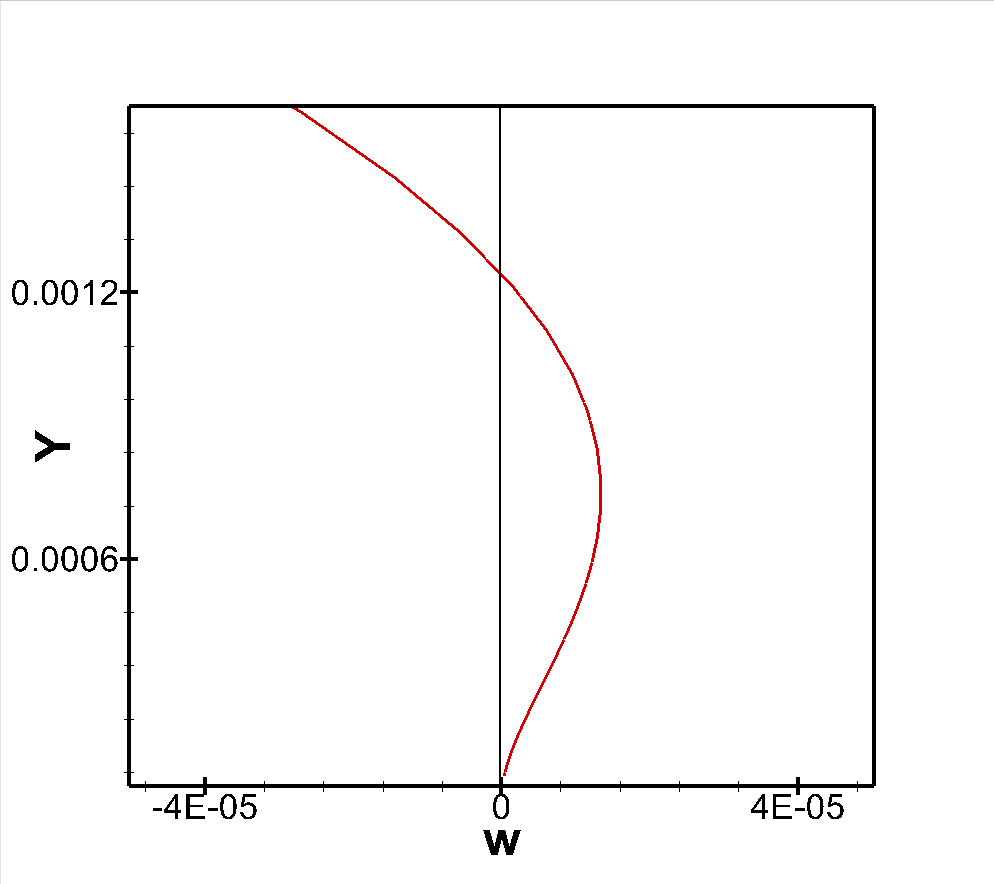} }}
    \qquad
    \subfloat[\centering{\label{figH5}}]{{\includegraphics[width=5cm]{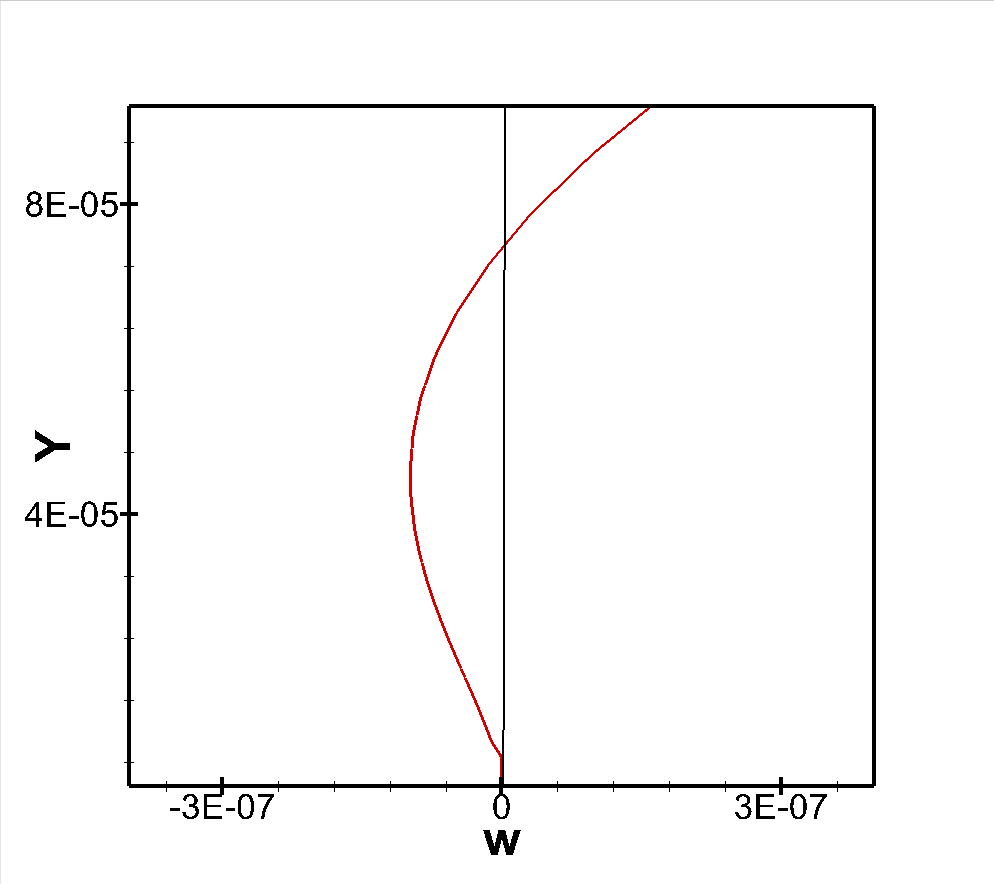} }}
    \qquad
    \subfloat[\centering{\label{figH6}}]{{\includegraphics[width=5cm]{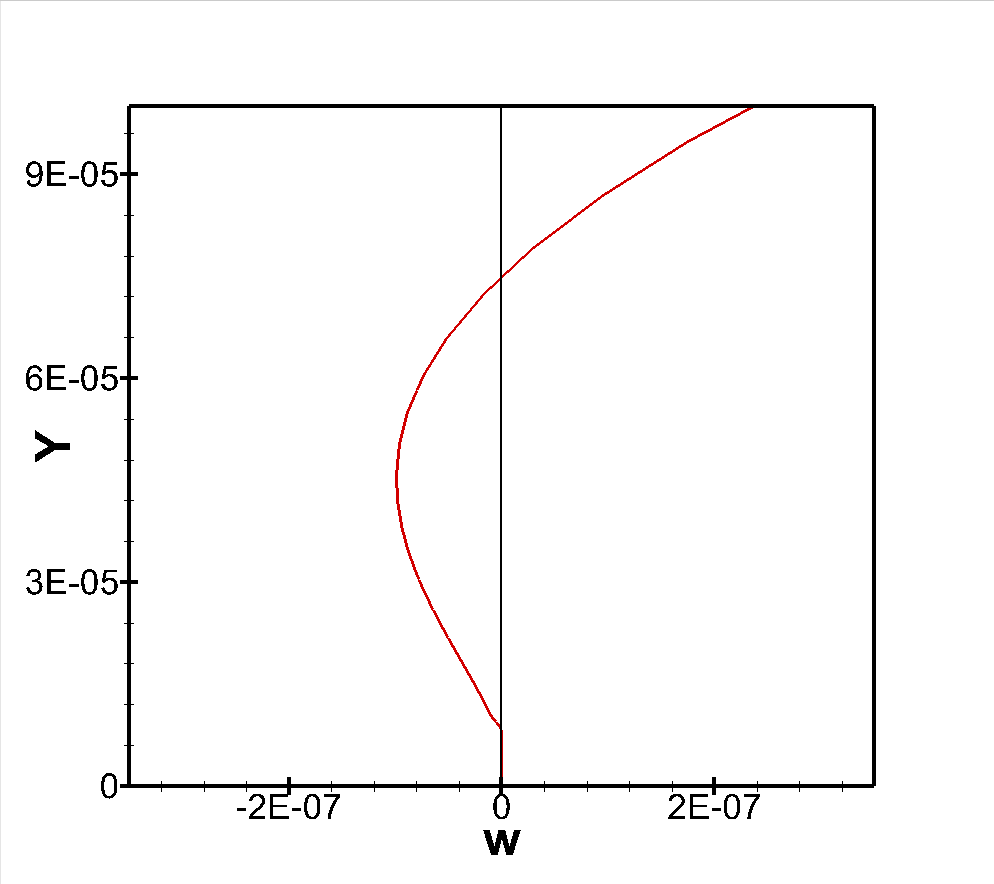} }}
\caption{Self-similarity of corner vortices for Stokes flow ($Re = 1$), illustrated by the vorticity distribution within the square cavity. The sequence of left corner vortices is shown in (a) BL1, (c) BL2, and (e) BL3, alongside the right corner vortices in (b) BR1, (d) BR2, and (f) BR3.}
\label{figH}
\end{figure*}

\begin{figure*}[!ht]
\centering
    \subfloat[\centering{\label{figHA1}}]{{\includegraphics[width=5cm]{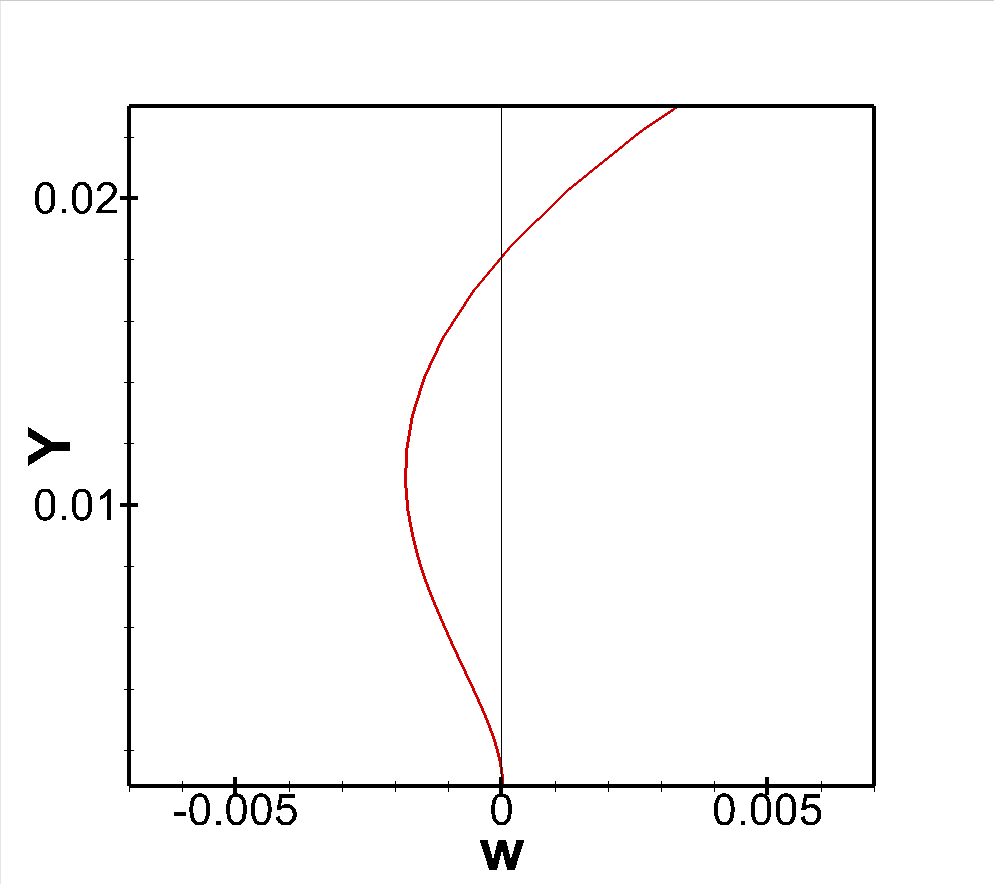} }}
    \qquad
    \subfloat[\centering{\label{figHA2}}]{{\includegraphics[width=5cm]{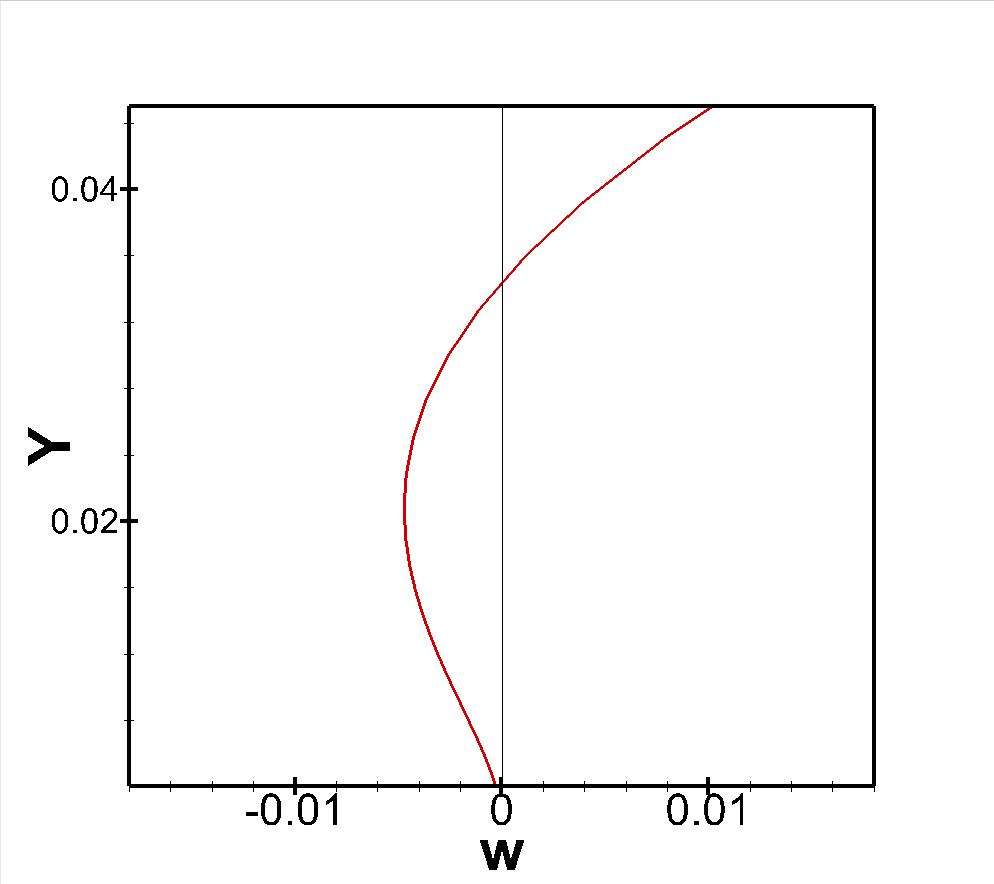} }}
    \qquad
    \subfloat[\centering{\label{figHA3}}]{{\includegraphics[width=5cm]{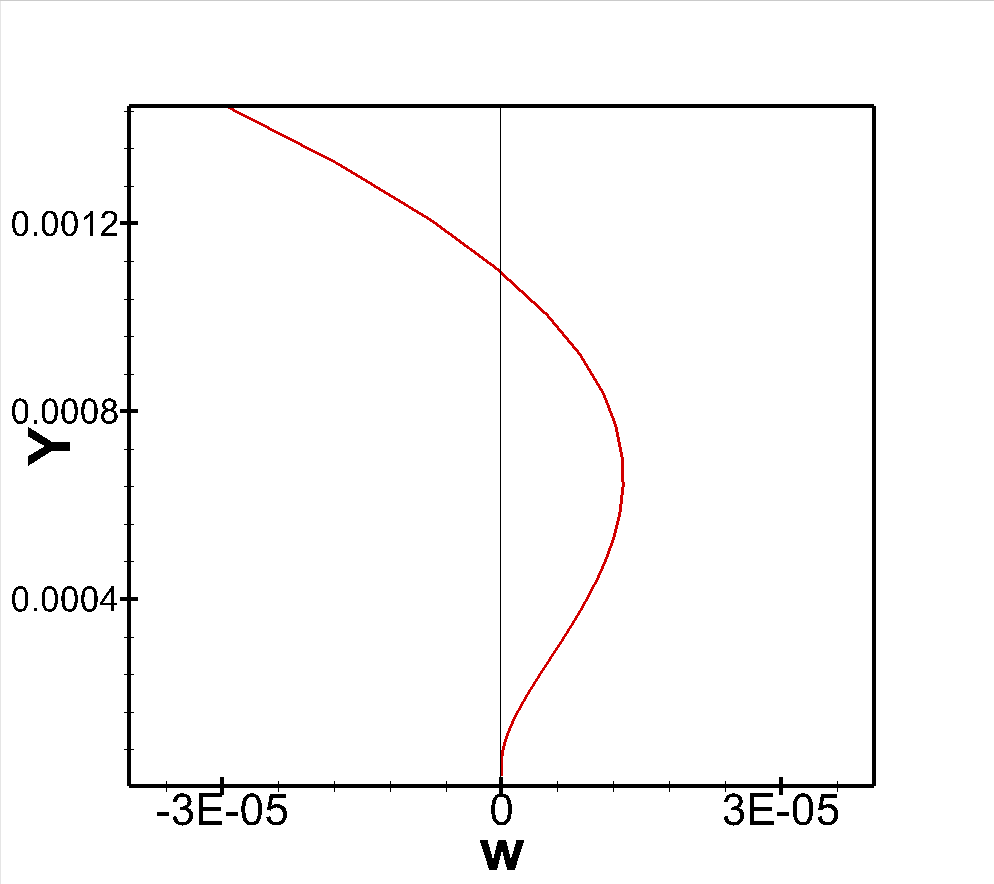} }}
    \qquad
    \subfloat[\centering{\label{figHA4}}]{{\includegraphics[width=5cm]{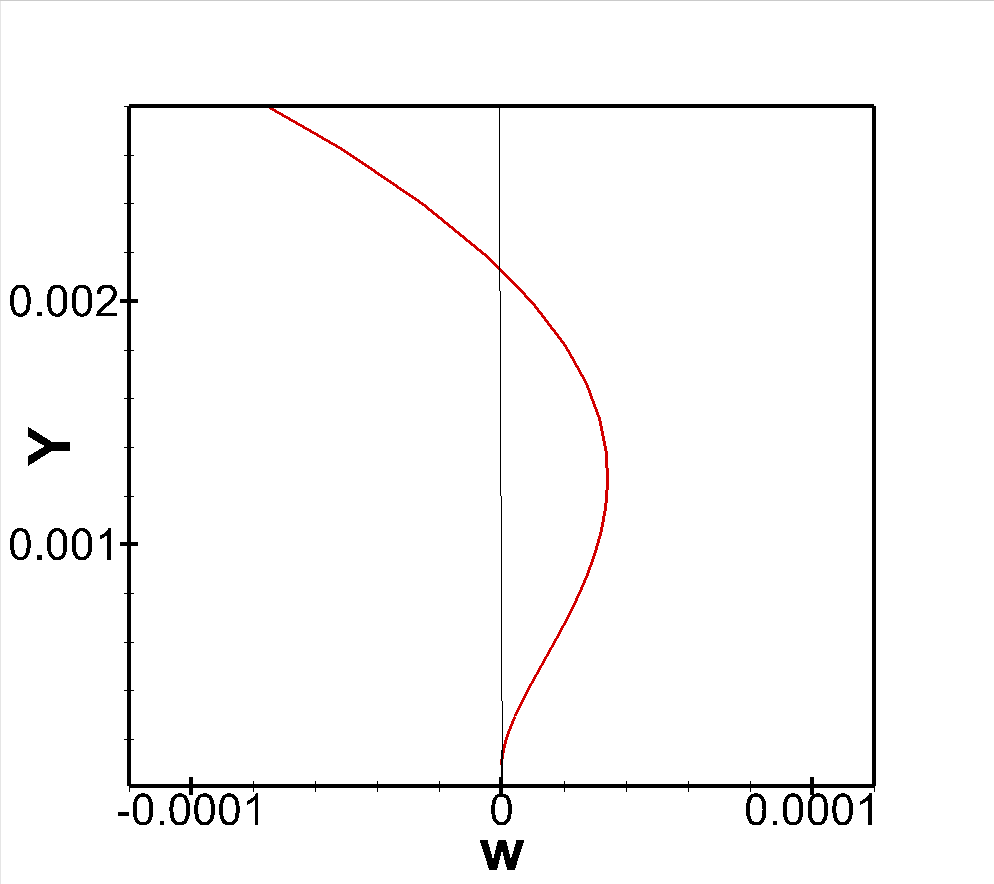} }}
    \qquad
    \subfloat[\centering{\label{figHA5}}]{{\includegraphics[width=5cm]{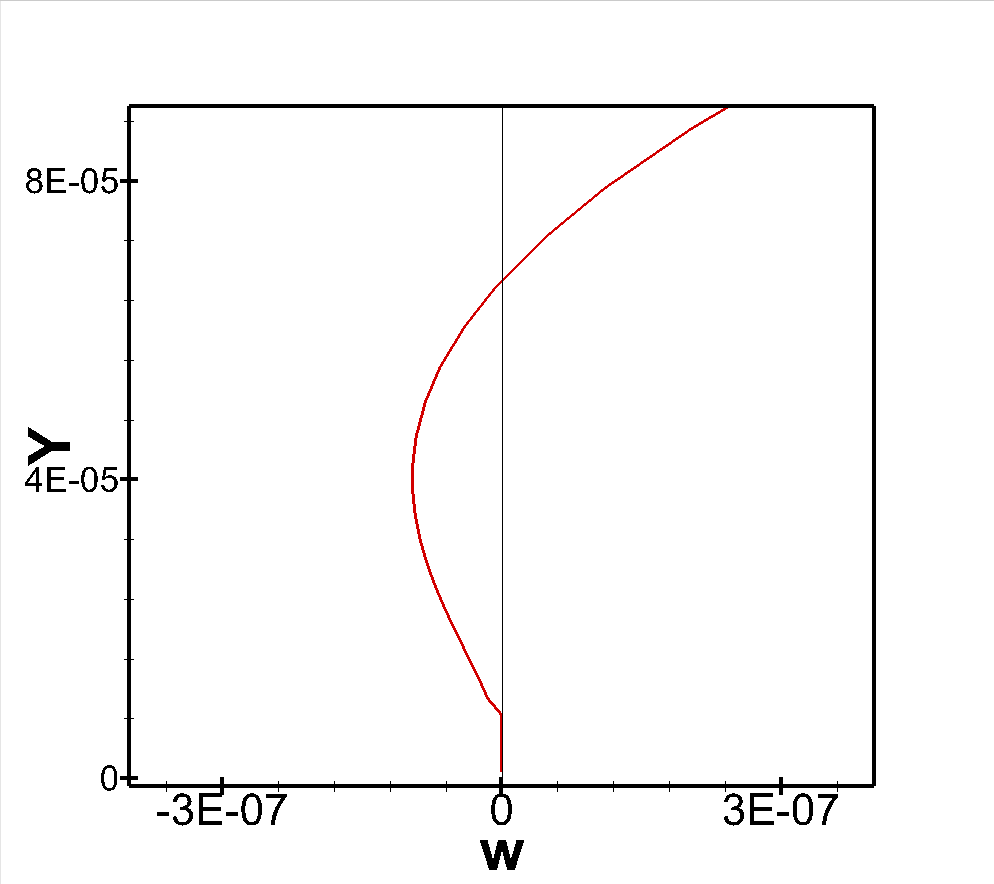} }}
    \qquad
    \subfloat[\centering{\label{figHA6}}]{{\includegraphics[width=5cm]{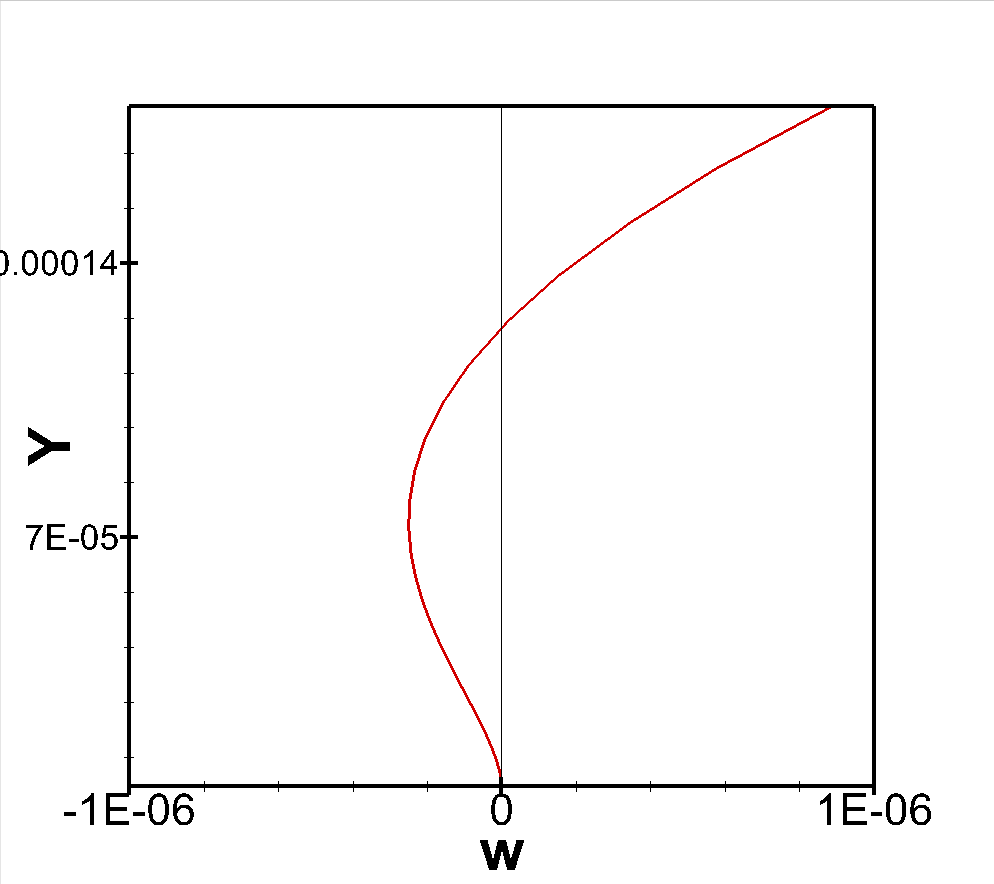} }}
\caption{Self-similarity of corner vortices at $Re = 100$, illustrated by the vorticity distribution within the square cavity. The sequence of left corner vortices is shown in (a) BL1, (c) BL2, and (e) BL3, alongside the right corner vortices in (b) BR1, (d) BR2, and (f) BR3.}
\label{figHA}
\end{figure*}

\begin{figure*}[!ht]
\centering
    \subfloat[\centering{\label{figHB1}}]{{\includegraphics[width=3.8cm]{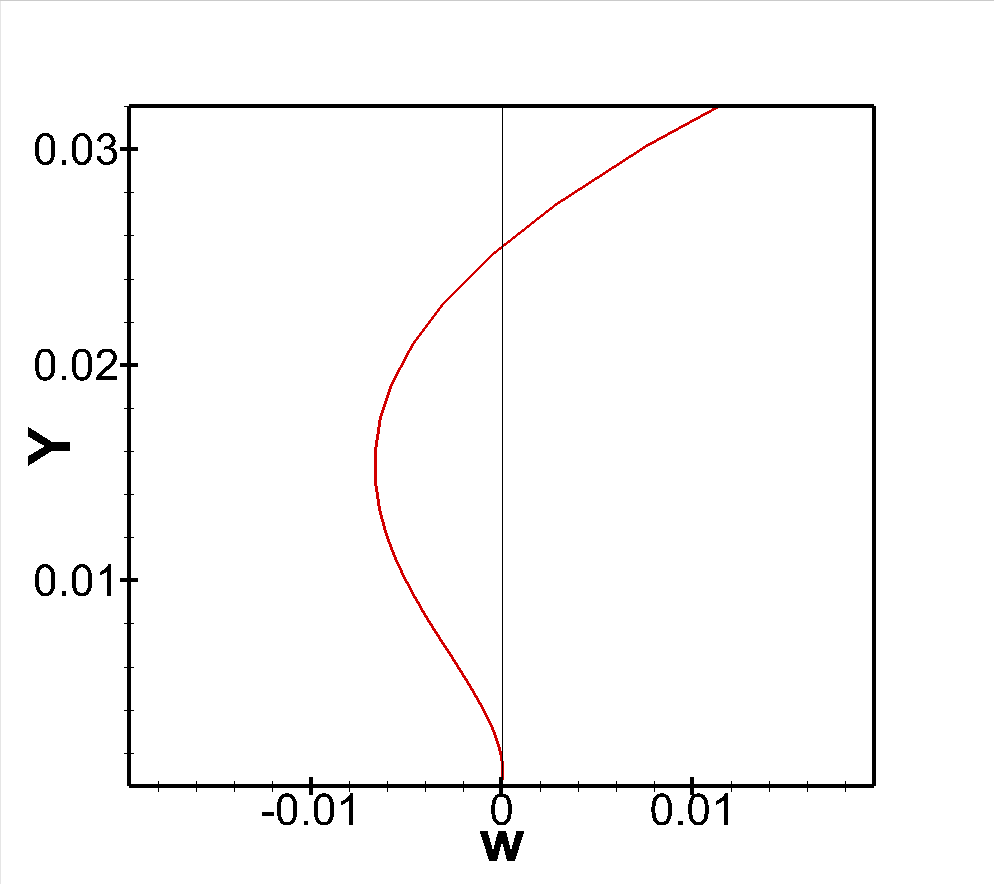} }}
    \qquad
    \subfloat[\centering{\label{figHB2}}]{{\includegraphics[width=3.8cm]{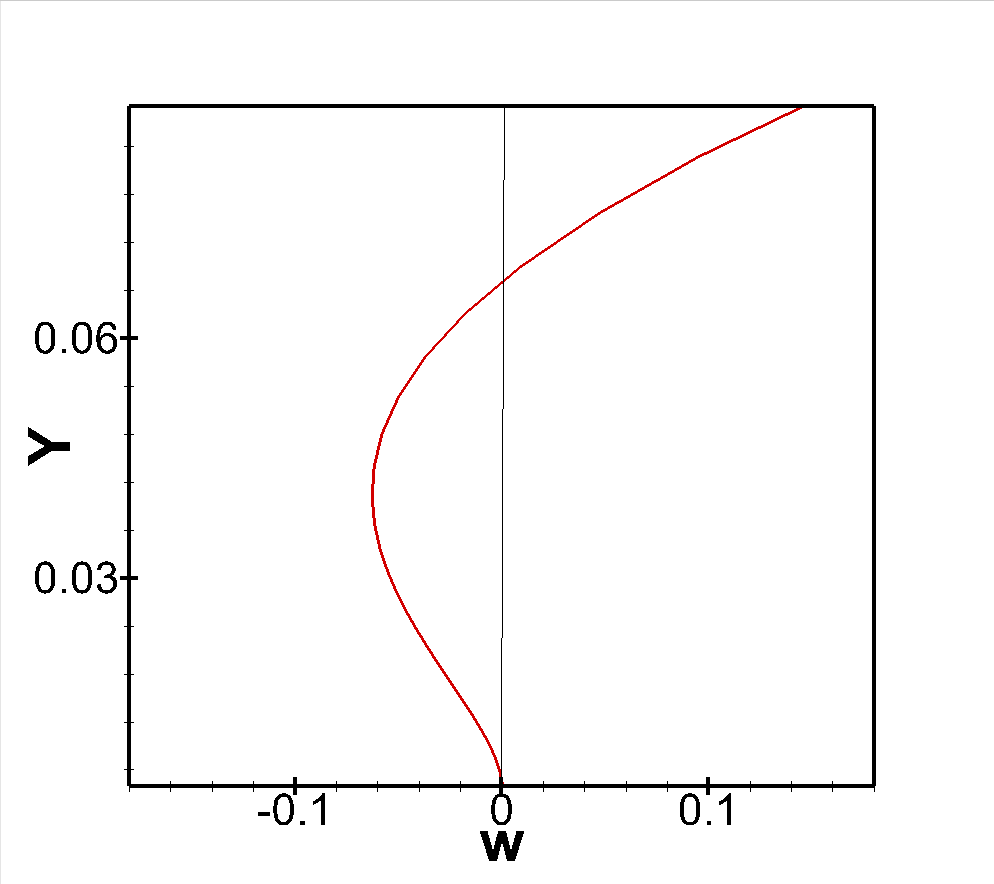} }}
    \qquad
    \subfloat[\centering{\label{figHB3}}]{{\includegraphics[width=3.8cm]{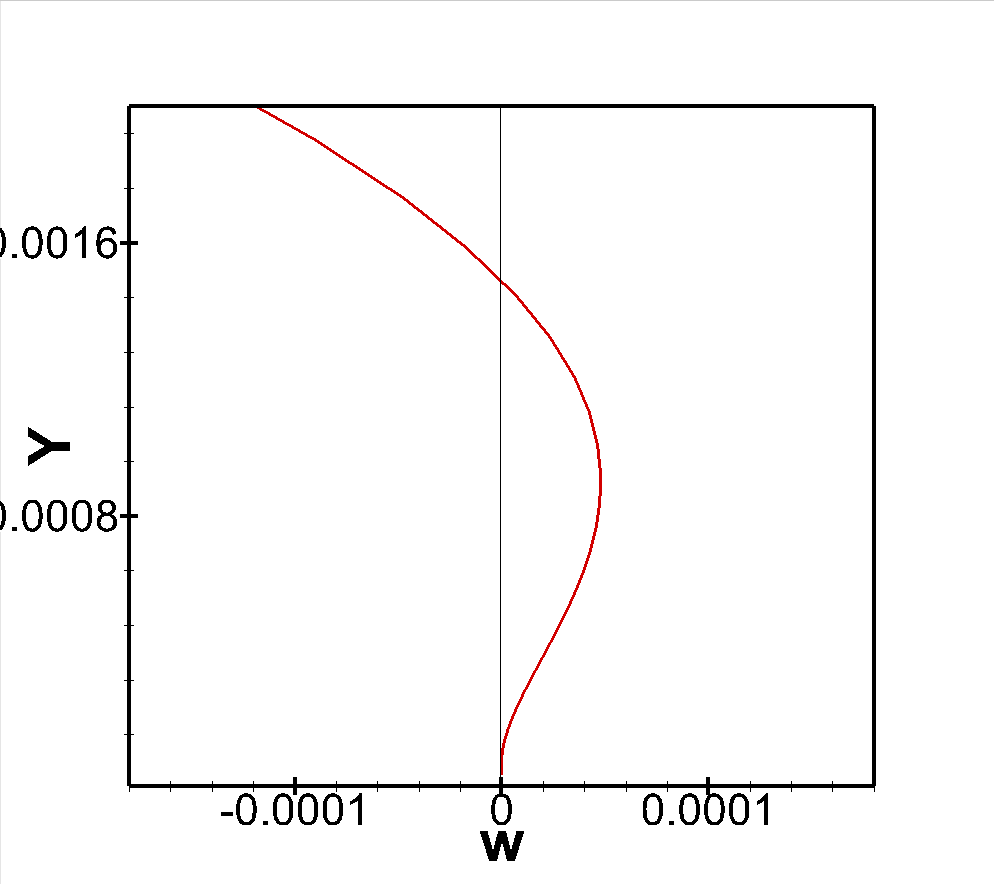} }}
    \qquad
    \subfloat[\centering{\label{figHB4}}]{{\includegraphics[width=3.8cm]{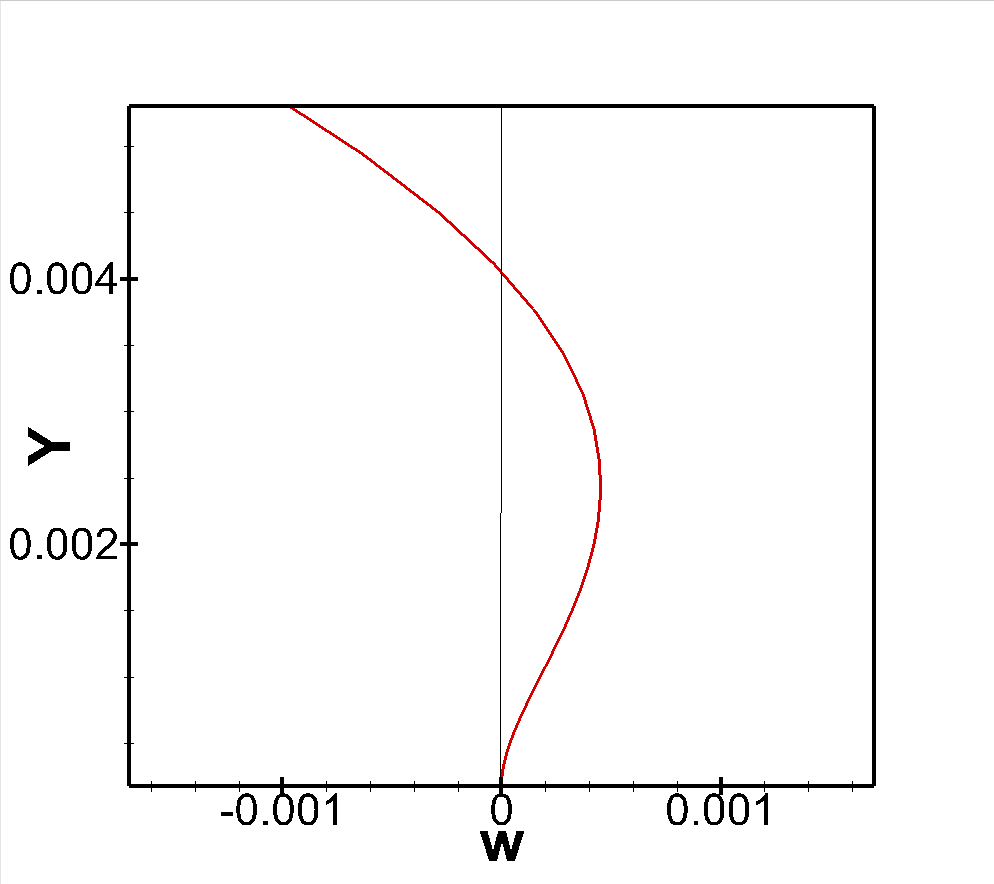} }}
    \qquad
    \subfloat[\centering{\label{figHB5}}]{{\includegraphics[width=3.8cm]{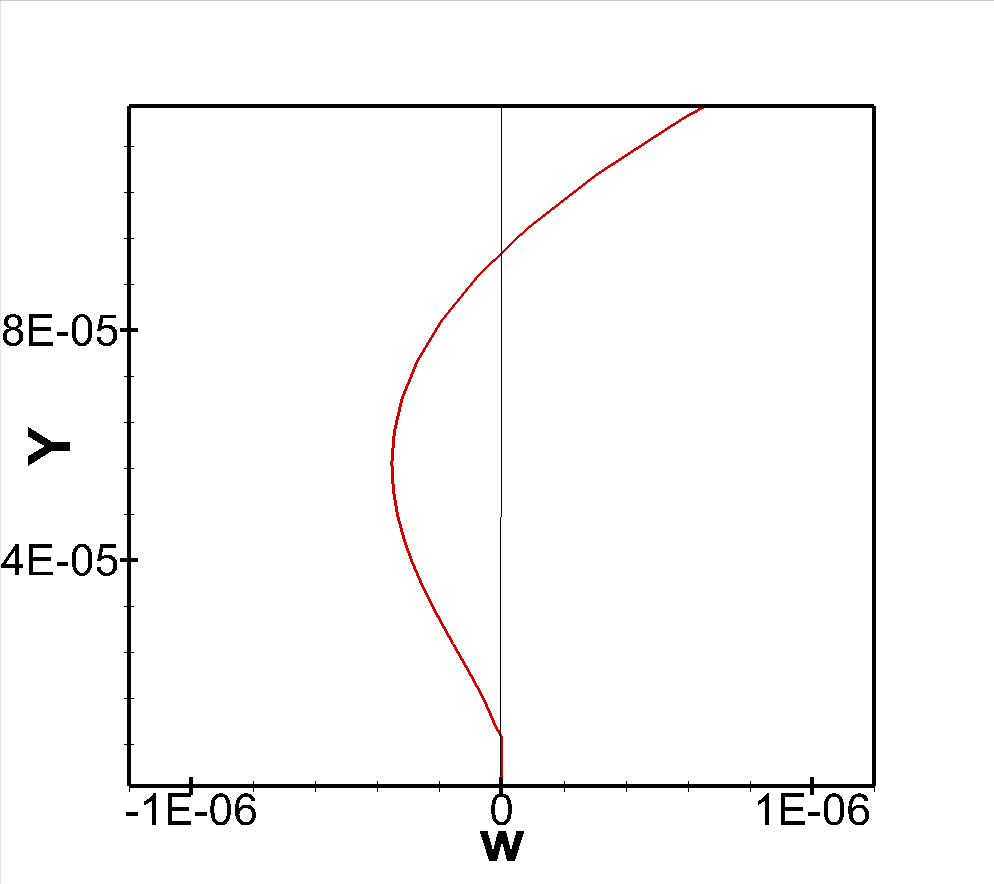} }}
    \qquad
    \subfloat[\centering{\label{figHB6}}]{{\includegraphics[width=3.8cm]{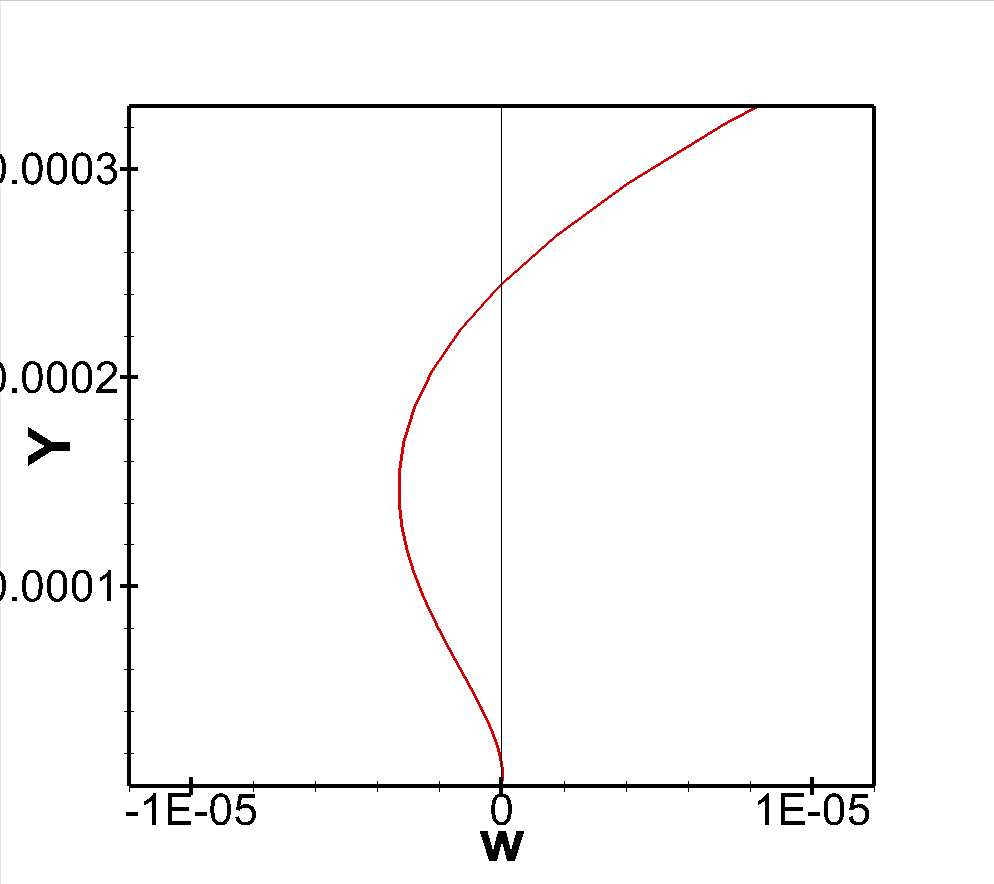} }}
    \qquad
    \subfloat[\centering{\label{figHB7}}]{{\includegraphics[width=3.8cm]{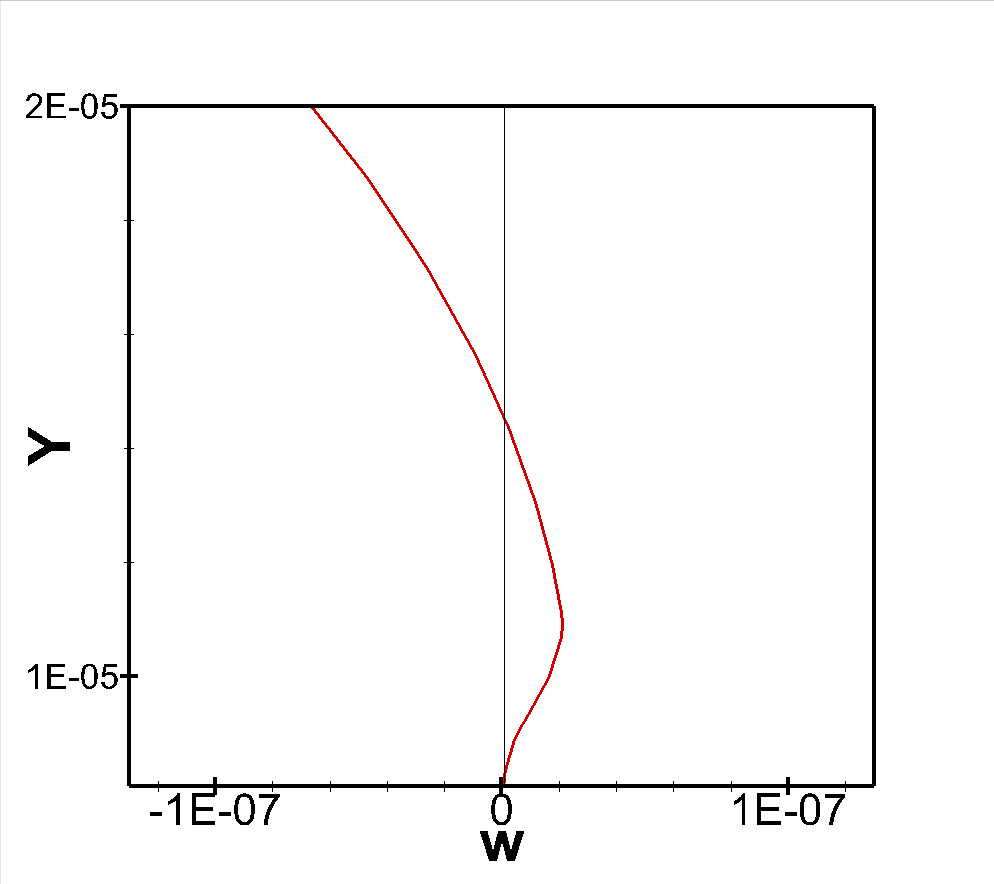} }}
\caption{Self-similarity of corner vortices at $Re = 400$, illustrated by the vorticity distribution within the square cavity. The sequence of left corner vortices is shown in (a) BL1, (c) BL2, and (e) BL3, alongside the right corner vortices in (b) BR1, (d) BR2, (f) BR3, and (g) BR4.}
\label{figHB}
\end{figure*}

\begin{figure*}[!ht]
\centering
    \subfloat[\centering{\label{figHa1}}]{{\includegraphics[width=3.8cm]{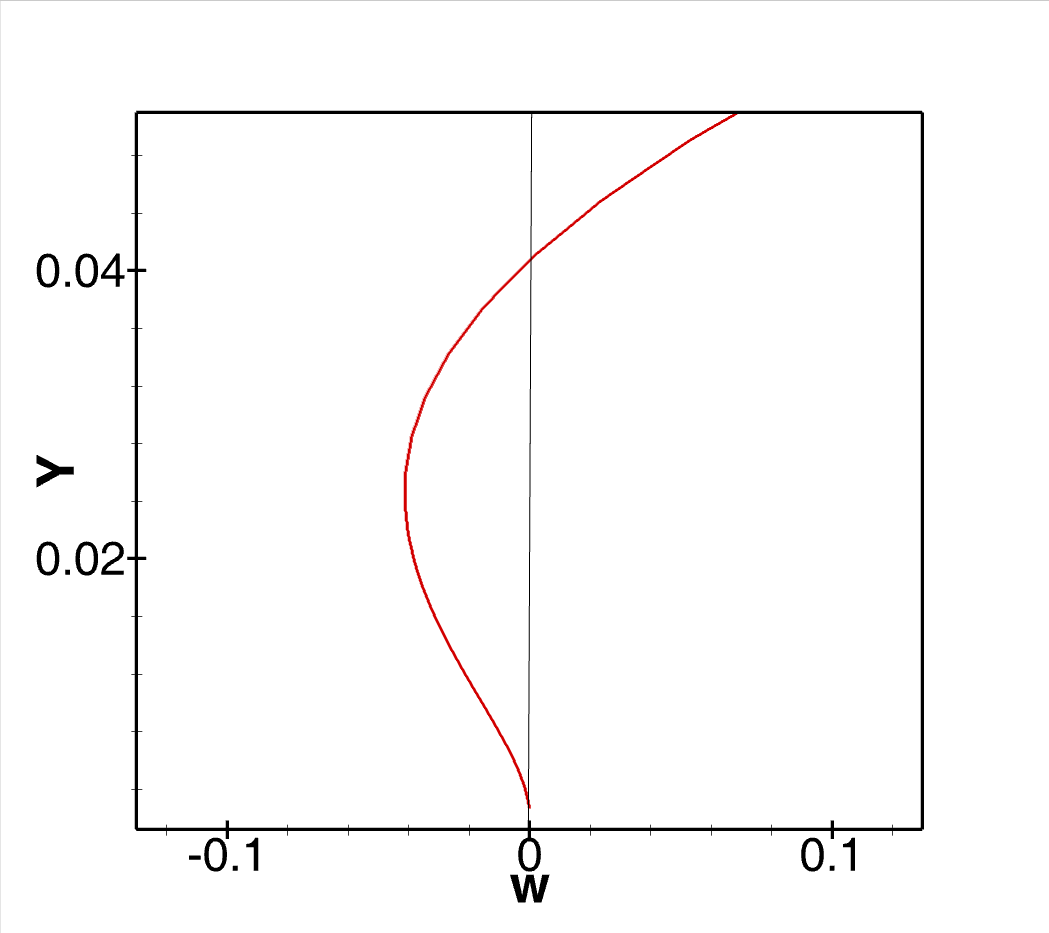} }}
    \qquad
    \subfloat[\centering{\label{figHa2}}]{{\includegraphics[width=3.8cm]{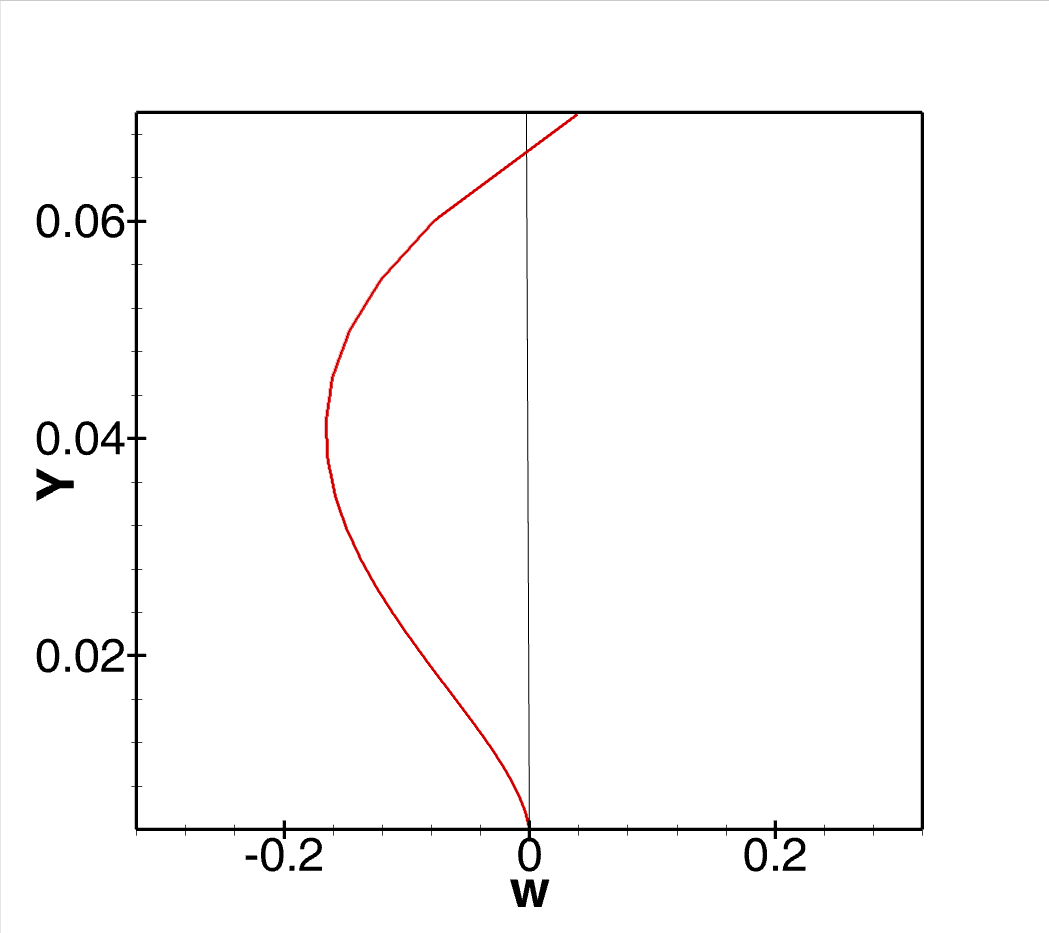} }}
    \qquad
    \subfloat[\centering{\label{figHa3}}]{{\includegraphics[width=3.8cm]{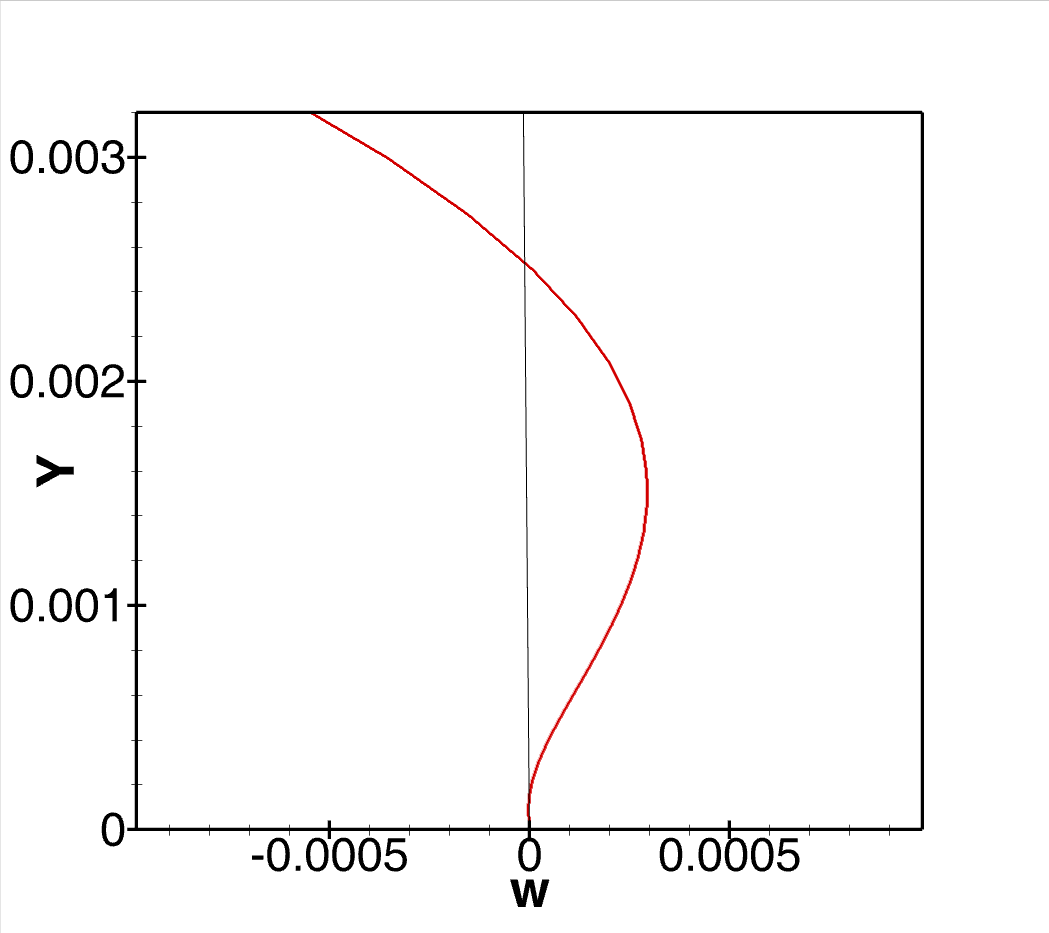} }}
    \qquad
    \subfloat[\centering{\label{figHa4}}]{{\includegraphics[width=3.8cm]{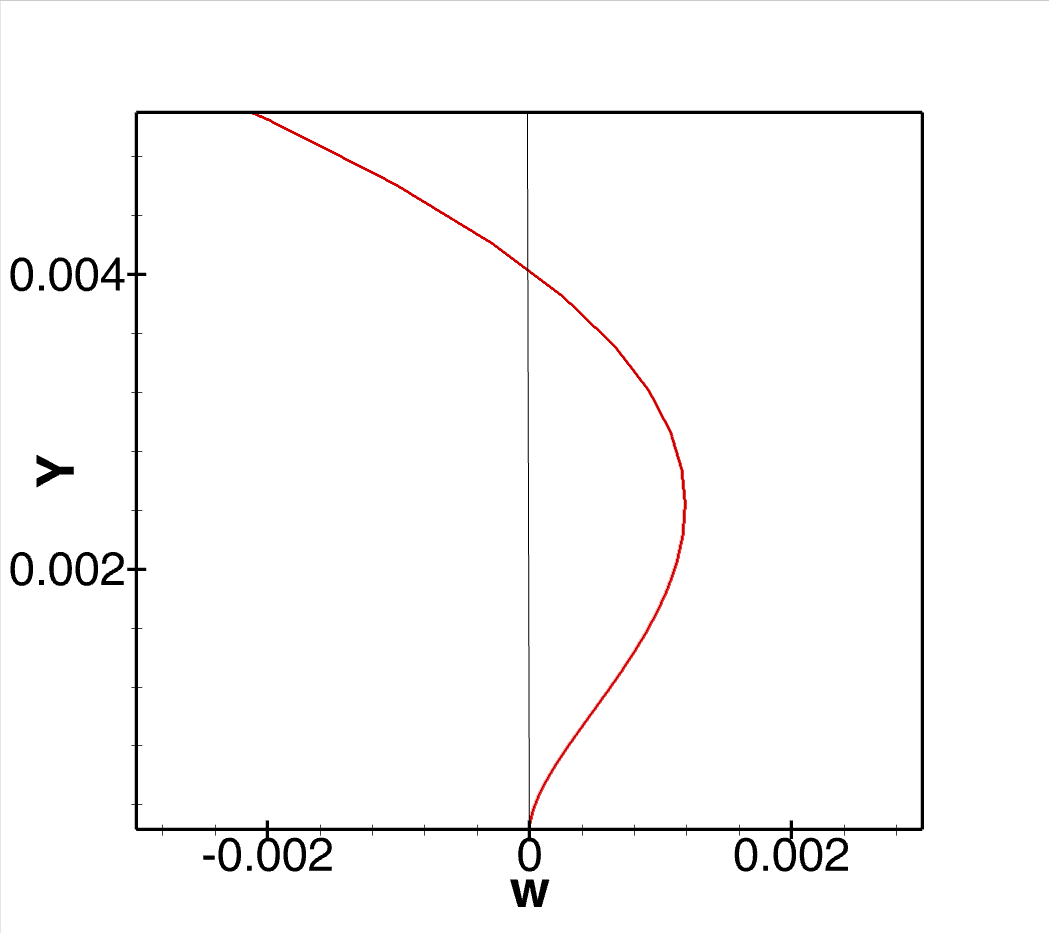} }}
    \qquad
    \subfloat[\centering{\label{figHa5}}]{{\includegraphics[width=3.8cm]{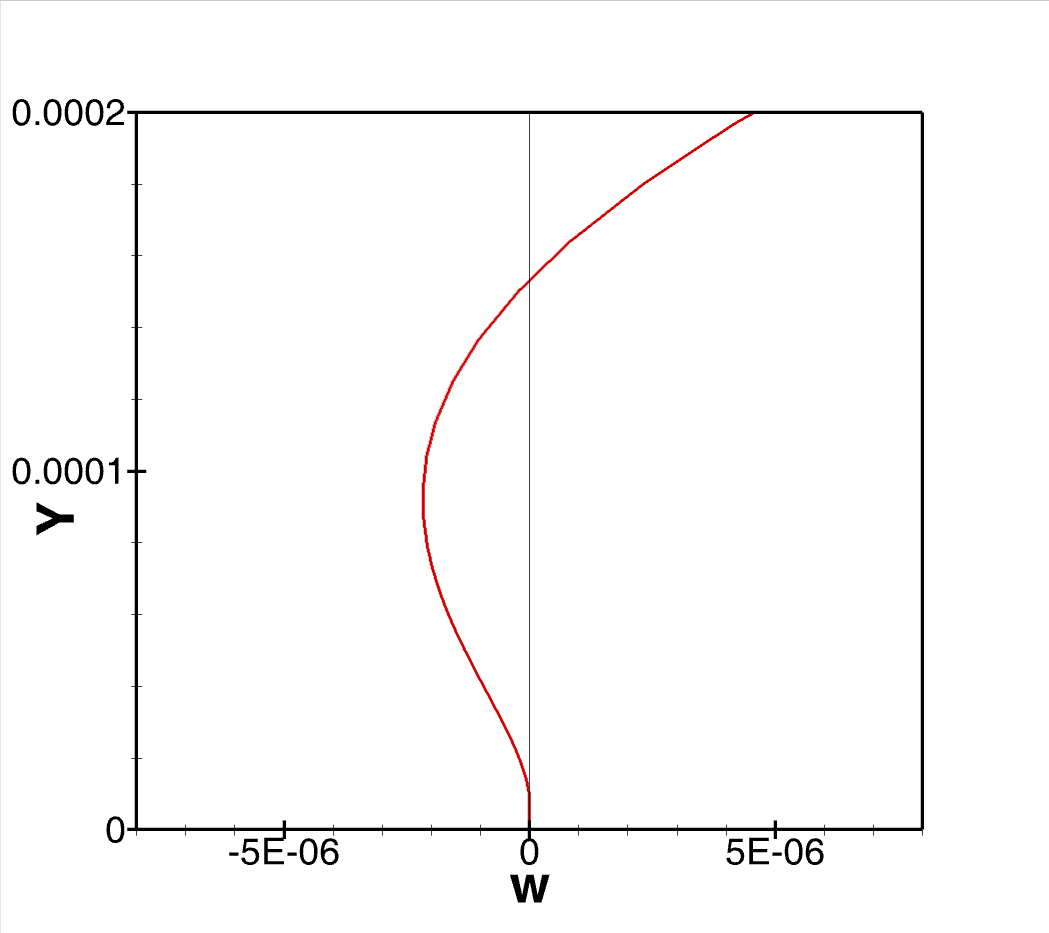} }}
    \qquad
    \subfloat[\centering{\label{figHa6}}]{{\includegraphics[width=3.8cm]{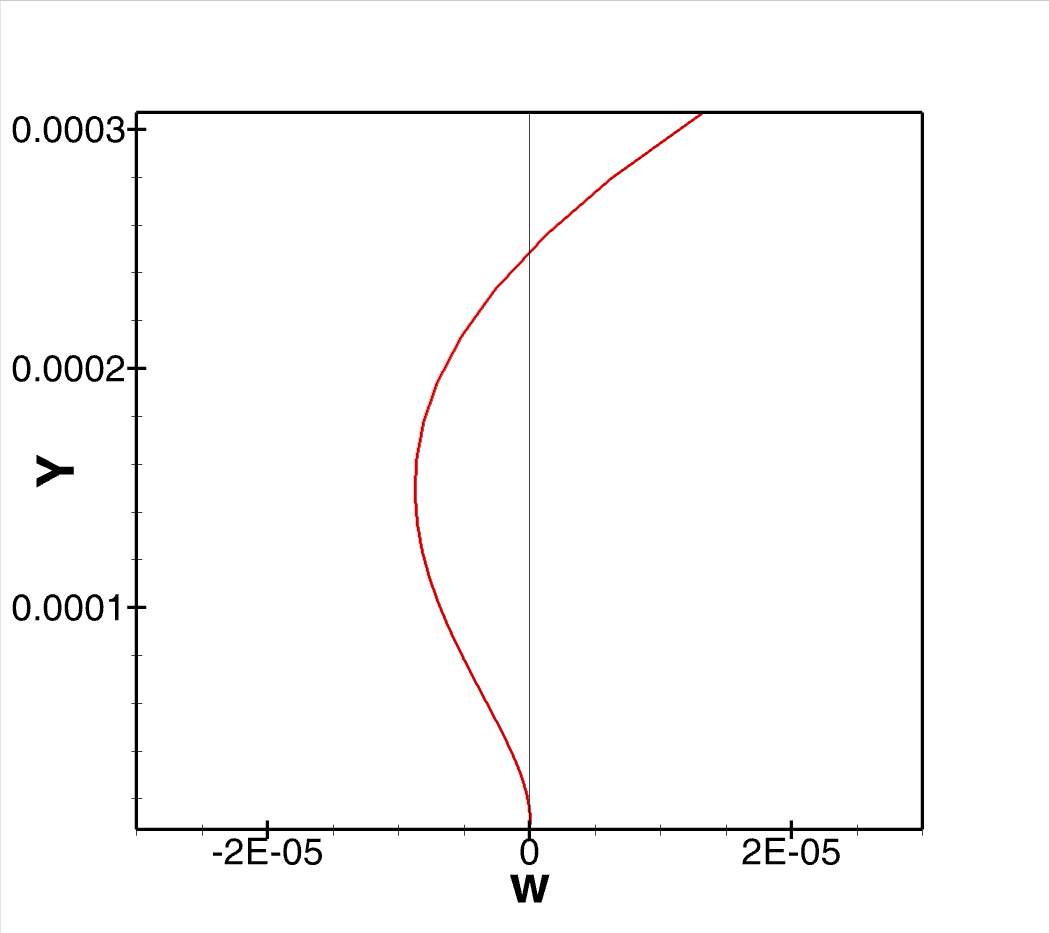} }}
    \qquad
    \subfloat[\centering{\label{figHa7}}]{{\includegraphics[width=3.8cm]{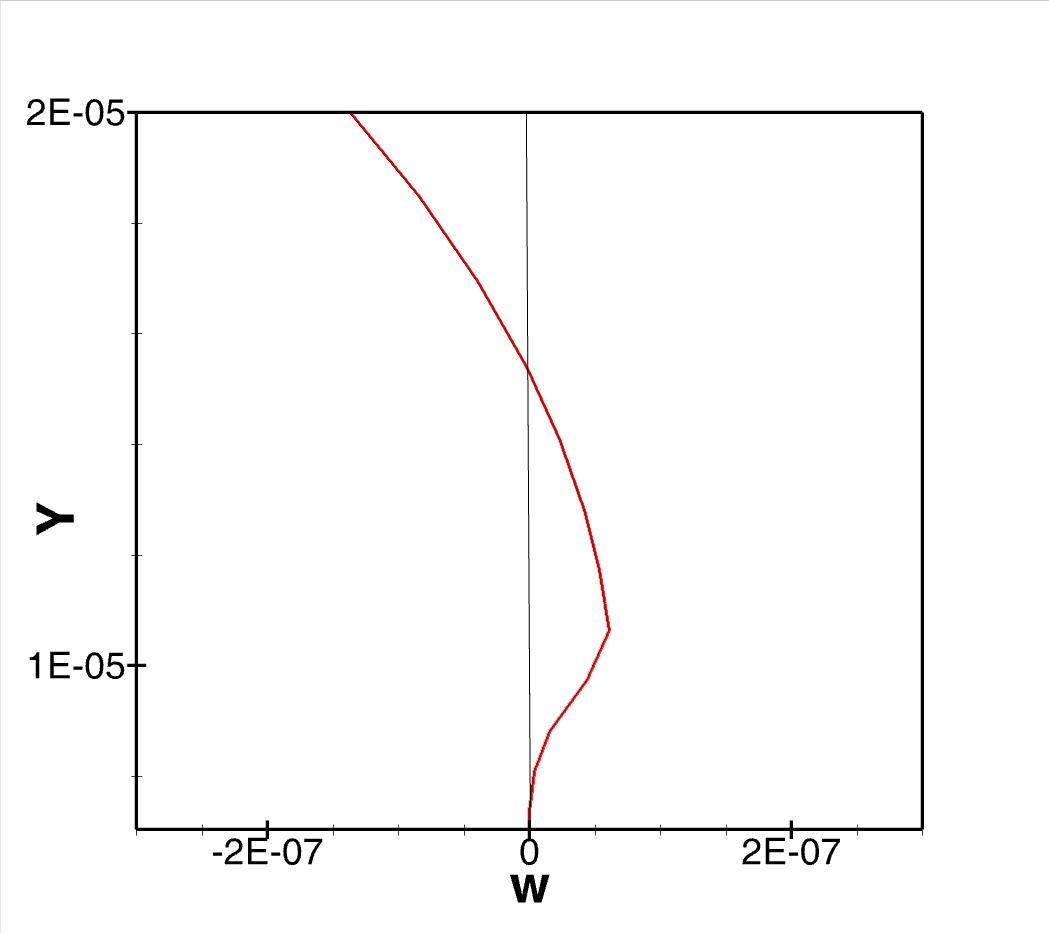} }}
\caption{Self-similarity of corner vortices at $Re = 1000$, illustrated by the vorticity distribution within the square cavity. The sequence of left corner vortices is shown in (a) BL1, (c) BL2, and (e) BL3, alongside the right corner vortices in (b) BR1, (d) BR2, (f) BR3, and (g) BR4.}
\label{figHa}
\end{figure*}

\begin{figure*}[!ht]
\centering
    \subfloat[\centering{\label{figHb1}}]{{\includegraphics[width=3.8cm]{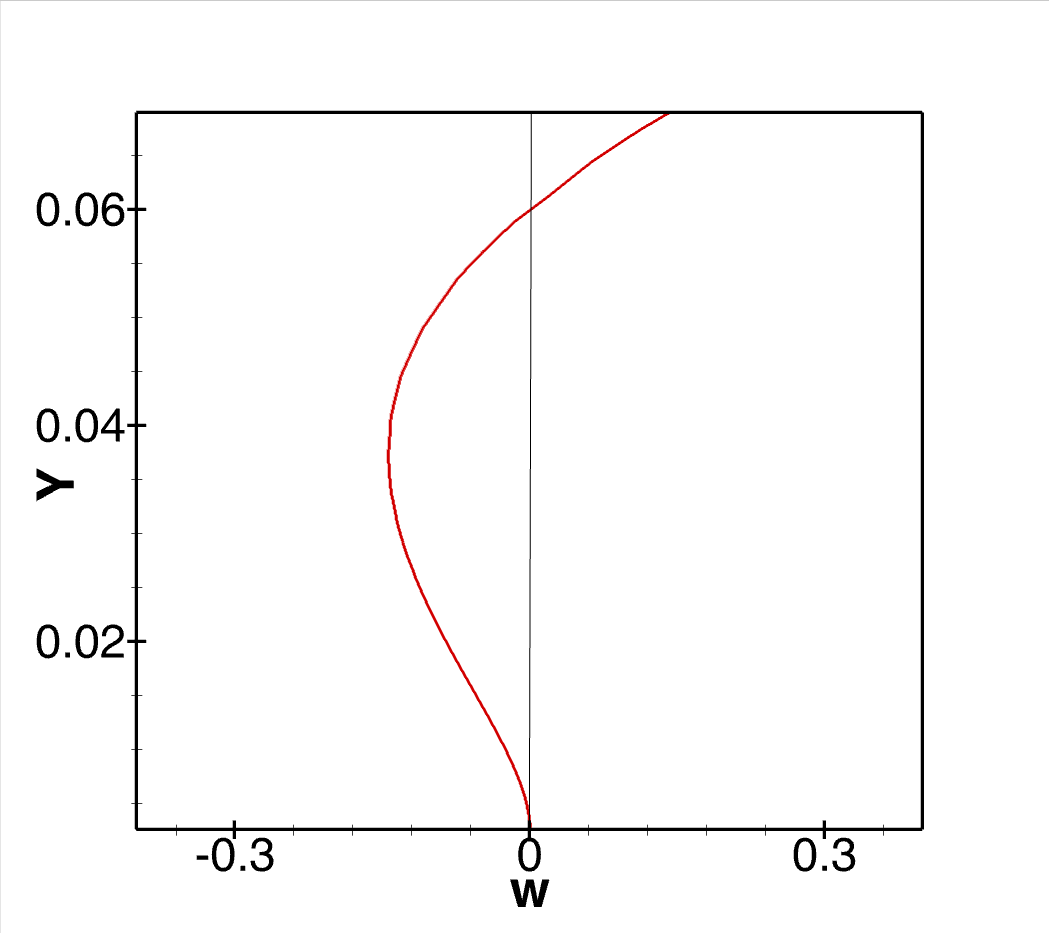} }}
    \qquad
    \subfloat[\centering{\label{figHb2}}]{{\includegraphics[width=3.8cm]{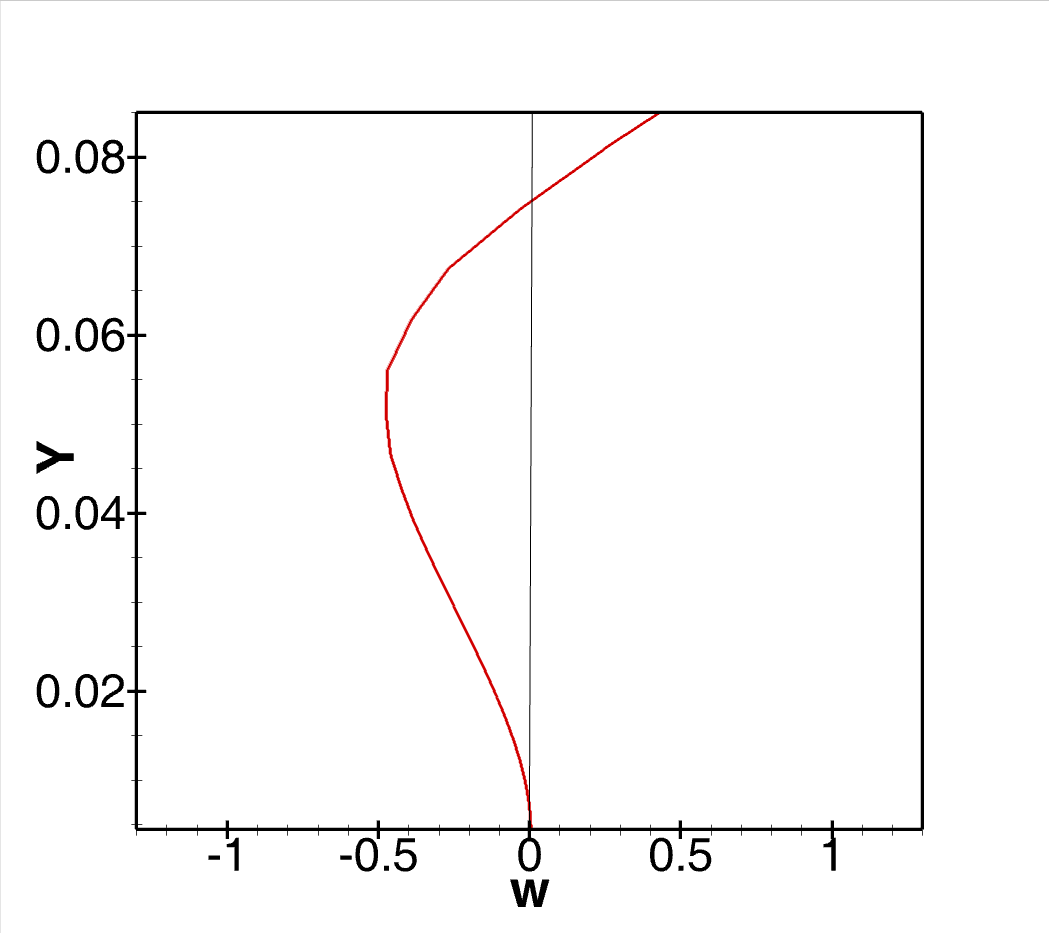} }}
    \qquad
    \subfloat[\centering{\label{figHb3}}]{{\includegraphics[width=3.8cm]{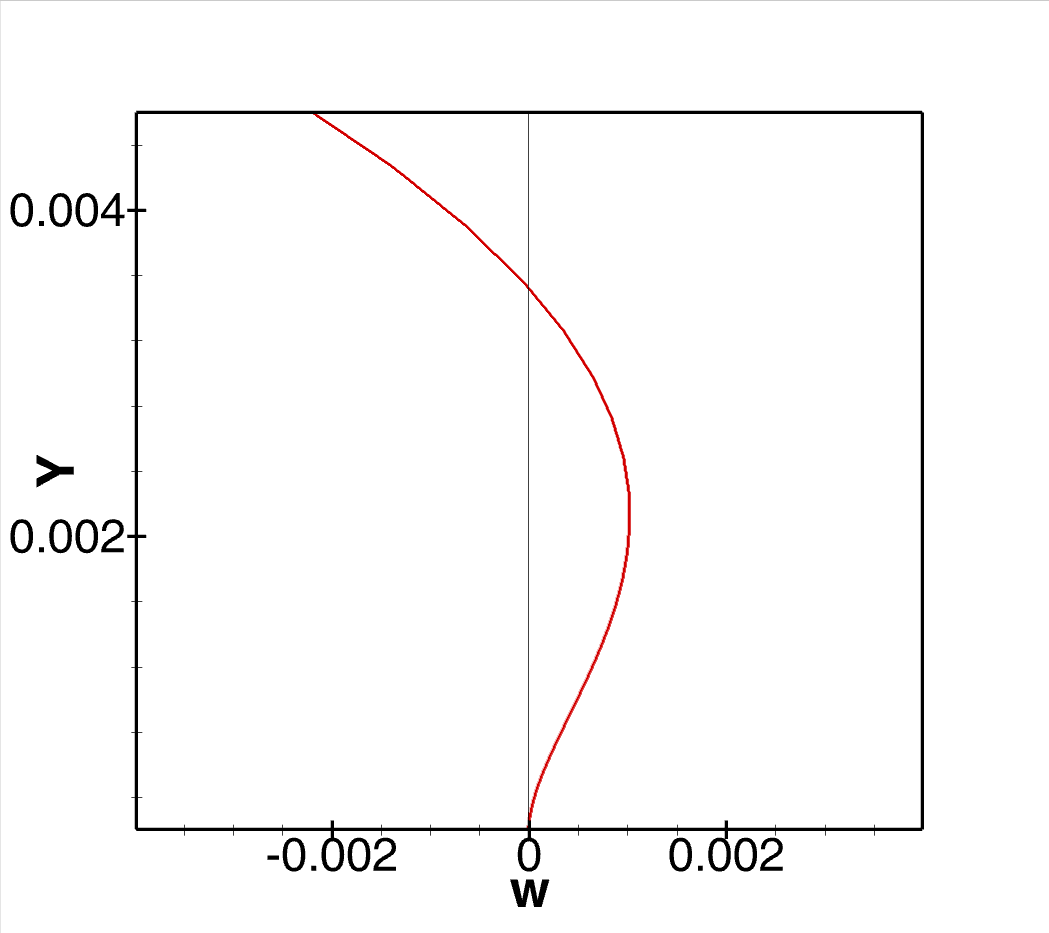} }}
    \qquad
    \subfloat[\centering{\label{figHb4}}]{{\includegraphics[width=3.8cm]{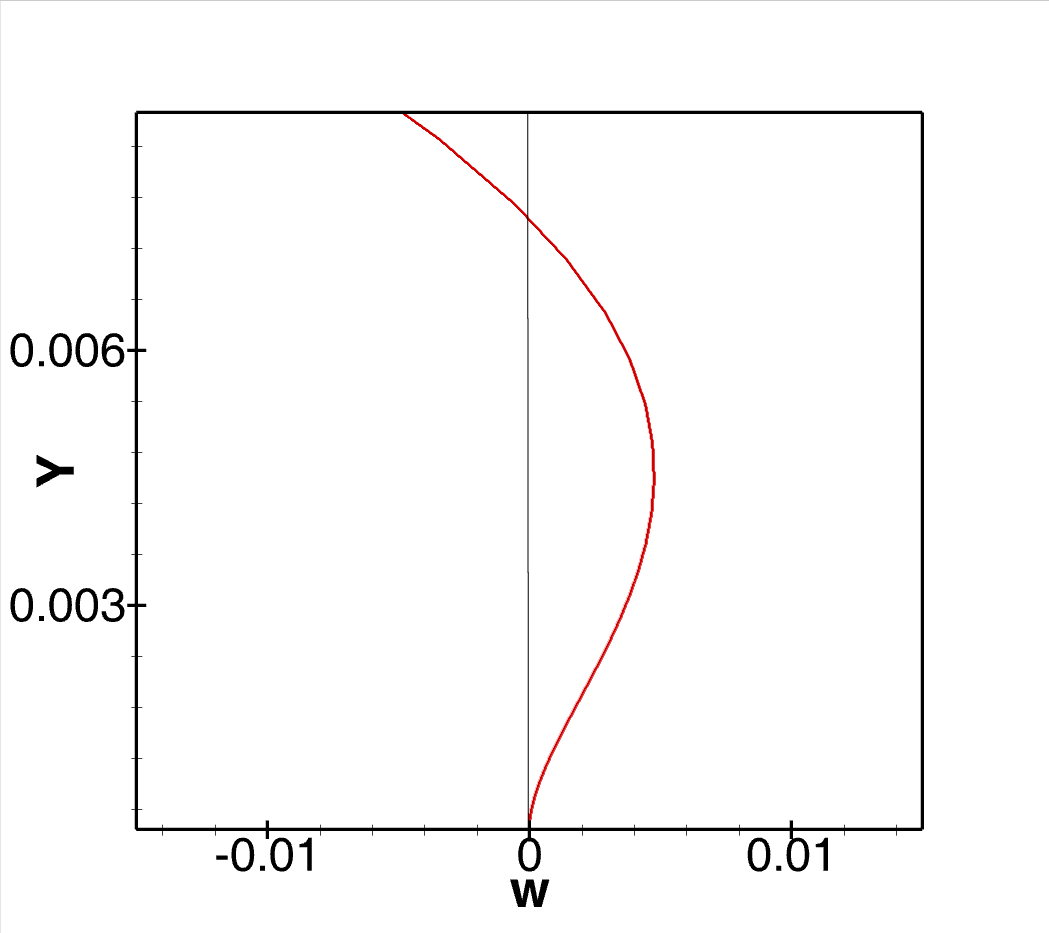} }}
    \qquad
    \subfloat[\centering{\label{figHb5}}]{{\includegraphics[width=3.8cm]{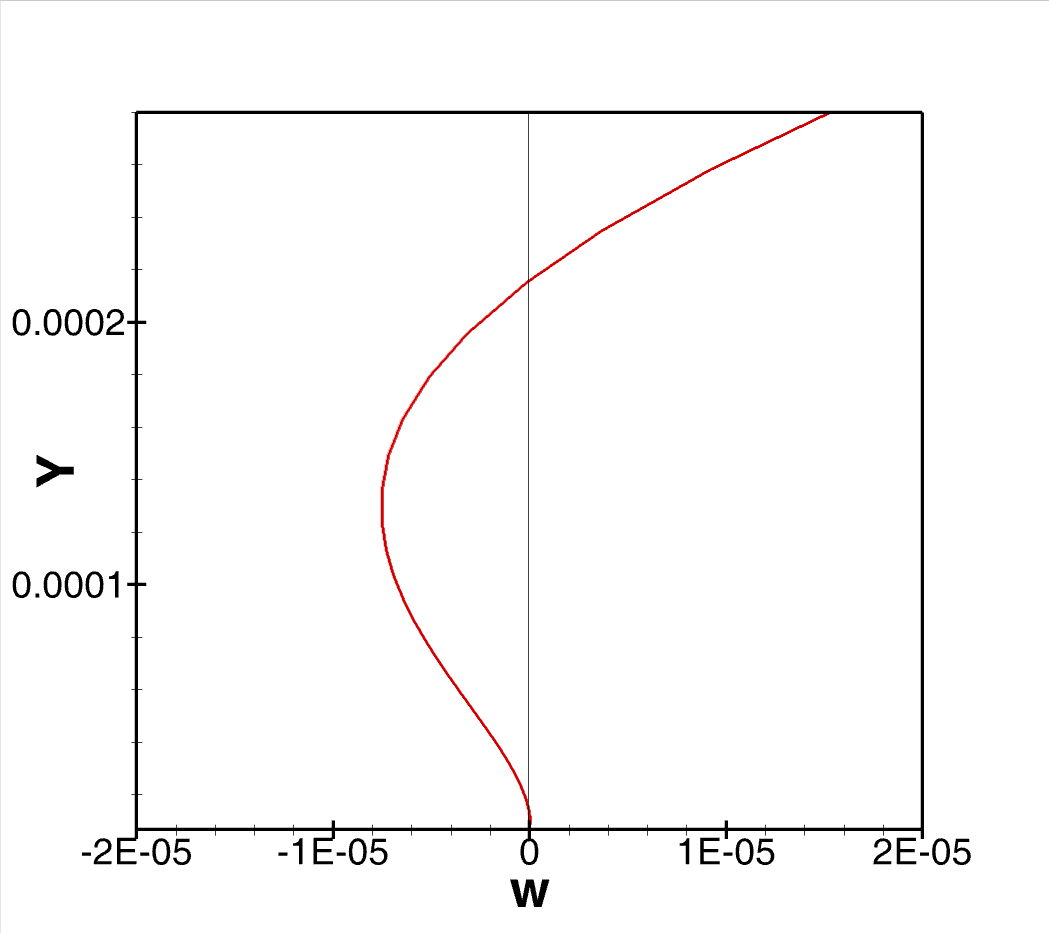} }}
    \qquad
    \subfloat[\centering{\label{figHb6}}]{{\includegraphics[width=3.8cm]{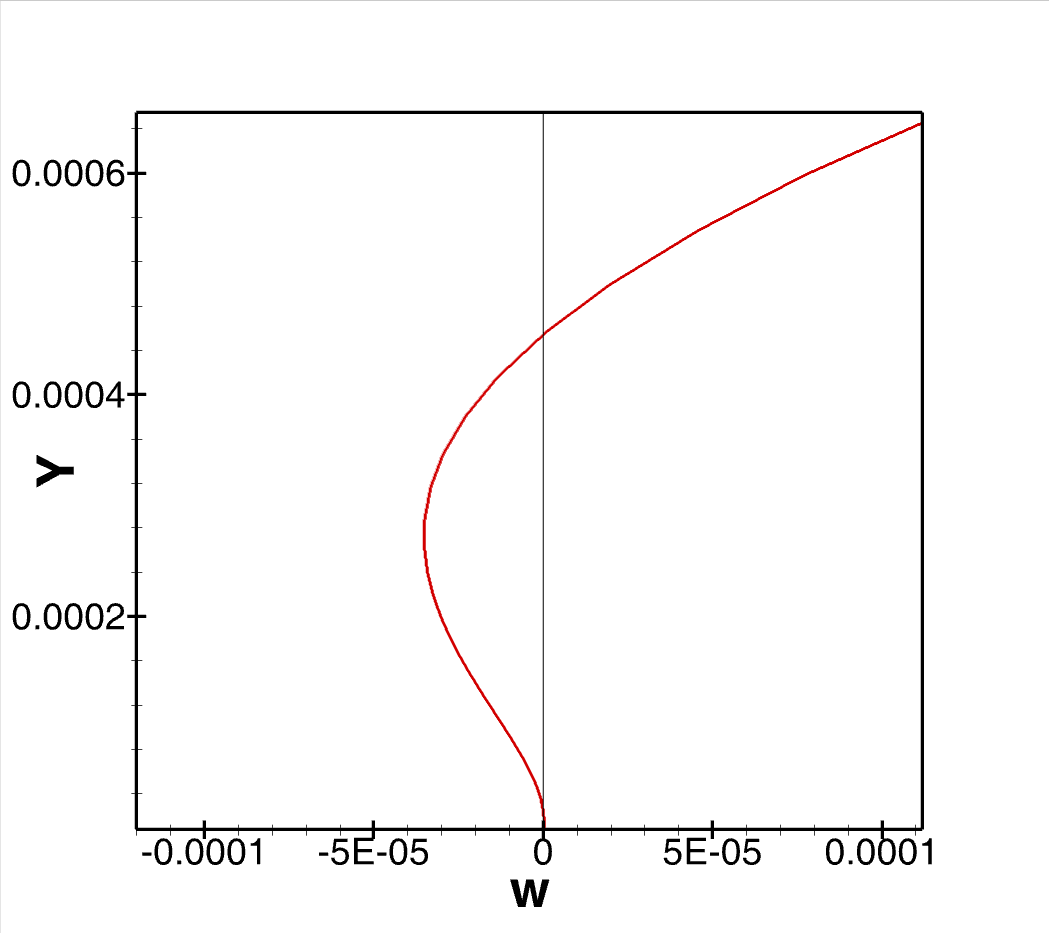} }}
    \qquad
    \subfloat[\centering{\label{figHb7}}]{{\includegraphics[width=3.8cm]{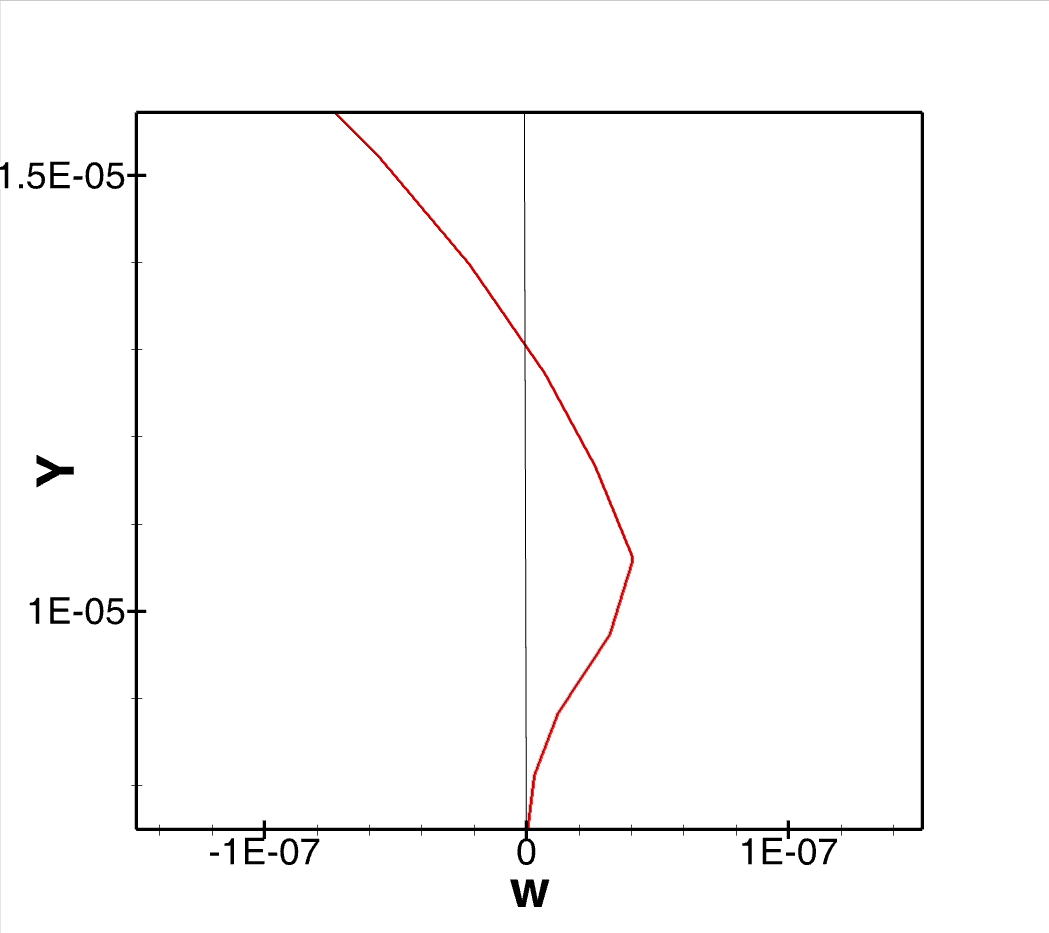} }}
    \qquad
    \subfloat[\centering{\label{figHb8}}]{{\includegraphics[width=3.8cm]{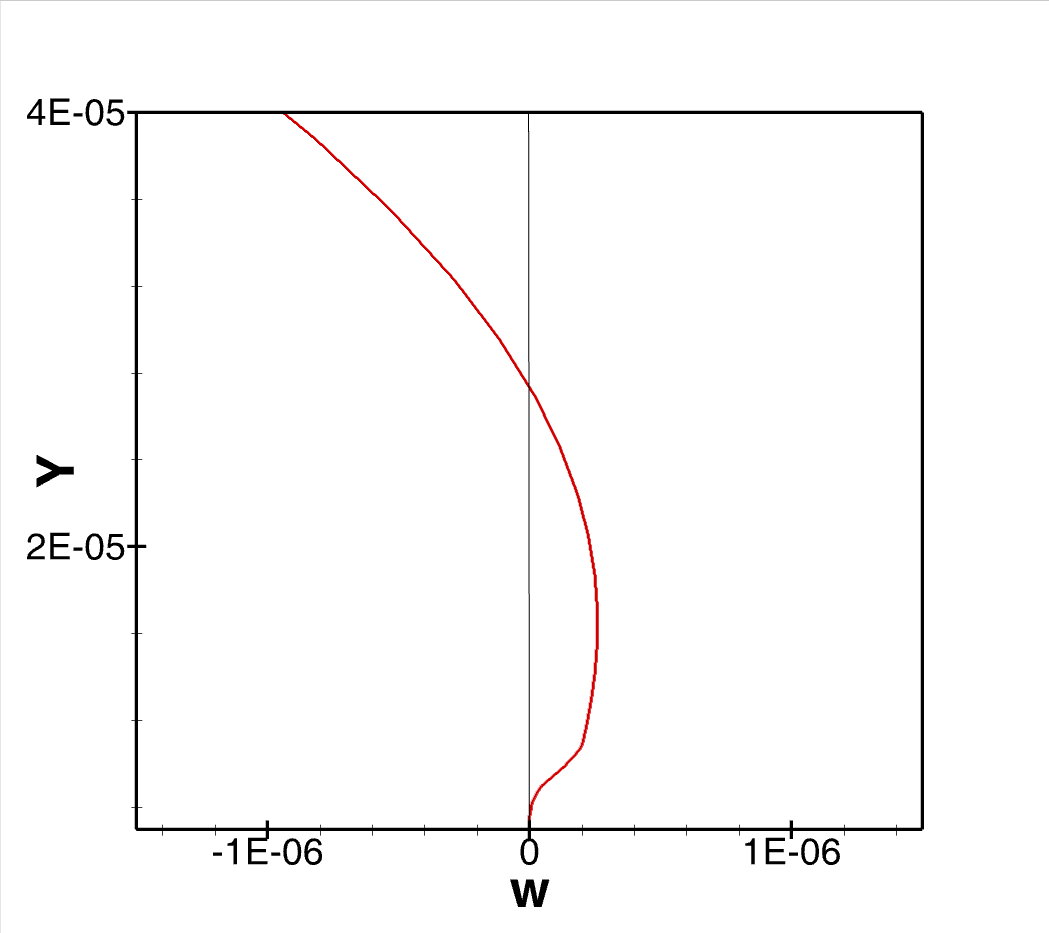} }}
\caption{Self-similarity of corner vortices at $Re = 3200$, illustrated by the vorticity distribution within the square cavity. The sequence of left corner vortices is shown in (a) BL1, (c) BL2, (e) BL3, and (g) BL4, alongside the right corner vortices in (b) BR1, (d) BR2, (f) BR3, and (h) BR4.}
\label{figHb}
\end{figure*}

\par

Finally, we observe the sign changes of vorticity within both the left and right sequences of the square enclosure for the Stokes region and moderate Reynolds number regime. Figures \ref{FigH1}, \ref{FigH2}, \ref{FigH3}, and \ref{FigH4} present the traces of all eight corner vortices observed at the left and right corners of the square cavity for $Re = 100, 400, 1000$, and $3200$, respectively. At $Re = 1, 100, 400,$ and $1000$, vorticity distribution curves display self-similar patterns identical to those demonstrated earlier for the Moffatt vortices in the triangular cavity, as illustrated in Figs. \ref{figH}, \ref{figHA}, \ref{figHB}, \ref{figHa}. The presence of these self-similar vorticity curves confirms the self-similarity of the left and right corner vortex sequences, establishing a key fractal characteristic. For a higher Reynolds number ($Re = 3200$), the vorticity curves in the square cavity exhibit nearly identical patterns across scales, as shown in Fig. \ref{figHb}. The vorticity profiles at the separation boundaries alternate between positive and negative directions; here, the left panels correspond to the left corner vortices, while the right panels represent the right corner vortices. This behavior verifies that, much like in the triangular cavity, the corner vortices within a square cavity retain their structural self-similarity even at significantly higher Reynolds numbers. 

\section{Conclusion}
The present numerical investigation has resolved the nested corner-flow structures in a triangular cavity and established the existence of a self-similar vortex cascade consistent with Moffatt’s theory. The coupled solver captures six distinct corner vortices, while the finest grid resolution indicates the incipient formation of a seventh vortex near the cavity apex. The computed vorticity fields are in good agreement with the experimental observations of Taneda \cite{taneda1979visualization} and the analytical predictions of Moffatt \cite{moffatt1964viscous,moffatt1964viscous2}, thereby lending strong physical credibility to the resolved structures.

A major contribution of this work is the fractal interpretation of Moffatt vortices. Using the area--perimeter method, the successive vortices are shown to possess non-integer fractal dimensions, revealing persistent geometric complexity across scales. An empirical relation is formulated to estimate the fractal dimension of any successive vortex in the cascade, and this expression is further generalized to arbitrary grid resolution. This provides a practical and systematic framework for quantifying multiscale corner vortices in confined viscous flows. The analysis also shows that the fractal dimension varies with Reynolds number and is closely associated with changes in vortex size and intensity.

The comparative study of triangular and square cavities under Stokes and moderate-Reynolds-number conditions demonstrates similar self-similar scaling in the successive corner vortices. This indicates that the observed fractal organization is a robust feature of confined viscous flow rather than a geometry-specific artifact. At the same time, the present approach remains limited by the mesh sensitivity of the area--perimeter method, particularly for smaller and deeper vortices, and therefore requires sufficiently fine spatial resolution for reliable extraction of fractal dimension.

Taken together, these results establish that corner vortices in triangular and square cavities are not merely nested flow features but constitute a genuine fractal cascade with measurable geometric scaling. The proposed framework opens a useful route for characterizing multiscale vortex structures in confined viscous flows and may serve as a basis for future investigations in more complex geometries and higher-Reynolds-number regimes.
\section*{Acknowledgments}

The first author is thankful to the UGC, India, for providing financial support under the Senior Research Fellowship program (NTA Ref. No. 211610162054).

\section*{Data Availability Statement}
The data that support the findings of this study are available from the corresponding author upon reasonable request.

\section*{Conflict of interest}
The authors declare that there is no conflict of interest in the present study.

\bibliography{sn-bibliography}

\end{document}